\documentstyle[12pt,aaspp4]{article}
\def\deg{\arcdeg}
\begin{document}

\title{Deep WFPC2 and Ground-based 
Imaging of a Complete Sample of 3C Quasars and Galaxies\altaffilmark{1}}

\altaffiltext{1}{Based
on observations made with the NASA/ESA Hubble Space Telescope, obtained
at the Space Telescope Science Institute, which is operated by the 
Association of Universities for Research in Astronomy, Inc., under
NASA contract NAS 5-26555.}

\author{Susan E. Ridgway\altaffilmark{2,3,4} and
Alan Stockton\altaffilmark{3,4}}
\affil{Institute for Astronomy, University of Hawaii, 2680 Woodlawn Drive,
Honolulu, Hawaii  96822\\Electronic mail:  s.ridgway1@physics.oxford.ac.uk,
stockton@uhifa.ifa.hawaii.edu}
\altaffiltext{2}{Current address:  University of Oxford, Department of Physics,
Nuclear and Astrophysics Laboratory, Keble Road, Oxford, OX1 3RH.}
\altaffiltext{3}{Visiting Astronomer, W.M. Keck Observatory, jointly operated
by the California Institute of Technology and the University of California.}
\altaffiltext{4}{Visiting Astronomer, Canada-France-Hawaii Telescope, 
operated by the National Research Council of Canada, the Centre National de la
Recherche Scientifique de France, and the University of Hawaii.}

\begin{abstract}
\tighten
We present the results of an HST and ground-based imaging study of a
complete 3C sample of $z\sim1$ sources, comprising 5 quasars and 5
radio galaxies.  We have observed all of the sample
in essentially line-free bands at rest-frame 0.33 $\mu$m with 
WFPC2 and in rest-frame
1 \micron\ images from the ground; we have also observed
most of the sample in narrow-band filters centered on [\ion{O}{2}]. 
We resolve continuum structure around all
of our quasars in the high-resolution WFPC2 images, and in four of
the five ground-based $K^\prime$ images.
All of the quasars have some optical continuum structure that is aligned with
the radio axis. In at least 3 of these cases, some
of this optical structure is directly coincident with a portion of
the radio structure, including optical counterparts to radio jets in
3C\,212 and 3C\,245 and an optical counterpart to a radio lobe in 3C\,2.
These are most likely due to optical synchrotron
radiation, and the radio and optical spectral indices in the northern 
lobe of 3C\,2 are consistent with this
interpretation.

The fact that we see a beamed
optical synchrotron component in the quasars but not in the radio galaxies
complicates both the magnitude and the alignment comparisons.
Nonetheless, the total optical and $K'$ flux densities of the quasar hosts are
consistent with those of the radio galaxies within the observed dispersion
in our sample.
The distributions of $K'$ flux densities of both radio galaxies
and quasar hosts exhibit similar mean and dispersion to that found for other
radio galaxies at this redshift, and the average host galaxy luminosity
is equivalent to, or a little fainter than, $L^*$.
The formal determination of the alignment in the optical 
and infrared in the two subsamples yields no significant difference
between the radio galaxy and quasar subsamples,
and the quasars 3C\,196 and 3C\,336 have aligned
continuum and emission-line structure that is probably not due to beamed
optical synchrotron emission.

Very blue and/or edge-brightened structures are present in some objects 
within the probable quasar opening angle; these are possibly the
result of illumination effects from the active nucleus, {\it i.e.,} scattered
quasar light or photoionization.
In 3C\,212, we see an optical object that lies 3\arcsec\ 
{\it beyond} the radio lobe, but which
looks morphologically quite similar to the radio lobe itself.  This
object is bright in the infrared and has a steep spectral gradient along its 
length.  A striking, semi-circular arc
seen associated with 3C\,280 may possibly be a tidal tail
from a companion, enhanced in brightness by scattering or photoionization.

In the near-infrared, most of the radio galaxies have elliptical morphologies
with profiles that are well-fit by de Vaucouleurs r$^{1 \over 4}$-laws
and colors that are consistent with an old stellar population.
All components around the quasars have optical-infrared colors that
are redder than or similar to the colors of their respective nuclei; this
is more consistent with a stellar origin for the emission than 
with a dominant scattering contribution.

From the correspondence between the total magnitudes in the
galaxies and quasars and the detection of aligned components in
the quasars, we conclude that this study provides general support for the
unification of FR II radio galaxies and quasars.  
Some of the objects 
in the sample ({\it e.g,} 3C\,212) have properties 
that may be difficult to explain with our current understanding of
the nature of FR II radio sources and the alignment effect.

\end{abstract}

\section{Introduction}
\label{sec:intro}
Powerful radio sources, and the objects associated with them, provide
probes of
the Universe from $z \sim 1$ up to the highest redshifts at which
discrete objects have been detected.
Over the past decade, observations of radio sources at high redshift
have focussed primarily on the
galaxies, since
the extended hosts of quasars at those redshifts are much more difficult to observe.
We will present here the results of a systematic imaging survey of a 
complete sample of $z \sim 1 $ radio galaxies and quasars; first we will briefly
review the background and motivation for this work.

\subsection{High Redshift Radio Galaxies}
Observations of high-$z$ radio galaxies have been reviewed by McCarthy \markcite{mcC93}(1993); 
here we summarize and update the aspects most relevant to our investigation.
Radio galaxies at $z \ga 0.8$ often have both distorted, multi-component
optical continuum structures
and associated emission-line regions
extending many tens of kpc. 
Both the emission-line and the continuum shapes
tend to be elongated along the direction 
of the radio axis (Chambers, Miley, \& van Breugel \markcite{cha87}1987; 
McCarthy et al.\ \markcite{mcC87}1987), whereas similar alignments
have been observed only rarely in the lower-redshift 
($z<0.6$) 3C sample (Baum \&
Heckman \markcite{bau89}1989; McCarthy \& van Breugel \markcite{mcC89}1989). 

This ``alignment effect'' also seems to be
a function of wavelength; near-infrared morphologies of $z \sim 1$
radio galaxies are rounder and the strength of the aligned component
is generally less, though in many cases still significant (Rigler et al.\
\markcite{rig92}1992;  Dunlop \& Peacock \markcite{dun93}1993).
It cannot be assumed, of course, that the differences between the 
high and low redshift galaxies arise solely
from evolutionary effects.
Optical imaging of high redshift galaxies corresponds to the rest-frame 
ultraviolet, which is observable only with great difficulty from 
the ground for their low-$z$ counterparts. 
Another difference between the high-$z$ and low-$z$ samples
is that the radio power of the low-redshift sources in a flux-limited sample is systematically lower
than that of the high-$z$ objects. 
Observations of morphologies of samples of low radio
power galaxies at high redshift have found no evidence for significant
alignment, indicating a radio power dependence to the alignment effect 
(Dunlop \& Peacock \markcite{dun93}1993; Eales et al.\ \markcite{eal97}1997).  
 
A simple explanation for this correlation of optical and infrared morphologies
with the radio axis remains elusive, and observations 
probably rule out any single cause.
Early attempts were made to interpret the morphologies of high-redshift 
radio galaxies in terms of interactions ({\it e.g.,} Djorgovski et al.\ \markcite{djo87}1987),
which seemed an attractive and natural way to explain 
the observed distorted morphologies and close companions. 
Such explanations were largely abandoned 
with the discovery of the alignment effect, because it seemed unlikely
that the interaction axis and the radio axis would be naturally correlated
with each other (but see West 1994). 

Currently, two processes are generally considered as the most probable 
causes of 
the alignment effect: jet-induced star formation, and the scattering of
light from an obscured quasar. 
High polarizations ($\sim10$\%)
have been confirmed in 3C\,277.3, 3C\,324, 3C\,256, 3C\,265 and in a number of other
radio galaxies (Cimatti et al.\ \markcite{cim93}1993, \markcite{cim96} 
1996; Dey et al.\ \markcite{dey96}1996). Direct detections of 
broad \ion{Mg}{2} lines in polarized light from the aligned continuum in
3 high-$z$ radio galaxies
(di Serego Alighieri et al.\ \markcite{diS94}1994), and
detection of broad components to \ion{Mg}{2} $\lambda$2798 and
\ion{C}{3}] $\lambda$1909 in deep spectra of 3C\,324
(Dickinson, Dey \& Spinrad \markcite{dicm95}1995)
make it very likely that scattered quasar radiation 
is a significant component of the aligned optical light 
in at least these galaxies (Fabian \markcite{fab89}1989; 
Cimatti et al.\ \markcite{cim93}1993; di Serego Alighieri et al.\ 
\markcite{diS94}1994).

However, high-resolution HST WFPC2 imaging of $z \sim 1 $ radio galaxies
generally has not revealed a ``scattering cone'' morphology as is seen in 
low redshift AGN such as NGC 1068; the alignment comes in many cases from discrete
lumps that are 
closely confined to the track of the radio jet (Longair et al.\
\markcite{lon95}1995; Dickinson et al.\ \markcite{dicm95}1995;
Best, Longair \& R\"ottgering \markcite{bes96}1996).  
This morphological evidence tends to favor 
the suggestion that the aligned emission is dominated by star
formation induced by the radio jet
(Chambers et al.\ \markcite{cha87}1987; Begelman \& Cioffi 
\markcite{beg89}1989;
de Young \markcite{deY89}1989; Rees \markcite{ree89}1989; Daly 
\markcite{dal90}1990). 
Longair et al.\ \markcite{lon95}(1995) and  
Best et al.\ \markcite{bes96}(1996) find a variety of morphologies 
in their sample of 8 $z \sim 1 $ 
galaxies, 
and, though 
most exhibit some degree of alignment, the nature of the 
correspondence with the radio structure varies from source to source.
The smaller scale radio sources such as 3C\,324 and 3C\,368
have structures aligned close to the radio axis, while the larger scale radio
sources tend to have material that appears more relaxed, exhibiting less alignment.
Best et al.\ speculate that this is an age effect: if the very closely
aligned structures seen in the smaller, younger radio sources are the result of
radio-jet-induced star formation, this activity would cease 
after the radio lobes have passed outside of the host galaxy, allowing the stars formed to relax into
the compact morphologies observed in the larger, and presumably 
older radio sources. 

There is so far no
conclusive evidence for a case in which stellar emission in likely
to dominate an aligned component, in spite of the early 
claim of Chambers \& McCarthy \markcite{cha90}(1990)
for possible stellar absorption features in summed spectra of aligned regions
in two different radio galaxies.
The best example of an object believed to be dominated by stars formed 
by a radio jet is 
Minkowski's Object, a blue clump of line-emitting material in the path 
of the jet from
a $z \sim 0.018$ radio galaxy. Its broad-band colors, 
line ratios and emission-line morphology make jet-induced star formation
the most likely interpretation (Brodie, Bowyer, \& McCarthy 
\markcite{bro85}1985; van Breugel et al.\ \markcite{vanB85}1985;
Hansen, N{\o}rgaard-Nielsen, \& J{\o}rgensen \markcite{han87}1987)
but even in this low-redshift case it is difficult to
get conclusive proof of the existence of a young stellar population. 

Other possible contributors to the aligned UV continuum in these galaxies 
are thermal emission from a hot plasma, optical synchrotron radiation, 
and inverse Compton scattering of microwave background photons by relativistic
electrons (Chambers et al.\ \markcite{cha88}1988; Daly 
\markcite{dal92a} \markcite{dal92b}1992$a,b$).  Some of these mechanisms
require that the optical morphology directly trace the radio 
structure: such a correspondence
may be present in some smaller scale radio objects
such as 4C 41.17 
(Miley et al.\ \markcite{mil92}1992),
but these mechanisms cannot account
for the many cases in which 
there is no strong correlation between the radio and optical morphologies.
The aligned components of the blue, closely aligned $z \sim 1$ galaxy 3C\,368,
originally thought to be 
dominated by scattered quasar light, seem low in polarization in 
recent imaging polarimetry and spectropolarimetry
(Dey et al.\ \markcite{dey97}1997).  These components 
have Balmer-line--to--continuum ratios that indicate 
the thermal nebular continuum from the ionized gas
actually dominates the near-UV aligned continuum radiation
(Dickson et al. \markcite{dicr95}1995;
Stockton, Ridgway \& Kellogg \markcite{sto96}1996).

\subsection{QSO Host Galaxies}
Far less is known about the hosts of high redshift quasars than about
radio galaxies. 
Until recently, studies of QSO host galaxies have concentrated
primarily on the low redshift range. Evidence has accumulated that QSO activity
might be triggered by interactions or mergers:
the extensive ground-based work
on these $z < 1 $ QSOs has shown a significant fraction of
them to have asymmetries, distortions, tidal features, and a tendency to have
close companions. Their host galaxies have colors bluer than do normal 
ellipticals;
such colors would be consistent with interaction-triggered star formation
(Heckman \markcite{hec90}1990; Stockton \markcite{sto90}1990 and 
references therein).  In the past few years, several groups
have studied radio-loud and radio-quiet QSOs at $z$ = 2 to 3 (Heckman
et al.\ \markcite{hec91}1991; Lehnert et al.\ \markcite{leh92}1992; 
Hutchings et al.\ \markcite{hut94b}1994$b$;
Lowenthal et al.\ \markcite{low95}1995;
Arextaga et al.\ \markcite{are95}1995). They have been able to 
resolve extensions around a reasonable 
percentage of the high-$z$ QSOs they have studied in optical and/or the 
near-infrared ($\sim 50\%$); these groups find similar results in that
the extended portion contributes $\sim 20\%$ of the total quasar flux
at observed optical wavelengths.
Lehnert et al.\ \markcite{leh92}(1992) found the extended quasar 
light to have redder optical--infrared colors than
the quasar itself. 
This result is more consistent with the extensions coming from a stellar host
galaxy than from scattered nuclear emission.
They also found that the 5 quasars in their sample
have total host $K$ magnitudes significantly brighter than those of radio
galaxies in a comparable redshift range.
In the optical studies of the same sample, 
Heckman et al.\ \markcite{hec91}(1991) saw no evidence for 
alignment of the quasar extensions 
with the radio axis. These two results seem to pose difficulties for
theories in which radio galaxies and quasars differ solely in
orientation (Barthel \markcite{bar89}1989); radio galaxies at 
these redshifts are quite aligned in the UV. 
Recently, however, Bremer \markcite{bre97} (1997) has found that the
$z=0.734$ quasar 3C\,254 has continuum and emission-line
morphologies that are well aligned with the radio structure.

Of course, the study of QSO host-galaxy properties
is complicated by the bright nucleus which contaminates or swamps
any extended component that underlies the seeing disk. 
For this reason, the Hubble Space Telescope (HST) is well-suited 
to the study of QSO host galaxies, and even more so for these high redshift 
objects in which (at $z$ = 1) $\sim6$ kpc
of physical scale\footnote {$\rm H_{\rm o} = 75~ km~ sec^{-1}~ Mpc^{-1}, 
q_o = \onehalf $ throughout this
paper.}
would be hidden under a typical $1\arcsec$-diameter groundbased seeing disk.
A number of recent WFPC2 imaging studies 
of QSO host galaxies at low redshift have
been made.
These include the major GTO survey of $z \sim 0.2$ QSOs of Bahcall et al.\
\markcite{bah94}(1994, \markcite{bah95a}1995$a$, \markcite{bah95b}
1995$b$, \markcite{bah96}1996, \markcite{bah97}1997),
a survey of 4 QSOs by Disney et al.\ \markcite{dis95}(1995), 
and several low-$z$  QSOs studied by Hutchings et al.\
\markcite{hut94a}(1994$a$).
Early claims of discrepancies in host-galaxy magnitudes between the results 
of previous ground-based
imaging and the HST images (Bahcall et al.\ \markcite{bah94}1994, 
\markcite{bah95a}1995$a$) now seem to have been resolved (McLeod \& Rieke 
\markcite{mcL95}1995; Bahcall et al.\ \markcite{bah97}1997), and the
higher resolution of the HST images allows details of the galaxy morphology
to be determined.  QSOs in general are found to have a fairly wide range
of host galaxies, ranging from apparently normal ellipticals and 
spirals, to obviously interacting systems.  Radio-loud
quasars are, as expected, usually found in elliptical host galaxies, but
(contrary to general expectation) radio-quiet QSOs seem also to be
predominantly in elliptical host galaxies.

\subsection{Unified Models}
The optical spectra and extended radio properties of radio galaxies
and quasars have many similarities, inspiring attempts to explain the
differences between the two classes as due primarily to orientation
(Scheuer \markcite{sch87}1987; Barthel \markcite{bar89}1989).
In this view, FR II radio galaxies and quasars are
drawn from the same population; the non-thermal optical
continuum and broad spectral features seen in
the quasars are obscured in the radio galaxies
by a dusty torus oriented perpendicularly to the radio jet.
That radio sources have strongly beamed, relativistic jets
is well-established from the observed jet asymmetries, apparent
superluminal motion, and depolarization aymmetries between the
jet and counter-jet-side radio lobes of quasars (Laing \markcite{lai88}1988;
Garrington et al.\ \markcite{gar88}1988; Ghisellini et al.\ 
\markcite{ghi93}1993).
Orientation must play a part in how
we view and classify such beamed objects. 

Unified hypotheses of AGNs in general have been reviewed recently by 
Antonucci \markcite{ant93}(1993) and of radio-loud AGNs in particular 
by Urry \& Padovani \markcite{urr95}(1995).
Spectrophotometric studies of 3C\,234 and seven other
radio galaxies revealed
obscured broad-line regions
(Antonucci \& Barvainis \markcite{ant90}1990; see review by Antonucci 
\markcite{ant93}1993). These studies,
the recent probable discovery of broad \ion{Mg}{2} in emission in the
very nearby powerful FR II galaxy Cygnus A
(Antonucci et al.\ \markcite{ant94}1994), and detections of broad
Pa-$\alpha$ from some narrow-line radio galaxies (Hill et al.\ 
\markcite{hil96}1996), provide direct support for this unification
of radio galaxies and quasars.
Another way to test 
the unification hypothesis for quasars and narrow-line
radio galaxies is to compare properties that should be isotropic
in matched, complete, and 
unbiased samples of radio galaxies and quasars selected on the basis of
some other supposedly isotropic property, such as extended radio emission.
As summarized by Urry \& Padovani \markcite{urr95}(1995), a number of such 
tests can and have been
made, with results that either favor some form of unified model or at least
do not strongly contradict such models.

However, although
it is certain that viewing angle has affected our classification 
of radio galaxies and quasars, 
precise and unambiguous
tests of the hypothesis that all objects in
these separate apparent classes belong
to the same intrinsic population are difficult. 
The total sample of objects cannot be wholly homogeneous,
and the objects must have a certain dispersion in physical properties
such as opening angle; 
in addition, various insidious selection effects from 
the beamed properties of the objects may bias  
samples chosen to measure isotropic properties.

Recent successes at resolving quasars at $z >1$ have prompted us to undertake a
project to image a complete sample of 15
quasars at $z \sim 1$ in order to address
the issues of the relationship between radio galaxies and quasars
and, possibly, of the alignment effect in high redshift radio sources. 
We seek to minimize the bias in our samples by selecting on radio-lobe 
properties alone.
We have taken deep WFPC2 imaging 
of a small but complete 
subset of these quasars and the matched complete 
sample of radio galaxies in the hope that the higher resolution of 
HST will enable us to make our comparison of radio galaxy 
and quasar host properties less hindered by systematic effects than previous
ground-based studies.  We present here the results of the WFPC2
and groundbased imaging of this HST subsample; the remainder of the larger
sample will be discussed elsewhere.

\section{Sample Selection}

\label{sec:sample}

The revised 3C (3CR: Bennett \markcite{ben62}1962) catalog has long 
provided the only large sample of radio sources selected at low frequency
for which both optical identifications and redshifts are now
essentially complete (Spinrad et al.\ \markcite{spi85}1985). For this 
reason, we have concentrated on 3CR objects to form our samples.
Our larger complete sample of 3CR quasars is 
chosen to match the 3CR galaxy sample of
Rigler et al.\ \markcite{rig92}(1992) in extended radio
properties and in redshift range. We provisionally included all 
3CR quasars
in the range $0.8<z<1.25$ with steep-spectrum fluxes  
above the survey limit of 9 Jy at 178 MHz.
We define as ``steep-spectrum'' objects with spectral indices
$\alpha$ (computed at 2700 MHz; see note to Table 1)
$>$ 0.5, where 
$S_\nu \propto \nu ^ {-\alpha}$.
Our final sample was limited to those steep-spectrum quasars whose 
total flux is not brought above the survey limit by a flat spectrum 
compact core, in order to try to ensure
that we are selecting on an isotropic property. 
The extended, steep-spectrum lobes of FR II radio galaxies
are optically thin and probably mostly unbeamed. 
While Garrington et al.\ \markcite{gar91}(1991)
found that the spectral index is flatter on the jet-side, consistent
with a beamed contribution to the lobe flux, Blundell \& Alexander 
\markcite{blu94}(1994) have argued
that this spectral index asymmetry may be explainable simply by projection 
and light-travel-time effects, in that 
the radio emission from the more distant lobe should have age-steepened 
more than the emission from the nearer lobe. 
In any case, the contribution is
considered minor (Urry \& Padovani \markcite{urr95}1995), and it is 
probably not  a significant source of bias in our radio-lobe-luminosity 
selected sample.

For our HST observations, we have defined a complete subset
of the above $z \sim 1$ 3C sample. We include all
3CR objects that fulfill
the radio flux, morphology and spectral index constraints of the 
quasar sample, 
as well as having a redshift within the range 0.87$< z <$ 1.05 and a
$\delta < 60\deg$. 
The lower declination limit of the revised 3C survey is $-5 \deg$; our sample
could equally well have been drawn from the further-restricted sample
of Laing, Riley, \& Longair (LRL;\markcite{lai83} 1983), except for 
our inclusion of 3C\,2 and 3C\,237,
which have declinations below the LRL declination limit of $10 \deg$.

The redshift range for our sample was chosen so that the WFPC2 filters 
F622W and F675W give passbands
centered near rest-frame 3300 \AA\
with little or no contamination from emission lines. 
After deleting 3C\,22 because of high extinction ($A_B = 1.09$)
along the line of sight, we are left with
a sample of 5 radio galaxies and 5 quasars. 
These are 
listed in Table \ref{HSTtab} along with some information about their optical and radio
properties. The 10 sources all have double-lobed radio structure; eight
have largest angular sizes (LAS; from lobe hotspot to hotspot) $\ge7$\arcsec;
3C\,2 has LAS $\sim$5\arcsec\ and has sometimes been termed a ``compact
steep spectrum'' (CSS) source (Saikia et al.\ \markcite{sai87}1987). 3C\,237, 
with a LAS of $\sim$ 
1\farcs3, is included in most samples of CSS objects
({\it e.g.,} Fanti et al.\ \markcite{fan90}1990). 

\section{Observations and Reduction}

\subsection{HST WFPC2 Imaging}
Table \ref{HSTlog} gives the log of HST observations.
Total exposure times for each object in our sample were calculated to 
give roughly the same signal-to-noise level to a given proper 
surface brightness limit; actual total integration 
for our WFPC2 observations are given in Table \ref{HSTlog}. 
Our general technique was to take two, three, or four separate exposures 
of 900--1100 seconds at the same pointing to aid in removal of cosmic rays,
but to dither the telescope in a square pattern 10.5 pixels on a side
between subsequent sets of 
exposures to aid in removal of hot pixels and to improve our sampling of
the PSF.
The object was centered on WFC3, giving a scale of
0\farcs1 pixel$^{-1}$ (undersampling the HST point-spread function).
HST points accurately to  $\sim 3$ mas (0.03 pixel), and we found that using the
intended pointings to shift and combine the dithered images 
gave better final image quality
than re-assessing the offsets by centering on the undersampled stars.
 
We tried recalibrating the raw data with several different types of
STScI-provided bias---dark combinations and
found the best standard deviation in final combined frames
from using the pipeline-supplied biases 
(averaged from 40 individual bias frames
close in
time to the observations) and a super-dark (an average of 100 darks).
The recalibration routine generates bias-subtracted, dark-subtracted, and
flattened files, as well as associated bad pixel masks (data quality 
or ``DQF'') files.
We resampled the images by a factor of two to a size of $1600\times1600$,
then aligned by shifting in integer
(resampled) pixels. 
For combination and cosmic ray removal, we used primarily the 
STSDAS $gcombine$ routine.
We first rejected all pixels marked
as bad or saturated in the 
DQF files generated by
the recalibration routine. We then rejected pixels that were more
than 3$\sigma$
from the median, using the rejection algorithm ``rsigclip'' to compute a 
median and sigma that are robust against unidentified outliers. 
We then compute each output pixel value from the average of the remaining
unrejected pixels. 
From all summed frames in which we had unsaturated stars, we measured
an average stellar
full width at half maximum (FWHM) of 0\farcs14 with a 1$\sigma$
variation of 0\farcs02. 

We have derived the flux density in our images
from the calibration information provided by STScI, {\it i.e.,} from the 
PHOTLAM calibration keyword generated by SYNPHOT in the pipeline calibration.
These calibrations are determined 
from knowledge of the instrumental sensitivities and filter bandpasses,
supplemented by WFPC2 imaging of known standards. 
We normally will give our photometric results in terms of
flux densities for the optical images in this
paper, since we use non-standard filters.

\subsection{Ground-based Optical Imaging}

The ground-based observations are summarized in Table \ref{obslog}.  Our non-standard
optical filters are designated by \{central wavelength\}/\{FWHM\}, where both
quantities are given in \AA ngstr\"oms.  The
optical continuum images were taken at the University of Hawaii 2.2-m 
telescope with an
anti-reflection-coated, thinned Tektronix $2048\times2048$ (Tek2048) CCD.
While we have ground-based images of all of the quasars in our sample
obtained with filters very similar to the WFPC2 filters, we used these
images mainly for consistency checks on the HST calibrations; we will
make no further use of them in this paper.
For 3C\,2 we also obtained an image in 
a redder passband near Mould $I$ (our 8964/1063 filter). 

We also observed 4 quasars and 2 of the radio galaxies in narrow-band
[\ion{O}{2}] interference filters. These filters are high-transmission ($\sim$ 90\%),
square-bandpass filters with widths $\sim 30$ \AA, centered at redshifted
$\lambda$3727 \AA\ for each of the objects. 
The narrow-band images were
taken at UH 2.2m with 
the Tek2048 CCD and at the CFHT with SIS fast guiding and
the Orbit1 (Orb2048) CCD; specifics of the seeing conditions (FWHM) and filter 
positions can be found in Table \ref{obslog}. 

We followed standard CCD data reduction procedures, with a few adjustments 
for some of the peculiarities of the detectors. 
For the continuum images, we made sky flats from the median average 
of the dithered, bias-subtracted
raw images (while masking objects out of the median).
We compared the results of using sky flats versus dome flats in flattening
the raw frames;
in general, the
sky flats worked better.
In a few cases, bright stars left residuals that 
we could not successfully mask out, and in this situation we adopted the dome
flats.
Each frame was corrected for atmospheric extinction using mean extinction
coefficients.
The narrow-band images were treated similarly, except that,
because we needed long exposures 
to approach sky-noise-limited statistics in the background, we normally
obtained only 
3 dithered frames and could not construct good sky flats; 
we therefore used dome flats instead. 

To produce our final CCD images, we median-averaged the separate frames after 
aligning each to an accuracy of $\sim$0.1 pixel (0\farcs02), 
using the brightest unsaturated stars in each field. We then
flux-calibrated using Kitt Peak spectrophotometric standard stars
with data to 1 $\mu$m 
(Massey et al.\ \markcite{mas88}1988; Massey \& Gronwall 
\markcite{mas90}1990). To determine the standard 
star flux in our non-standard continuum and narrowband filters, we 
approximated our filters as square with width equal to the
FWHM of the filter profile, and integrated the stellar
spectral flux within this region. 

\subsection{Infrared Imaging}

The near-infrared images were taken at the UH 2.2 m, CFH, IRTF, and Keck 
telescopes.
All objects were observed at
Mauna Kea $K^\prime$ (Wainscoat \& Cowie \markcite{wai92}1992).
This filter is
centered at 2.1 $\mu$m (shorter than standard $K$) 
in order to reduce thermal background.
A few objects were observed at standard $H$ as well, which is centered at 
1.65 $\mu$m.
At the UH 2.2 m, we used the NICMOS3 $256\times256$ (Nic256) HgCdTe
infrared array at f/10, with a 
pixel scale of 0\farcs37 pixel$^{-1}$, and at f/31, with 0\farcs12 
pixel$^{-1}$.
We also used the $1024\times1024$ QUIRC (Qrc1024) HgCdTe camera, with a scale 
of 0\farcs18 pixel$^{-1}$ at f/10.
At the CFHT, we used the UH NICMOS camera in March 1992,
with a scale of 0\farcs3 pixel$^{-1}$
and the CFHT Redeye Camera, (also a $256\times256$ NICMOS3 device),
in November 1993 with a scale of 0\farcs2 pixel$^{-1}$.
The IRTF observations were made with NSFCam, a $256\times256$ InSb array.
NSFCam
has an adjustable pixel scale; we chose to use 0\farcs15 pixel$^{-1}$
and 0\farcs3 pixel$^{-1}$ on separate occasions.  The Keck Near-IR Camera
(NIRC) also uses a $256\times256$ InSb device.
Seeing and photometric conditions were variable; best seeing 
was 0\farcs5 FWHM and worst $\sim$1\farcs3. We give the specifics
of each observation in Table \ref{obslog}. 
We offset 
the telescope slightly between each exposure on a field
to facilitate creation of sky flats and removal of bad pixels.
Dome flats were made by subtracting observations of the
incandescent-illuminated dome from observations of the dome with the lights
off. Except at Keck, the standard readout procedure for these devices leaves little or no bias.

We outline our reduction procedure for the UH 2.2m NICMOS3 observations,
and afterwards indicate variations made for other devices.
An iterative process was used to flatten the
data and replace the bad pixels. First a bad pixel mask was created
from the dark frames and dome flat field frames.
The raw object frames were then normalized by their median sky value
and combined to create a 
sky frame.
(Every calculation of a median sky value excluded
the masked-off regions associated with that frame).
This sky frame was then scaled to each raw frame and
subtracted.
The subtracted
frames were flattened with the dome flats. These rough flattened
images were aligned to the nearest pixel using stars or the quasar itself
and median-averaged to create a rough combined image. A mask was made from this
combined frame of the positions and extents of the objects. The portion of 
this object mask that was associated with each individual frame was
added to the
bad pixel mask to create a combined mask for that frame. The process
was then repeated; using the object mask to mask objects out of the 
sky frame, we made superior flattened frames, recalculated the
centering and alignments, and made 
a better combined image. This combined image was then 
used to replace the bad pixels in each flattened frame with the 
median of good pixels from the rest of the frames.
The bad-pixel-corrected flattened frames were then interpolated and 
resampled; since most of our infrared data
are undersampled, the magnification 
factor was generally in the range 4--8.
The centering was recalculated (bad pixels may badly skew centering),
as were the fluxes of a number of photometric reference objects 
we had previously specified. Using these photometric fluxes
to scale each frame to
the median flux value, we made a final combined frame, using sub-pixel 
alignment with an accuracy of $\sim$0.2 pixel ($\sim$0\farcs07 for the UH
NICMOS images). 
In cases where we had variable extinction from clouds, we
scaled the individual frames to the maximum flux recorded.
This process treats incorrectly any contribution from the dark current,
which was 
scaled along with the sky, but the effect is negligible as long
as the dark contribution is a small percentage of the total background. 
For the NICMOS chip, the uncertainty in the level of dark to subtract 
is greater than any resulting uncertainty this process may add.

The 1024$\times$1024 QUIRC chip (at the UH 2.2m) has a much stabler
dark pattern, and we generally subtracted
a dark frame from the object exposures prior to creation of the sky flat.
With NIRC at Keck, twilight sky flats were used instead of dome flats.
A bias must be subtracted, which we obtained from
dark frames taken at the 
beginning and end of each night. The scale is 0\farcs15 pix$^{-1}$,
resulting in a field of view of 38\arcsec. This meant that we were not able
to have a star on the frame in all cases.
Otherwise reduction procedures followed were similar to that for the NICMOS 
at the UH 2.2m.

We calibrated our near-infrared data with standards from the 
UKIRT faint standard list (Casali \& Hawarden \markcite{cas92}1992)
and from the Elias et al.\ \markcite{eli82}(1982) list 
of moderately bright standards. 
We corrected for the color difference between $K$ and $K^\prime$ as
prescribed by Wainscoat \& Cowie \markcite{wai92}(1992); the resulting flux difference was 
$\sim 2$--3\%, which is generally less than the error in our
absolute calibration. 
A secondary check on our absolute calibrations can be made by comparing fluxes 
of stars from separate observations of the same field. Since we have 
Keck observations for all objects except for 3C\,2, all other
observations are referenced to our Keck data.  The 
average percentage difference in absolute 
calibration is 0.4\%, 
with a standard deviation of 5.1\%, and a maximum mismatch of 12\%;
this standard deviation flux difference 
corresponds to a 
magnitude difference of 0.05. 
\section{Analysis}
\label{sec:psfdet}
\subsection{Point Spread Function Determination}
To determine the morphologies and magnitudes of the extended
material underlying the quasar nuclei, we must remove the contribution
of the nuclear component.
We describe here some of the details
of how we determine the point spread function (PSF)
that we use to subtract off this nucleus.

\subsubsection{The HST Point Spread Function}
For the HST data, we observed a PSF star in association with each of the 
quasar observations. We planned the exposures to try to match 
as closely as possible the observational procedure for the 
quasars themselves; {\it e.g.,} we followed the same 
dithering procedure we discuss in \S 3.1 and
calculated the PSF exposure times in order to saturate the PSF
to the same extent that we saturated on the quasars themselves.
In Table \ref{psftab}, we give the PSF star we observed in conjunction with
each quasar, the filter in which each was taken, the integration times,
and the greatest extent of saturation in any dimension for both
the QSO and PSF. As can be seen, our predictions of QSO and PSF
magnitudes were not quite perfect, and the saturation diameters
range from 0 to 0\farcs25. 
We note here that these saturation regions are the intersection
of whatever pixels were marked as A-to-D-converter-saturated
by the DQF files; we mask 
out entirely these saturated pixels and replace those data with 
other data values from the stack of images if other pixels
are not saturated. Therefore, because we dithered
our exposures by half-pixel steps, and saturation may vary slightly
from exposure to exposure, we may end up with a saturated diameter of
1 resampled pixel (0\farcs05), which is equivalent to half an 
original resolution element of the WFC.

The PSF varies across the field, so we checked the 
observed QSO and PSF positions on the chip. This positioning was very 
stable: the variation in the distance from the mean column position
for all PSF stars and QSOs was $\sigma_{column}=0\farcs35$, and
the maximum deviation was 0\farcs7, for the 3C\,245 PSF star. The 
row positioning varied a little more ($\sigma_{row}$=0\farcs6),
but the maximum deviation from
the mean position was 1\farcs2 for the PSF star of 3C\,336. 
As we have more than one measurement of the PSF in each filter, and
our mean positions on the frame do not vary much, we can now
consider averaging the PSFs in each filter to increase our signal-to-noise.
In order to assess how much the PSFs differ from each other, we determined 
their relative centers and scalings in annuli outside of their
respective saturated regions and subtracted them.  We use the same 
centering
and subtraction techniques (discussed in \S\ref{psfsub}) that we used for subtracting the
PSFs from the quasars, though these cases are of course 
not complicated by the existence of extensions.
The PSFs in the same filter subtract well from each other with few
systematic residuals. The inner 0\farcs35--0\farcs45 radius 
is very noisy, and we find in the subsequent quasar-minus-PSF 
subtractions that we are unable to recover much information from this
region, even when it is not saturated. We note that the diffraction
spikes (especially in the region close to the center of the PSF)
may leave some residuals. 
We show some examples of these PSF-minus-PSF subtractions in
Fig.\ \ref{psffig}$A$--$C$.  
We then average the 2 F622W PSFs, associated with quasars
3C\,196 and 3C\,336, and the 3 F675W PSFs, associated with
quasars 3C\,2, 3C\,212, 3C\,245, using the normalizations and centers 
determined from the subtraction technique.
We mask out the saturated regions in making 
the combined PSFs, and replace saturated values with values
from unsaturated stars in the average if available.

We display in Fig.\ \ref{psffig}$D$ the combined PSF for the F622W filter
(the average of the 3C\,196 and 3C\,336 PSF stars). 
We have also checked the residual of the difference between the two average 
PSFs.  As expected, we
find that there are significant differences between the two 
filters; we did not, therefore, average all of the PSFs together.

\subsubsection{The Infrared Point Spread Function}

The ground-based PSFs are, of course, very dependent 
on the details of the atmospheric conditions throughout the observations, and
they must be determined from stars taken as nearly simultaneously as possible. 
For both the Keck and CFHT images that we present here, the observational field 
is small enough that we must usually take
exposures of a star (of comparable brightness to
the quasar) interleaved with the object observations
in order to determine the PSF. The one exception in the quasar
sample is 3C\,212, which has a star of sufficient brightness within
the NIRC field. We also obtained PSF stars for the Keck observations of 
the radio galaxies. (For the radio galaxies, the star need not be so
bright;
there were therefore unsaturated stars of sufficient brightness
on each radio galaxy field except for 3C\,280 and 3C\,217).

When we must interleave observations of a PSF star with the actual
object integrations, we want to sample as well as possible
the seeing conditions and any PSF field variations. We therefore
bracketed the object observations with observations of the PSF,
and alternated between the PSF star and object as often as 
efficiency considerations allowed (generally within 15 minutes).
We used the same dithering pattern in the two sets of integrations to
make sure that any PSF variations in the field or systematic effects
caused by our dithering and 
centering technique are similar in the two cases; we also used 
the same effective individual integration times where possible. 
Despite these precautions, natural variations in the ground-based seeing
conditions result in some systematic PSF residuals;
these are generally obvious and confined to the 
inner region of the seeing disk. 

\subsection{Point-Spread-Function Subtraction}
\label{sec:psfsub}
To determine the morphologies and magnitudes of the extended
material underlying the quasar nuclei, we must remove the contribution
of the nuclear component. 
We will discuss some of the uncertainties in this process, and
some differences between the ground-based and HST data.
(All of the PSF subtraction was done in the original combined
images, before subsequent rotations and transformations that
would smooth the images and affect their pixel-to-pixel
noise characteristics). 

Some information is not recoverable from the PSF subtraction process; 
for example, a compact peaked host galaxy might be indistinguishable 
from extra flux in the PSF and would be subtracted. We can, however, make 
reasonable assumptions about the probable behavior of the host galaxy
and use these to estimate the total magnitude of the nebulosity.
The simplest technique is to fit the PSF to the quasar nuclear component
in a specified inner region,
subtracting the quasar flux in this region to zero. 

Subtraction to zero should give a lower limit to the magnitude
of the extended material if any exists but will oversubtract
if there is any extended flux in the inner region.
Some simulations of ground-based observations
of low-redshift quasars, where the host galaxies
were assumed to be normal spirals and ellipticals, found
that subtraction to zero decrease derived magnitudes by $\sim0.5$ mag
or more
(Abraham et al.\ \markcite{abr92}1992). 
A more realistic criterion 
would be to require that the host galaxy 
increase monotonically or at least remain constant from the outer
regions that are mostly unaffected by the PSF subtraction into the interior 
dominated by the PSF.
Even this monotonicity criterion is likely to underestimate the total 
host galaxy flux, if the galaxy peaks at all in the center, as would a 
normal spiral or elliptical.
We calculate two subtraction limits for our quasars, both
a subtraction-to-zero lower limit, and a monotonic-across-inner-region 
best estimate. In those objects in which we find little or lumpy 
extended flux, these two cases end up essentially the same.

In the interests of objectivity, we have tried to automate the
subtraction process, though
visual inspection of
the subtraction residuals remains a useful cross-check on
the process. 
We have developed IRAF scripts that allow us to center and scale 
the PSF to the quasar, display the residuals in a defined annulus, and 
calculate the reduced $\chi^2$ of the PSF fit to the quasar. We
first estimate the proper centering and scaling by
calculating the fluxes and the $x$ and $y$ first moments of both the PSF
and the quasar in an inner
defined annulus, correcting for partial pixels.
The annulus used is one of the most subjective 
parts of this process: it is chosen to exclude any saturated interior region, 
to be small enough to minimize the host 
galaxy contribution, yet large enough to provide decent statistics.
For the HST data, the inner radius we use varies from 
$0\farcs0$--$0\farcs35$, and the outer from $0\farcs2$--$0\farcs45$.
Our HST data are only minimally saturated (several pixels or 0\farcs25 at
the most; see Table \ref{psftab}). 

We then optimize the centering by calculating the 
the $\chi^2$ values of the difference in the defined annulus
over a grid of $x$ and 
$y$ shifts; we reject pixels from the $\chi^2$ sum that deviate by 3$\sigma$
from the mean difference value to reduce the effect of intrinsic PSF/QSO
shape differences. 
We fit a quadratic to each of
the $\chi^2$ vs. $\delta x$, $\delta y$ plots, and we take the minima as
the optimum $\delta x$, $\delta y$.
Though this procedure incorrectly treats the $\delta x$, $\delta y$ 
and the QSO-to-PSF scaling as independent,
the centering does not change significantly with any reasonable scaling.
In addition, 
the centering is well-determined, while the scaling is a much
more arbitrary and subjective quantity.
When we center by hand, by inspecting the residuals, the best center
is determined to about 0\farcs01, and it matches the results of
automatic centering to within this tolerance. 

To obtain the best scaling, we use a minimization technique similar to that 
we use for the centering:
we vary the scaling around the initial value, using the already determined
best center, then fit a quadratic. In Fig.\ \ref{psfsub} we show
the determination of the $\chi^2$ minimum scaling for the HST image of 3C\,2.
The minimum $\chi^2$ corresponds to the ``best fit'' between the QSO
and the PSF two-dimensional distributions; as discussed above it is 
probably generally an oversubtraction of any extended flux, but provides
a lower limit to the host galaxy flux.
Further evidence of this is the fact that the minimum $\chi^2$ 
is usually reached
at different
scalings depending on what annulus is chosen; using annuli including
data at a greater radial distance from the QSO center causes the QSO:PSF scaling
ratio to increase, as may be expected if the QSO is contributing more extended 
material than the PSF star at these radii.
In part, we have adopted this approach of estimating the $\chi^2$ minimum of
the fit in order to provide a good comparison to the
Bahcall et al.\ \markcite{bah94}(1994,\markcite{bah95a}\markcite{bah95b}\markcite{bah96}\markcite{bah97}1995$a,b$,1996,1997) results,
in which the $\chi^2$ fit is minimized
with respect to 3 variables: the $x$ and $y$ shifts between the quasar and PSF,
and the
QSO:PSF scaling. 
These subtractions correspond, therefore, to ``subtraction-to-zero'' lower
limits; similar conservative 
limits are used by Heckman et al.\ \markcite{hec92}(1992), and 
Lehnert et al.\ \markcite{leh92}(1992).
We also wish to automate an objective version of the 
``monotonicity'' constraint; this is similar to 
the approach used in a variety of studies of low-$z$ quasars, such as the HST study
by Disney et al.\ \markcite{dis95}(1995) and the groundbased studies
of McLeod \& Rieke \markcite{mcL94a} \markcite{mcL94b}(1994$a$,$b$). 
To establish the monotonicity of 
the profile difference requires averaging the difference profile in annuli with 
some width; generally this will be equivalent to requiring our interior 
annulus to have a mean value equal to the annulus immediately exterior to it.
We therefore define an annulus 
exterior to our inner annulus, and record the mean and median of the values
in both the inner and outer annuli 
(after 3$\sigma$ rejection, to minimize the
effects of PSF residuals in skewing the mean and median).
We once again desire to sample as narrow a width as possible
of the profile that will still give signal-to-noise adequate to prevent the
scaling being dominated by statistical noise. These outer annuli had widths
from 0\farcs1--0\farcs2 for the HST images. Any montonicity constraint applied
requires some such assumption; the wider the annular width used,
the more host galaxy is subtracted. We check the average values in the profile
difference in annular 
steps of this width for monotonic behavior, but it is the inner two
areas that set the scaling. In Fig.\ \ref{psfsub} we show how we automate this 
constraint; we plot the decreasing average values in the two inner
annuli and take as the scaling the point at which lines fit to these
two basically linear functions intersect. 
This gives us a better estimate of the proper subtraction,
and if there is no extension, the $\chi^2$ minimum 
scaling and the monotonic limit scaling are the same. In addition,
much of the residual structure is lumpy and irregular, and when we
take a radial
radially, the mean value in the annulus
is much less than the intrinsic surface brightness of the clumps.

We subtracted all of our ground-based quasar images in an identical
fashion, forming both the $\chi^2$ minimum limit and monotonic limit.
As the ground-based observations were never saturated, 
we were able to use the very central portion of the profiles,
and varied the annular widths according to the seeing.
To give a specific example of radii chosen, for the CFH $K^\prime$ 
observations of
3C\,2, we used the circular region within 0\farcs4 to calculate the $\chi^2$
minimum scaling and an annulus with width 0\farcs4 exterior to that to
calculate the scaling which satisfies the monotonicity constraint. 
We have checked our subtraction techniques by subtracting PSF stars from 
each other, as mentioned in \S \ref{sec:psfdet}.  
We have made other consistency checks on our PSF subtraction results:
for the HST data, we have subtracted each quasar both
with the average PSF for its filter and with the PSF observed close in time,
and found no systematic differences. 
Detailed discussion of the individual objects will be presented in \S 
\ref{chap:indiv}.

\subsection{Magnitudes and Colors}
\label{sec:mag1}
In this section, we outline our general approach for determining magnitudes
and colors.  Details of our magnitude and moment analysis are given in
Appendix \ref{chap:moment}.  Magnitudes and colors of discrete components
in the fields of our sample are discussed under the individual object
descriptions in \S\ref{chap:indiv} and in \S\ref{sec:colcomp}, 
and global properties of the extended material are given in 
\S\ref{sec:magextend} and \S\ref{sec:morphextend}.
\subsubsection{Total and Isophotal Magnitudes}
For the WFPC2 imaging, we planned our exposure times to achieve similar 
detection limits in all of our 10 images, after normalization for redshift and 
reddening. In Table 2, we give the
1 $\sigma$ detection limit we actually achieved in our WFPC2 images
for each of our 10 objects, after normalization to $z$=1 and E(\bv)=0.
To correct for Galactic extinction
we obtained E(\bv) values
for our sources from Burstein \& Heiles \markcite{bur84}(1984) 
and converted these 
to Galactic extinctions at the appropriate HST filters (F622W and F675W) 
using the values given in Table 12A of Holtzmann et al.\ 
\markcite{hol95}(1995).
For each object, we aligned all our ground-based images to the HST scale
and reference frame, using the IRAF tasks $geomap$ and $register$. 
This reference frame is in general rotated from
standard astronomical position angle. 

Since our primary goal is to make an objective comparison between
the radio galaxy and quasar sample properties, ideally we would
compare magnitudes and morphologies at identical 
isophotal flux limits (after normalization to a common reference redshift).
In practice this procedure is complicated by the very irregular nature
of the extended material we have resolved around the quasars; unresolved 
or linear features do not lend themselves so well to isophotal analysis
as smoothly varying galaxies. For this reason, we also
will calculate simple aperture and annular magnitudes.

After subtraction of the PSF, 
we normalize 
the images to $z=$ 1 and make
a series of  masks from the normalized
images corresponding to isophotal levels of interest. We investigate 
magnitudes and moments in apertures with no flux limit,
and in apertures defined by standardized surface brightness
cutoff levels.  
When we discuss the properties of a PSF-subtracted quasar, we generally 
refer to the results from the monotonic-subtraction-limit;
however, we use the $\chi^2$ minimum fit as useful check, both 
as a means of error estimation and as an indication of
how dependent on our PSF subtraction method our results may be.

We have also PSF-subtracted the HST radio galaxy images with
the best combined PSF from the appropriate filter, and verified that
outside of the excluded central region, the wings of the PSF do not
significantly contribute to the radio galaxy annular fluxes.
For the near-infrared radio galaxy images, the extended flux
is mostly quite dominant, and we estimate an upper limit to 
the unresolved contribution by subtracting until the 
the residual galaxy profile obviously is no longer monotonic.
In the ground-based images, the amount of flux removed is very 
seeing dependent, and this upper limit is probably an 
overestimate of the unresolved contribution, nonetheless
the wings of the PSF contribute little to the annular fluxes outside
of the seeing radius. 

\subsubsection{Colors and Spectral Energy Distributions}
Since we are dealing here with images having quite different resolutions,
we take a two-step approach to determining the magnitudes and colors of
various components in the images. Inspection of the original
HST image can be used to identify components that occur close
to the quasar nucleus and may not be easily distinguishable at the resolution
of the ground-based data. We then smooth the HST image to match the
resolution of the ground-based image in question, so that an 
aperture will encompass a similar portion of the component's flux.
We will then use the same apertures (simple or tailored) in each 
bandpass. 
There is generally a straightforward correspondence between the 
HST data and the ground-based data, but sometimes components that
may be resolved from others in the HST images cannot be identified
clearly in the ground-based images, which differ not only in
resolution but in bandpass.
Thus we cannot get a $K'$ magnitude for every component for which we have
HST fluxes; in cases for which the residuals from
the subtraction process obscure
the component at ground-based seeing, we cannot even obtain
a useful upper limit. 
The core fluxes and limits are derived as a result of the PSF-subtraction
process, and represent the monotonic limit.

\section{Discussion of Individual Fields}
In this section we present and discuss 
the HST optical and ground-based $K'$
imaging of our 10 sources.
We will discuss their morphologies and give photometry
for components of interest, but we will reserve the discussion 
of the quantification of the alignment effect and the properties
of the sample as a whole for later. 
In Figs.\ \ref{3c2fig}--\ref{3c336fig},
we show the HST image and the $K'$
image at the same scale in the upper left and upper right panels,
respectively. In the lower left panel we display a magnified version of
the HST image, sometimes slightly Gaussian smoothed to bring up
lower surface-brightness features. In the lower right panel, we display
the continuum-subtracted [\ion{O}{2}] image, if we have one for that object;
if not, we generally
display another version of the HST image, scaled or smoothed
differently, or with radio map contours overlaid.
In the top left panel, we give the positions of the radio 
lobes as white crosses (or show the direction to
them as white arrows, if they lie outside the frame). 
Throughout the images, insets will display 
different intensity scalings or a non-PSF-subtracted version of
a quasar.  In all cases, N is at the top, and E is to the left. 
In Tables \ref{photomtab} and \ref{3c2sedtab}, we give photometry for the components labelled
in the figures. 
\label{chap:indiv}
\subsection {3C\,2}
\subsubsection{Morphology}
3C\,2 has 
(relative to its nucleus)
the most obvious extension of the quasars in 
our sample; it shows clear signs of ``fuzziness'' even in 
groundbased unsubtracted images, especially in the $K'$ image. The flux of
the extension itself is roughly comparable to those of the 
other quasars, 
but it has the faintest nucleus of the 5 
quasars we observed; the ratio of the optical flux of extension to that 
of the quasar nucleus is $\sim$0.15 for 3C\,2, and from 0.01--0.03 for the
other quasars.
In Fig.\ \ref{3c2fig}$A$ and $B$, we display the HST and CFHT $K'$ images; 
the PSF-subtracted images show the monotonic limit. 
We label as $a$, $b$, and $c$ the components within a few arcseconds 
that can be identified above the general nebulosity. 
Object $a$ has a nearly stellar peak at a distance of 0\farcs83
and a position angle (PA) of $11\deg$. 
We find it to be precisely coincident with the northern radio
lobe of 3C\,2.

The radio structure of 3C\,2 consists of this northern lobe
and a southern lobe at a PA of $-160 \deg$ and a mean distance of
4\farcs5 (Saikia et al.\ \markcite{sai87}1987).
From Saikia et al.'s 2-cm map of the northern component, which has a similar
resolution to our HST images ($0\farcs16 \times 0\farcs10$) 
we can see that even the details of the morphology of the northern radio 
component and the optical component $a$ agree. The WFPC2 image shows a similar 
extension directly north, then a turn to the east to a bright peak.
When we align the optical quasar and the radio core, the 
bright optical peak and bright radio knot align to within 0\farcs02.
The optical image has the same fainter north-eastern extension that is 
seen in the radio map.
(We display in an inset in Fig.\ \ref{3c2fig}$C$ two contours associated 
with the northern lobe from the 2 cm radio map of Saikia et al.\ [1987]).

Though object $a$ has a separate, unresolved
peak in the WFPC2 data and therefore might have been
considered a companion object on the basis of the optical
data alone, the close optical-radio lobe coincidence
and the ``bridge'' of connecting nebulosity imply a 
close association with the quasar.
Object $c$, on the other hand, seems morphologically
discrete (though
it is also enveloped
in the general nebulosity surrounding the nucleus).
Object $b$ is diffuse and seems to connect into the noise-dominated
inner region, and it shows no correlation with the radio structure.
Of the three objects we have labelled, it is the only one that
might be a feature of the host galaxy itself.

The sky level for the 3C\,2 HST observations was lower than expected; we  
reached a greater depth for 3C\,2 and for 3C\,217 than for any other fields.
The extended nebulosity we observe in 3C\,2 is mostly at a lower
surface brightness than
our lowest standardized contour level; it is possible therefore 
that a similar nebulosity might exist and have escaped detection in
the objects in our sample that have a lower signal-to-noise background.

Our CFHT $K$-band data shows the three major components visible in the HST image
as well as some extended nebulosity;
our IRTF $H$ image shows similar structure 
but is not deep enough to show the general extended nebulosity.

\subsubsection{3C\,2 Component Spectral Energy Distributions}
For 3C\,2, we have two extra passbands: $H$, and our 8964/1063 filter,
which is centered slightly redder than the $I$ passband. 3C\,2 is the only
object in our sample for which we have 4 passbands spanning the
rest-frame range from 3300 \AA\ to 1 $\mu$m.
We will discuss here the spectral energy distributions of various 
components: the lobe structure labelled $a$ in Fig.\ \ref{3c2fig}$A$,
the faint extension $b$, the nearby companion $c$, and the underlying
nebulosity (seen best in the highest-contrast HST image panel,
Fig.\ \ref{3c2fig}$D$). This faint extended structure appears relatively 
symmetric and has an angular extent of about 3\arcsec\ radius, as 
measured at the 2$\sigma$ sky level in the $K'$ image, where 
it is brightest relative to the other components. 
In addition, we derive the SED of the quasar nucleus alone, with
the contributions from the above components subtracted.

For the SED of the nebulosity, we adopt an annulus with inner
radius 1\farcs75 and outer radius 3\arcsec.
The fluxes within this radius, given in Table \ref{3c2sedtab},
certainly underestimate the total fluxes of the nebulosity in each passband,
but they should give the best color estimates for the extension.
The signal-to-noise in the $H$ and $I$ band 
images is insufficient to make spatial maps of the color distribution. 

For the discrete components we use a similar procedure.
Components $a$ and $c$ are quite compact, and we measure their
fluxes in apertures with radii equal to the FWHM of the degraded
seeing,
excluding, of course, the central 0\farcs5 radius around the nucleus that
is dominated by PSF residuals.
We subtract
off the background extension in the HST image before smoothing,
then Gaussian smooth to the poorer resolution. This process may lead
to systematic errors, so we estimate the background contribution
in several different ways and use the dispersion as an estimate
of our errors. The images with the poorest resolution have 
the greatest systematic uncertainty from the background subtraction
process. 
We present the result of this photometric analysis, with errors, for
all the components in Table \ref{3c2sedtab}, and give there the
exact apertures used for each component.

\subsubsection{Discussion}

Because of the detailed close correspondence between the optical 
and radio morphologies of the northern component $a$, it appears that 
the physical mechanism producing the optical radiation is directly 
related to the radio emission.
To address the nature of this relation, and to investigate the likely origin of
the other components we have resolved, we plot
the flux values for each component values versus the rest wavelength (assuming
that all objects are at the narrow-line redshift of the quasar)
in Fig.\ \ref{3c2sedfig}$A$ and $B$.
The fluxes are normalized to the flux value
at $\lambda_0 = 3300$ \AA\ (from the F675W HST image).
We show both the subtracted and unsubtracted fluxes since this background
subtraction process could be prone to large systematic errors.  
Comparison of the two plots shows that, although the details of each SED value
change, each component's SED shows the same qualitative behavior
with or without the background subtracted.

Though broad-band colors alone are a notoriously inaccurate way to derive
information about the nature of any single object, we may still
gain some indication of the probable origin of the radiation
we observe and of the differences between the components. 
We see that the nebulosity and component $c$ are both redder than the QSO.
This result is consistent with domination
by a stellar population rather than by scattered quasar light. 
Lehnert et al.\ \markcite{leh92}(1992) found a similar result for the
extensions in $z \sim 2$ quasars. 
In panel $B$, we plot models representative of several stellar systems.
None of these simple stellar models 
is a perfect fit, but component $c$ 
has a 4000 \AA\ break that is fit very well by the Bruzual \& Charlot 
\markcite{bru93}(1993) 
models. This is evidence that it is at the redshift of the quasar and that it
is stellar in origin. Reddening by dust, either internal or from the surrounding
nebulosity, could explain the slightly redder colors in $c$ than in the 4 Gyr
stellar models.
The colors of the nebulosity are a poorer match for these basic 
stellar models; this could result 
from a significant
contribution to the extended emission from a younger, bluer stellar population
or perhaps from scattered quasar light.
Component $b$ is of much lower flux, and very close to the nucleus; it was
thus difficult to estimate its flux values, particularly 
for the two lower signal-to-noise images. Nonetheless, the
colors are consistently bluer than either the nebulosity or $c$.
Though we have suggested that it may be a feature
of the host galaxy,
its colors are consistent with a scattered light origin
as well as with a young stellar population. 

We also plot a short-dashed line in panel $B$; it is a power-law fit to 
component $a$ with $\alpha = 1.36$, where $S_{\nu} \propto \nu^{-\alpha}$.
(We weighted the fit with the errors on each point; for this reason,
the $\lambda_0=0.44$ $\mu$m contributes little to the fit). This optical
$\alpha$ is similar to that seen in optical synchrotron counterparts of 
radio hot-spots (Meisenheimer et al.\ 
\markcite{mei89}1989). In Fig.\ \ref{3c2radfig},
we show our photometry of the northern lobe (component $a$) with
the radio photometry of Saikia et al.\ \markcite{sai87}(1987) for the same component (crosses).
We plot the power-law
fit to our data (with $\alpha$=1.36) with a solid line; 
a linear fit to the Saikia et al.\ points gives $\alpha$=0.83, and is
plotted as a dashed line.  As we shall discuss more fully in
\S\ref{sec:optsynch}, this break in the spectral index is more likely
to point to optical synchrotron radiation than to competing explanations,
such as inverse Compton scattering of microwave background photons.

\subsection {3C\,175.1}

\subsubsection{Image Editing}
In the 3C\,175.1 field, there are several bright stars near the galaxy,
and these caused various artifacts that had to be removed manually 
prior to further analysis. In the HST image shown in Fig.\ \ref{3c175fig},
a diffraction spike from a bright star to the south has been subtracted
out. Diffraction spikes are among the features in the HST PSF that 
vary most with position in the field, and we were 
more successful in subtracting an azimuthally averaged version of the
star from itself rather than subtracting the observed PSF, which is
accurate only for the center of the field.
This process should have only added slightly to the noise, and there
is no morphological feature in the radio galaxy that could be due
to unsubtracted flux from the diffraction spike. 
In Keck NIRC images, bright stars bleed along rows.  A bright star in
the 3C\,175.1 field has bled through the radio galaxy; we subtract the 
extra flux by fitting a low-order polynomial along the row prior to rotating 
the images. 
\subsubsection{Morphology}
Our HST image of this radio galaxy has an elongated component, in two 
parts, but it also has a strong nuclear source.
The $K'$ band data (both
from Keck and CFHT) show a round component that coincides
with the HST nuclear component, with more elongation
at the lower isophotal levels. 
In the Keck image, the radial profile between 0\farcs65
and 1\farcs75 is consistent with a
de Vaucouleurs $r^{1 \over 4}$-law with an effective radius $r_e$ = 1\farcs1
$\approx$ 7 kpc.
In addition to this symmetric profile, we barely detect 
a discrete component at the position of $a$, to the southwest. 
We show the HST and Keck images in Fig.\ \ref{3c175fig}$A$ and $B$ respectively.
In panel $D$, we show the HST image with the contours of our 3.6 cm VLA map
superimposed; we make a firm detection of an unresolved core and align this
with the unresolved HST peak and near-infrared center.
This 3.6 cm core does not correspond in position to the emission 
found at 6 cm by Neff, Roberts \& Hutchings
\markcite{nef95}(1995) that they identify as a possible core; we accordingly
find that the source is less bent than they suggest.
Ignoring the core, the PA between the two radio lobes is 70\deg; with the 3.6 cm
core detection, the PA to the NE lobe is 78\deg\ and the PA to the SW lobe is
$-122\deg$. The bending angle of the axis (as defined in Barthel \& Miley 
\markcite{bar88}1988)
is 20\deg.
The HST image shows no obvious alignment of the radio galaxy 
components and the positions of radio lobes. Nonetheless, some of the 
components of the optical image are interesting. The extension to the NE of
the optical core is aligned approximately at the PA of the SW lobe; the 
linear feature
$a$ is connected to the core by a ridge of emission which leads into the 
jet-contact point on the SW lobe.  However, there is no detected radio emission
underlying these features; most plausibly this optical morphology is 
the result of quasar light within a large opening angle being scattered
off of shreds of material. 
Our moment analysis of the radio galaxy at $K'$ gives a PA
that varies from $70\deg$ to $90\deg$ with isophotal cut-off level;
the axial ratio over the same range is 0.93--0.98. 
The result of Dunlop \& Peacock \markcite{dun93}(1993) 
that the the $K$ band light is 
elongated with an aspect ratio of 1.48, aligned at a PA of
$70\deg$ (well-aligned with previous assumptions of the radio PA)
is clearly influenced by seeing conditions, causing them to include 
the seemingly discrete red companion objects to the NE.
They conclude that possibly 3C 175.1 is more aligned in the IR than in the 
optical.
This situation illustrates the difficulty of deciding what is
appropriate to include in analyses of the alignment effect. However, whether
or not these objects are actually associated with the radio galaxy,
our WFPC2 image and VLA detection of the radio core indicate
that there is no longer reason to conclude that the
infrared is better aligned than the optical.

\subsection{3C\,196}

\subsubsection{Morphology}
We display the PSF-subtracted HST and Keck images of
this $z=0.87$ quasar in Fig.\ \ref{3c196fig}$A$
and $B$.
This quasar has an obvious barred spiral galaxy to the southeast,
also seen in the WFPC2 image of Cohen et al.\ \markcite{coh96}(1996)
and the PC image of Le Brun et al.\ \markcite{leb97}(1997).
This galaxy is probably the source of both the prominent 21-cm
absorption line system and the metal line absorption systems
at $z=0.437$ in the quasar spectrum
(Cohen at al.\ \markcite{coh96}1996; Le Brun et al.\ \markcite{leb97}1997;
Boiss\'e \& Boulade \markcite{boi90}1990, hereafter BB). 
As this is almost certainly a foreground object,
we will exclude it
when analyzing the other extensions we see around the quasar. 
This procedure is not so much of a 
problem with the excellent resolution of the WFPC2 data
but is difficult in the analysis of the ground-based data.

3C\,196 has 
a LAS (lobe hotspot to hotspot) of 5\farcs6 (Reid et al.\ 
\markcite{rei95}1995) but shows a fairly wide, extended lobe structure.
The optical extension that we see to the north proceeds almost directly
north for several arcseconds, then extends to the east at the lowest
detectable level on our image. The lowest level extension has a slight knot
of emission that is coincident with the north-eastern radio lobe (best
seen in panel $C$).
At higher isophotal levels, as seen in panel $A$ or the inset in panel $C$,
component $a$ appears semi-discrete and is elongated
in the direction of the northern radio lobe. 
In addition, as noted by BB, there is a
non-stellar object ($d$) 6\farcs9 to the northeast that lines
up approximately with the radio lobe direction and 
is elongated perpendicular to the radio axis.
We see a similar object to the other side of the radio source, 
at a PA of $-172\deg$, elongated basically parallel
to $d$. We will call this second companion $e$; although
it is out of the field included in panel $A$, it is quite similar in appearance
to $d$.

Our [\ion{O}{2}] image (in panel $D$) shows a bright, asymmetric
emission-line region whose morphology 
(inset in panel $D$) corresponds fairly well to the optical structure.
This correspondence tends to confirm that 
the peak of the $z =0.87$ line emission matches the location of $a$ and
provides secondary confirmation that $a$ is at the redshift of the quasar,
rather than the $z = 0.437$ absorption line system.
At the lower isophotes, the morphology is curved symmetrically to the
east both north and south of the quasar.

\subsubsection{Discussion}

3C\,196 represents one of the best examples in our sample
of quasars of aligned structure in 
that is unlikely due to optical synchrotron emission.
Components $a$ and $b$ are probably both associated with the
quasar. They are fairly blue and are present in both 
the continuum and [\ion{O}{2}] emission.
This type of alignment is similar to that seen in high-$z$ radio galaxies.
Possible contributors for the material seen here
are thermal continuum and scattering of the quasar light off of ambient 
material.

BB bring up the possibility that 3C\,196 
could be lensed by the galaxy $c$.
Though the [\ion{O}{2}] morphology and the continuum morphology
are distorted and arced, there is no report of a multiply imaged radio
core. 
BB suggest that a secondary image of the quasar
may lie within the image of the spiral; this suggestion is based on
the elongation of the companion galaxy and a difference
in morphology between their $B$ and $R$ band images. The elongation
they note is probably just the bar of the spiral that is obvious 
in our higher resolution HST image; however, BB have
color information and claim evidence for an extra blue component to the 
east of the center of the companion galaxy. We see a faint component
in the bar of the spiral that might plausibly be associated with this nucleus
but could equally well be intrinsic to the spiral galaxy.
An inset in the NW corner of panel $A$ gives 
the unsubtracted quasar image at a lower contrast level, showing the
bar of the spiral galaxy and the faint peak. This peak is most likely
simply a knot or nucleus in the bar of the spiral galaxy.

Could there be lensing of the radio source itself?
Lensing on this scale would probably require a large 
mass associated with a many-$L_*$ galaxy or 
a foreground group or cluster, but the spiral 
galaxy appears to be $\sim L_*$ if at $z =0.437$ (from our
data and that of Cohen et al.\ \markcite{coh96}1996).
We estimate using an isothermal sphere model with the AIPS task $glens$
that the geometrical distortion induced
in the lobes from lensing by an $L_*$ spiral is minor, and the
resultant magnification factors are at most a few tens of
percent.
This makes it unlikely that $d$ is a gravitational
arc, as also consisidered by BB. On our images, 
both  $d$ and $e$ are elongated, but we see no strong
morphological evidence for gravitational distortion.

BB find companion $d$ to be very blue in the optical and
favor the interpretation that it 
is associated with the radio source (through jet-induced
star formation), though it lies beyond the radio
lobe. This suggestion is particularly interesting, given
that we have an example of a somewhat similar situation
in 3C\,212 (\S5.4): there we have an object (component $f$ in Fig.\ \ref{3c212fig}$A$)
that falls directly on the radio
axis and is morphologically similar (in the optical) to the radio lobe,
yet lies 3\arcsec\ beyond it.  
However, the 3C\,196 companions $d$ and $e$ 
have similar and very blue optical-to-infrared colors; 
the optical-to-$K'$ flux ratios are 10.9 and 8.5, respectively, while the QSO
itself has a flux ratio of only 4.
As these objects appear very similar morphologically and in color,
it is tempting to try to ascribe them to a similar origin. 
In this case,  though the southern object is off of the radio axis by $20\deg$,
this is well within a typical quasar opening angle (Barthel 
\markcite{bar89}1989; Saikia \& Kulkarni \markcite{sai95}1995), and 
the blue objects may result from quasar light being scattered in 
dusty galaxies (this possibility was also considered by BB).  
In qualitative support of this explanation, 
the peak surface brightness in the more distant companion is $\sim$0.25 the surface
brightness of the nearer. 
In 3C\,280, there may also be morphological evidence of quasar radiation
impinging on and scattering from an existing structure. 
However, in 3C\,196, we see no [O\,II] emission from these
objects at the redshift of the quasar, making this explanation much less
likely.

\subsection{3C\,212}
  
\subsubsection{Morphology}

Even a casual inspection of the ground-based images of 3C\,212 
shows an unusual clustering of objects around the quasar. 
With the high resolution of the HST image
we see that several of these objects have morphologies
and locations that make it likely that they are directly associated
with the quasar and its radio jet.
Here we have a clearcut example of a quasar with optical and near-infrared
continuum structures that are aligned with the radio axis; 
although we see some degree of alignment in each of our 5 quasars, 3C\,212
is certainly our most spectacular example.
In Fig.\ \ref{3c212fig}$A$ and $B$ we show the HST and Keck $K'$ images;
in panel $C$ we show the HST image magnified 2$\times$, and
in panel $D$ the same image Gaussian smoothed to bring up 
low surface brightness features. 
The crosses in Fig.\ \ref{3c212fig}$A$ indicate the positions of the
radio lobes in the 6 cm MERLIN map of Akujor et al.\ \markcite{aku91}(1991).

Close to the nucleus, we see a general nebulous extension, but the
most striking features
are the three discrete blobs $a$, $b$, and $c$, which extend
almost directly toward the NW radio lobe.  They are compact but resolved 
($b$, the brightest, has a FWHM
$\sim$ 0\farcs25) and fairly blue.  Though they are not resolved 
in the Keck image, there is a narrow extension in the $K'$ band structure
at the positions of $b$ and $c$. 

At larger distances from the nucleus, features $f$ and $g$ are
also closely aligned with the radio axis, but they lie just
{\it beyond} the radio lobes.  Object $g$ is fairly regular and
compact, though with some wispy extensions we will discuss in
more detail later.  Object $f$, however, shows a structure that is
extremely unusual (even in the context of the range of morphologies
exhibited by objects in other deep WFPC2 fields) and reinforces the 
impression that it may be associated with the radio source.
In fact, its optical image
looks very much like a VLA image of a radio lobe:  it even has
a ``hot spot,'' barely resolved (FWHM = 0\farcs17), near its tip 
and closely aligned with the 
three inner knots $a$, $b$, and $c$.  
The more diffuse component of $f$ has a tail-like shape, with
a sharp outer boundary but a gradual fading 
away in the direction of the quasar.  
When we ratio the images of the component $f$ (after smoothing
them to the same resolution), we find that 
$f$ exhibits a steep color gradient. This effect can be seen qualitatively 
by comparing the optical and $K'$ images:  the $K'$ $f$ component
peaks $\sim$1\farcs8 south of the optical point source. 
The $K'$-to-optical flux ratio is 0.18$\pm$0.04 at the optical peak
and 0.98$\pm$0.09 at the $K'$ peak. 
Object $g$ is the reddest component, with a $K'$-to-optical flux ratio of 1.9;
the colors of other components are fairly blue. 

Thus, 3C\,212 shows evidence for optical/IR components aligned with the
radio structure both within and beyond the radio lobes.  In view of
the importance this object may have for our understanding of the
nature of powerful FR II radio sources and of at least one version
of the alignment effect, we have obtained deep multifrequency VLA
maps and longslit spectroscopy with the Keck Low-Resolution
Imaging Spectrometer (LRIS) for this field.  Details will be given
elsewhere (Stockton \& Ridgway \markcite{sto97}1997), but we 
summarize the most
important conclusions from the present stage of the work here.

In Fig.\ \ref{3c212xop} we show contours from the 3.6 cm 
VLA map overlain on the HST image; the radio and optical
frames were registered with the quasar nucleus position.
The bright optical objects $a$, $b$, and $c$ are seen to coincide precisely
with knots in the radio jet.  
The close morphological tracking of these features implies an
emission mechanism that links
the radio and optical radiation directly, such as synchrotron radiation. 
In spite of the higher dynamic range of our new maps, there is still no
evidence for radio emission beyond the lobes seen by Akujor et al.\ 
\markcite{aku91}(1991).
The 3.6 cm map shows a fairly broad, normal hotspot/lobe to the 
SE side of the quasar. The NW side has a straight jet; 
the NW lobe resolves into a chain of peaks, continuous on
one side with the jet, which look like 
hot spots produced by the radio jet precessing over small angles.
Interestingly, the morphology of this NW radio lobe appears
to mirror closely that of the optical component $f$ lying 3\arcsec\
exterior to it.  The morphological correspondence between the SE radio 
lobe and the optical component $g$ lying just beyond it is less
striking, but the curved wisp extending
N from the outer edge of $g$ closely parallels the outer contour
of the radio lobe.

For our LRIS spectra, we placed a 1\arcsec-wide slit through the
quasar and components $f$ and $g$.  In a total integration of 
2 hours, only a faint continuum 
was seen for the SE component $g$, but a pair of
faint emission lines is present in the spectrum of the NW component
$f$.  From their separation, these can be unambiguously identified
as the [\ion{O}{2}] $\lambda$3727 doublet at a redshift of 0.927.
In the rest frame of the quasar ($z=1.05$), the radial velocity
difference is over 18000 km s$^{-1}$.  If this gas is in fact
associated with 3C\,212, we require a mechanism that can eject it
coherently at this velocity while keeping the internal velocity
of the gas quite low.  Note that the fact that we see a blueshift
relative to the quasar frame is consistent with the NW radio lobe
being the closer to us, as inferred from the radio jet being visible
on that side.

\subsubsection{Discussion}

The coincidence of optical components $a$, $b$, and $c$ with peaks in
the radio jet indicates that these features
are most likely due to optical synchrotron emission.
The plausibility of this conclusion is reinforced by our observation of
similar features in 3C\,2 and 3C\,245.

Understanding the nature of the aligned components that lie
beyond the detected radio structure is much more
difficult.  Are $f$ and $g$ truly
associated objects? The evidence in favor of association is entirely 
morphological, but if confirmed it could be very important for our
understanding of FR II radio sources and the alignment effect in
high-redshift radio galaxies and quasars.
Another object that appears to have a red aligned component outside
of its radio lobes is the $z \sim 0.7$ radio galaxy 3C\,441.
In this case, the red knot is only a couple of arc seconds
beyond the radio lobe and 
strong emission lines are seen from the region just within the radio
lobe (McCarthy et al.\ \markcite{mcC95}1995; 
Lacy \markcite{lac97}1997).
However, HST imaging and optical and infrared 
spectroscopy make it clear that
this is probably simply 
the result of the radio jet impacting a companion spiral galaxy,
in which the enhanced line emission comes from an interaction of the spiral
disk with the radio jet (Lacy \markcite{lac97}1997).

We will present a more detailed discussion of 3C\,212 in 
our upcoming paper, but we will briefly outline the nature of the
problems and some possible options here.

The problem is this:  some features of the morphologies 
of the optical components $f$ and $g$ suggest that their emission
has resulted somehow from radio jet interaction with the ambient material,
but the absence of any detected radio emission coincident with them
makes it very difficult to understand how any plausible scenario of
this type could work. 

Object $g$ has a projected distance from the lobe ``edge'' (in the 
high-resolution 3.6 cm map) of $\sim0\farcs5$, corresponding to
only $\sim3$ kpc. 
The narrow arclike wisps seen in the optical image
could conceivably result from the bow shock expected to form at
the head of a radio lobe propagating into the ambient medium. 
While the radio emission comes primarily from particles accelerated at
the hotspot and the beam shock interior to the contact
discontinuity, the broad bow shock itself is expected to be radio-quiet 
(Williams \& Gull \markcite{wil85}1985; Williams \markcite{wil91}1991).
This view is supported by rotation measure (RM)
observations of Cygnus A, in which
a radio-quiet bow shock associated with a hotspot is observed through a RM
discontinuity the shocked gas induces in the projected radio lobe emission
(Carilli, Perley, \& Dreher \markcite{car88}1988). The projected 
distance from the bow shock 
(RM discontinuity) to hot spot is $\sim$ 3$\arcsec$ or $\sim$ 3 kpc
for Cygnus A. 
The detection of the expected bow shock in Cyg A
means that
the association of the rest-frame UV emission in $g$ with a bow 
shock outside the radio lobe cannot be ruled out
on considerations of distance from the lobe alone. 
While continuum and line emission in the wisps associated with $g$ 
could result from thermal emission from shock-heated gas,
it would be difficult to explain the bulk of the
emission from $g$ with this mechanism.  Not only is the morphology
of $g$ as a whole unlike that expected for a bow shock, its color
is correct for a companion galaxy with an old stellar population and 
is much too red to be consistent with thermal emission from a shock.

Object $f$ is at a (projected) distance from the NW radio 
component of $\sim$20 kpc, or at least several times the
apparent lobe width. This emission cannot plausibly result from
a bow shock directly associated with the visible NW radio component.
We briefly enumerate various possible explanations for $f$, along
with some of their problems:

1) $f$ is an unassociated foreground object
at $z = 0.927$.  In this case, the continuum is naturally explained
as due to stars at the emission-line redshift, but the alignment and
the apparent morphological connection is purely fortuitous.  

2) $f$ is dominated by optical synchrotron emission
associated with undetected radio synchrotron emission.  Our upper limits
to radio emission at the position of $f$ at 3.6 and 20 cm mean that
the radio---optical spectral index must be flatter than 0.44, which
would be unusually low for an extended radio lobe. 
In the 
$z = 0.162$ radio galaxy 3C346, Dey \& van Breugel \markcite{dey94}(1994)
found an optical synchrotron knot associated with a radio knot with
radio spectral index $\sim$0.4. However, subsequent HST imaging established on
the basis of the high resolution morphology that
this optical synchrotron was probably from a jet rather than a hotspot 
(de Koff et al.\ \markcite{deK97}1997)

3) $f$ is dominated by
inverse Compton scattering of microwave
background photons associated with undetected or relic radio emission.
This option would require a large total energy to be present in
low-energy relativistic electrons (Lorentz factors of 10--100), which
would be expected to emit synchrotron radiation around 1 MHz.  While
in general this explanation would be consistent with $f$ being a
``relic'' lobe from either a previous position of a precessing jet
or a previous outburst of the radio source, in the sense that synchrotron
losses would eliminate the higher-energy relativistic-electron population
over time, the absence of a comparable level of inverse Compton
emission from the current radio
lobe is difficult to understand unless the relic lobe originally had
far greater radio power.

4)  The continuum in $f$ is due to jet-induced star formation from
a previous position of the radio jet or a previous outburst.  This
interpretation deals with the lack of detection of radio emission,
but it presents an additional problem:  assuming that the stars
share the velocity of the ionized gas (since the two components
appear to be cospatial), how could the pre-stellar gas be accelerated
to $\sim18000$ km s$^{-1}$ while presumably retaining a velocity
dispersion comparable to that we now see in the [\ion{O}{2}] line?
This last option seems very unlikely.

None of these options seems very satisfying at this time. 
We have recently obtained additional deep spectroscopy and imaging that may
eliminate some of the possibilities or suggest others when fully
analyzed.  In the meantime, we can only emphasize that we regard
the nature of the exterior aligned components
in 3C\,212 as possibly one of the most important questions raised by this
investigation.

\subsection{3CR\,217}

\subsubsection{Morphology}
We display in Fig.\ \ref{3c217fig}$A$ and $B$ our HST and $K'$ images
of the radio galaxy 3C\,217. 
This is the only object in our sample without a detected radio core
available in the literature. In this case we cannot
align the
radio and the optical frames by aligning the optical/infrared
nucleus with the radio nucleus. 
Therefore, the radio lobes we indicate in Fig.\ \ref{3c217fig}
by the white cross and the white arrow show the approximate
positions of the radio lobes, referenced to the measured optical core position
(Pedelty et al.\ \markcite{ped89}1989). The lack of a radio core 
and the positional uncertainty make 
interpretation of the radio/optical morphologies somewhat ambiguous.

The optical image consists of an elongated galaxy with a very
faint nuclear peak (the faintest in our sample); this
peak is best seen in the inset to the magnified version of
the HST image, shown in panel $C$. $D$ shows  
the same magnified version, slightly Gaussian smoothed.
The near-infrared component appears round, and is centered 
on the optical peak. 
The seeing for the $K'$ image was poorer than average at
0\farcs83, however, and the infrared component is
not well resolved. Nonetheless, subtracting the best PSF leaves some diffuse
flux in the near-IR, and, for the $K$-band flux
to be entirely due to the unresolved component seen in the HST 
image, it would have to be extremely red ($\alpha = 2.9$).
There are also two semi-discrete components or 
companions (objects $a$ and $c$ in 
Fig.\ \ref{3c217fig}$A$). 
Both objects lie within the lowest envelope of 
extended emission in the HST image; the optical structure
of the central component extends towards $a$ to the NE, while the
object $c$ is elongated towards the SW side of the central component.
Component $a$ is rounder, more compact and much redder than $c$.
If these are included in a PA determination, the whole ensemble has a PA
of 63\deg. 
They are oriented at close to the angle of
the elongation of the rest of the galaxy; however, if they
are excluded on isophotal cutoff or aperture considerations, the 
galaxy PA is 79\deg. Therefore the dominant 
morphology of the  galaxy differs in orientation by 
$\sim30\deg$ 
from the PA between the radio 
lobes (105\deg) regardless of the details of what is included in
the moment analysis. 
This galaxy is therefore globally not very well aligned with the radio lobe
direction.
There is an extension ($b$ on Fig.\ \ref{3c175fig})
that is very linear at its highest isophotal level 
and has a PA (relative to its own
center) that matches the radio lobe PA
to within a degree. This extension is seen best in panel $C$;
it points directly towards the eastern radio lobe with the radio-optical 
frame centering we have chosen, but its end 
is offset by 0\farcs5 from the optical peak. 

\subsubsection{Discussion}

This radio galaxy is relatively unaligned and has a faint optical nucleus. 
Nonetheless, the radio jet could conceivably be involved
in the creation of the linear feature $b$, but only if 
the jet has moved 
through some fairly large angle ($\sim$ 20--30\deg).
The component $a$ is the reddest object 
among the nearby companions
or semi-discrete components that we measured in any of our 10 sources, 
while component $c$ is among the bluest.
This difference 
is unexpected from the appearance of the HST image, in which the two 
companions look fairly symmetric in distance and alignment with the rest of
the galaxy.  Assuming they are both associated, they must be 
quite different in nature; the color and morphology of $a$ is consistent 
with a companion galaxy, 
while the blue color of $c$ is similar to those of the
companion objects 3C\,196 $d$ and $e$.
Without a better radio map of this object,
particularly one with a radio core detection, it is difficult 
to determine whether 
the radio activity is related to the morphology of extension $b$ or to 
the colors of companions
$a$ and $c$. Certainly, $a$ resembles the many other red companion galaxies 
commonly seen around high-$z$ radio galaxies (Rigler et al.\
 \markcite{rig92}1992).

\subsection{3C\,237}
\subsubsection{Morphology }
This radio galaxy is a compact steep spectrum source, with an 
angular size of 1\farcs3, and an unresolved,
flat-spectrum core (van Breugel et al.\ \markcite{vanB92}1992).
We display the HST and Keck $K'$ images of the galaxy in 
Fig.\ \ref{3c237fig}$A$ and $B$; in panel $C$ we show a magnified 
version of the HST
image, with the fairly symmetric radio lobes of van Breugel et al.
marked with white crosses in an inset.
As can be seen in Fig.\ \ref{3c237fig}$B$, the HST image shows a slightly
elongated galaxy
with several close companion galaxies. Our Keck $K'$ band image shows
a fairly round galaxy (axial ratio $b/a$ $\sim$ 0.93)
whose center coincides with the optical peak to within 0\farcs05.
Despite its low ellipticity, the IR position angle is stable and aligned to the
radio axis within 1\deg\ at the lower contour levels. 
(We exclude the nearest discrete companion galaxy in our magnitude and 
moment analyses; it is at a distance of 3\farcs9.) 

The 15 GHz VLA map of van Breugel et al.\ has an angular resolution
$\sim$0\farcs15, close to the resolution of our HST image. 
The optical image
has a fairly dominant optical nucleus, with two
lobes in a morphology roughly similar to that of the radio. If we align
the radio and optical cores, the E lobe radio and optical peaks are 
fairly well aligned; the western radio lobe, however, is more
extended and peaks past the optical emission. 

\subsubsection{Discussion}
Despite the fairly close correspondence
in scale and overall alignment between the radio and optical morphologies, 
there does not seem to be a close point-to-point coincidence. 
We therefore do not find for this object a compelling morphological case
for the aligned, extended structure to be non-thermal radiation, as we
do for 3C\,2. There is, however, an unresolved core in both the
HST image and  the van Breugel et al. \markcite{vanB92}(1992) radio map, and
its radio-optical spectral
index is 0.72; there is no problem with ascribing the optical core
to synchrotron emission.
The radio-optical spectral indices for the ``peaks'' of the
extended material is in the range of 1.0--1.3 for the two lobes
(depending on details of what flux is included in the estimate).
Regardless of how much of the aligned emission is from synchrotron
(or other non-thermal process) the kind of alignment shown is 
similar in some ways to what is seen in other small-scale radio 
sources, which tend to be closely aligned.
The radio and optical structures are roughly co-spatial, as are
those of 3C\,368. 3C\,237 is much redder, however, presuming the bulk of
the near-IR is coming from the same origin as the optical flux, which is
consistent with the close alignment of the near-infrared with the radio
axis. 
It is less likely, therefore, to be dominated by
nebular continuum, as is 3C\,368.

\subsection{3C\,245}

\subsubsection{Morphology}
We show in Fig.\ \ref{3c245fig}$A$ and $B$ our PSF-subtracted
HST and Keck images of the quasar 3C\,245. 
This quasar had the brightest nucleus in our sample,
making detection of underlying
extension more difficult. We were also hampered in our $K'$ imaging
by an unfortunate period of poor seeing ($\sim$ 1\farcs3).
Both images 
clearly resolve the companion galaxy 2\farcs3 southeast of
the quasar nucleus, studied 
by LeF\`evre \& Hammer \markcite{leF92}(1992). 
Their spectra show this to be a fairly normal elliptical, at $z = 1.013$.

In the HST image, other than this elliptical galaxy, the highest surface
brightness extension is a linear feature (containing
components $a$ and $b$) to the west of the
nucleus. If we align the quasar with the core of the 5-cm
map of Laing \markcite{lai89}(1989), this optical feature coincides exactly
with the radio jet. The direct correspondence, even in the position and
relative brightnesses of the 
radio and optical hotspots, is shown in Fig.\ \ref{3c245fig}$C$.
The inset at the lower right shows the central 3\arcsec$\times$4\arcsec\
region of the PSF-subtracted quasar image, with contours from
the 5-cm map of Laing \markcite{lai89}(1989) superimposed. The radio map
contours are offset vertically by 1\arcsec\ so that the optical image
is easily visible for comparison.

In addition, there is low-surface-brightness extended flux to the east of
the nucleus (including a peak at $d$ and extending down to the 
elliptical galaxy $e$) lying in the diffuse, extended eastern
radio lobe of 3C\,245;  the approximate
center of this lobe is indicated by a cross 
(Laing \markcite{lai89}1989; Liu, Pooley, \& Riley \markcite{liu92}1992).
The 5-cm map of Laing shows 3 faint peaks centered around the cross
position; we see a corresponding 3 peaked structure in the optical
emission. This structure could be due to nebular thermal continuum
in our HST bandpass, if there is $\sim10^4$ K ionized gas in this region. 
Unfortunately, our [\ion{O}{2}] image (Fig.\ \ref{3c245fig}$D$; 
continuum subtracted) has insufficient signal-to-noise to 
show whether there is [\ion{O}{2}] at a correspondingly low surface brightness
level at this location. However, we do see resolved [\ion{O}{2}] line emission
extending for about 3\arcsec\ to the north, and at the location of
the elliptical galaxy. In fact, the PA of the [\ion{O}{2}]
emission line region is $\sim$ 3\deg\ (measured in a 2\arcsec\ radius
aperture, but excluding the elliptical); this orientation is within 2\deg\
of perpendicular to the radio axis (as defined by the lobes; the jet direction
is rotated by 10\deg).

\subsubsection{Analysis and Discussion}
We give fluxes for the components seen in the HST image in Table 
\ref{photomtab}.
The $K'$ image has far too low resolution to allow us to get 
colors (or even useful limits)
for the extensions near to the quasar, other than the bright elliptical. 

The linear feature containing components $a$ and $b$ is certainly 
associated with the radio jet, and the small-scale correspondence 
between the radio and the optical makes it very likely that the optical emission
is synchrotron radiation.
From the Laing \markcite{lai89}(1989) contour map, we find
the approximate surface brightness ratio (at 6 cm) of
component $a$  to component $b$ is 2:1. In our optical image, $a$ has a 
surface brightness $\sim3$ times that of $b$. 
From the plot of radio spectral index between 5 and 15 GHz, $\alpha^5_{15}$, of 
Liu et al.\ \markcite{liu92}(1992), $a$ has $\alpha^5_{15}$ $\sim$ 0.8 and $b$
has $\alpha^5_{15}$ $\sim$ 0.9. 

If there were no high-frequency turnover to the spectral energy
distribution, the surface
brightness ratio between the two components would 
increase by a factor of 3 from observations at 6 cm to 6700 \AA.
This is a greater increase in the surface brightness ratio
than we observe; however, high-frequency turnovers are usual
in synchrotron spectra where optical synchrotron has been 
observed. 

\subsection {3C\,280}
\label{sec:3c280}

\subsubsection{Optical Morphology}

Under ground-based seeing conditions, the radio source 3C\,280 has been
observed to be 
associated with an elongated galaxy that is well-aligned with the radio axis
(within 1\deg : McCarthy et al.\ \markcite{mcC87}1987; 
within 20\deg: Rigler et al.\ \markcite{rig92}1992). As seems to be 
the norm, with HST's resolution this elongated galaxy separates
into several discrete components. 
However, even by the standard of previous HST images of high-redshift
galaxies, 3C\,280 is unusual.
As shown in Fig.\ \ref{3c280fig}$A$ and $C$, the elongated, aligned
UV component is resolved into a two-component central
region (peaks $a$ and $b$), connected to a narrow structure ($c$) 
by a semi-circular arc ($d$).
This arc is also clearly seen in the WFPC2 image of Best et al. (1996).
The length of the arc is $\sim$1\arcsec;
at the redshift of the radio galaxy this corresponds to $\sim$6 kpc. 
We see no similar structures in the rest of our sample,
nor are we aware any similar features found in any other
high-redshift sources.

The objects $a$, $b$, and the peak of $c$ are
collinear, at a PA of 80\deg, within 10\deg\ of the radio axis, as 
determined from the lobe hotspots. (Unfortunately, we have no high resolution
radio data available to use in this morphological comparison).  Object $c$ 
is elongated almost perpendicularly to this axis.  Object $a$
is unresolved, with an intrinsic width of 0\farcs14 on the HST image, and
presumably represents the position of the active nucleus and thus 
the radio core.

The arc $d$ can also be seen in our continuum-subtracted
CFHT [\ion{O}{2}] image shown in Fig.\ \ref{3c280fig}$D$);
the arc is best seen in the inset.
(This image had an
initial resolution of 0\farcs7, and its resolution was improved
by deconvolution with the Lucy restoration algorithm and subsequent 
reconvolution with a Gaussian).

Components $c$ and $d$ appear to be edge-brightened,
along the width of each feature, and well aligned with the radio jet
direction as can be seen clearly in Fig.\ \ref{3c280fig}$C$.
The brightened region in the edge of the arc has
about twice the surface 
brightness of the rest of the curve and extends out to 
a PA of about 30\deg\ from the radio axis, or 40\deg\
from the optical axis. This is consistent with 
the probable opening angle of the quasar opening cone 
derived by Barthel \markcite{bar89}(1989) and 
Saikia \& Kulkarni \markcite{sai95}(1995).

\subsubsection{The Near-Infrared Morphology}

In the near-infrared (rest-frame 1 $\mu$m) the galaxy is fairly
round and resolved with no obvious
sign of the discrete aligned components seen at rest-frame 0.33 $\mu$m.
Its center is coincident
with the brightest HST peak, the unresolved component $a$, to within 0\farcs06. 
(The $K'$ image has an intrinsic  FWHM
of $\sim$0\farcs58).

We fit the elliptical galaxy using the STSDAS routine $ellipse$,
which 
fits elliptical isophotes using the iterative algorithm of 
Jedrzejewski \markcite{jed87}(1987).
If we let the PA, ellipticity, and center vary, we obtain 
the profile shown in Fig.\ \ref{3c280ellfig}. This is 
well fit by a de Vaucouleurs $r^{1 \over 4}$-law profile 
between 0\farcs5 and 1\farcs46 
(the radial region in which the elliptical isophotes were good fits, and 
outside of the inner, seeing-flattened region). The effective 
radius is $r_e$ = 0\farcs86 $\approx$ 5 kpc. 
The PA of the elliptical isophotes varies from about 65\deg\
to $80\deg$ and the axial ratio ($b/a$) varies from 
about 0.87 to 0.91 in this radial region. We show the PA as a function of
the semi-major axis in the lower panel of Fig.\ \ref{3c280ellfig}. 
The PAs derived from these elliptical isophotal fits are consistent
with the results of the moment analysis discussed in 
\ref{sec:morphextend}; the dependence on radius translates roughly to
the isophotal cutoff and size of the apertures used in the formal
moment analysis. 
We find that the $K'$ light is aligned with the radio PA
within $25\deg$ at all isophotal levels, and more closely 
with the axis defined by the optical components $a$ and $b$.

To investigate the contributions of the 
discrete aligned components, we fix the axial ratio ($b/a$ = 0.9)
and PA ($\sim$ 70$\deg$), derive an intensity profile, and
use this to make a 2-dimensional model of the galaxy. We subtract 
the model from the infrared image of the 
radio galaxy, varying the center and scaling
slightly to reduce the residuals.
There is evidence from the best-fit residual for some flux
at the position of the component $c$ in the HST image. 
However, a discrete contribution from this
component cannot have caused the elongations
along PA$\sim$ 70--80$\deg$ that we observe, 
since the elliptical isophotal fits have relatively
consistent PAs and axial ratios through all the isophotes.
To further verify this, we subtract the galaxy's azimuthal 
average from itself and find
the obvious bimodal, large-scale residuals  
(along the axis defined by the best PA of the elliptical
isophotes), that are consistent with a non-circular morphology. 
(We also see again the extra flux at the position of the component $c$).
In addition, to make sure this elongation we see is not the result of
an elongation in the observed PSF, we have checked the position angle 
and axial ratio of the star on the $K'$ frame
and seen no preferential elongation. The stellar image is very circular; 
position angle determinations vary randomly over various apertures
and isophotes. 
This best-fit elliptical galaxy has a total $K'$ magnitude of 
17.0 within an aperture of 8\arcsec, consistent with the $17.1 \pm 0.3$
$K$ magnitude found by Dunlop \& Peacock \markcite{dun93}(1993). 

\subsubsection{Photometry}

We now use the elliptical model to scale and subtract from the
smoothed version of the HST image to try to determine an
optical-infrared color for this extended elliptical component.
We smooth the HST image with a Gaussian until the star that falls in
both the $K$ frame and the HST frame matches in FWHM; we then 
determine the colors of any shared components. We subtract off the 
$K$ elliptical galaxy model in steps until we have started to over-subtract
the wings; this defines the upper limit to how much of the elliptical
flux could be present in the HST $\lambda_0=3300$ \AA\ image. We make
this estimate from the unsmoothed HST image as well, as the wings of the 
elliptical outside of $r=0\farcs6$
should not be affected by the seeing, and the unsmoothed image will provide
better contrast against the nucleus. We find consistently from the
smoothed and unsmoothed images
that the ratio
of the HST to $K$ extended envelope is between 0.2 and 0.4, with 0.3 providing
the best apparent fit. 
Assuming all this flux were stellar, from a
single generation of stars, and without significant reddening,
comparison with Bruzual-Charlot (1993) models shows that this color would fit
a population with an age of between 3 and 4 Gyr, comparable to the age of
the old stellar population found for the $z \sim 1$ radio galaxy 3C\,65
(Lacy et al.\ \markcite{lac95}1995; Stockton, Kellogg, \& Ridgway
\markcite{sto95}1995)

We use these elliptical-subtracted HST images to obtain photometry
of the aligned components alone, using isophotal cutoff levels
of about 3--4 $\sigma_{sky}$. (This level is
comparable to the lowest grey level in the upper-right inset in
Fig.\ \ref{3c280fig}$C$). We give the results of this photometry in
Table \ref{photomtab}.

\subsubsection{Spectroscopy}
Although we have used essentially line-free filter
bandpasses to maximize contributions from stellar continuum radiation,
our bands could still be
dominated by emission from an ionized gas.
For our HST imaging, which covers the rest-frame region near
3300 \AA, 
the most likely potential contributors are thermal continuum
emission processes ({\it i.e.,} free-free, free bound, and 2-photon emission)
and the [\ion{Ne}{5}] $\lambda$3346,3426 lines.  
For 3C\,280, we see in our HST image
a relatively fainter echo of the [\ion{O}{2}] emission-line morphology
associated with the eastern radio lobe (Rigler et al.\
\markcite{rig92}1992; Fig.\ 10$D$).  This similarity increases the likelihood
that at least some of the structure we see in our HST image
may be dominated by gaseous emission.

We have recently obtained deep Keck LRIS spectroscopy of 3C\,280, which
will be reported in detail later (Ridgway \& Stockton \markcite{rid97a}1997a).
Here we restrict ourselves to an evaluation of the ionized gas
contribution to our HST images.
We consider three regions:  the central region, comprising $a$ and $b$ in
\ref{3c280fig}$A$, object $c$, and the faint emission just N of the E 
radio lobe.  We have evaluated the
contribution of the nebular thermal emission by scaling a model at
$10^4$ K to the observed H$\delta$ line, and we have measured directly the
equivalent widths of the [\ion{Ne}{5}], \ion{O}{3} $\lambda3133$, and
\ion{He}{2} $\lambda3203$ lines.  We find that the emission lines in all
regions account for 5\% or less of the total flux in the F622W filter.
The thermal continuum from the ionized gas is more significant:  it comprises
about 20\% of the total flux from the central region and about 50\% of
the total from object $c$ and the optical features in the E radio lobe.
In all cases, we clearly seem to have a continuum component beyond that
from the nebular thermal emission.  Although we cannot isolate a spectrum
of the arc $d$ alone in our data, its [\ion{O}{2}] $\lambda$3727 to 
continuum ratio is very
similar to that of $c$, and we assume that our conclusions for $c$ apply
to $d$ as well.

\subsubsection{Discussion}

3C\,280 is one of the more classical cases of the alignment effect
in our sample; i.e., one
in which a number of discrete components are quite linearly arranged, close
to the presumed radio jet direction. 
We know that this aligned morphology is traced by both the line and continuum
emission, and that there is probably a significant nebular
thermal continuum contribution to the main peak ($a + b$) and a
larger, possibly dominant, one to $c$ and $d$,
as well as to the extended material around the eastern radio lobe. 

One of the more intriguing and unusual
features of its high-resolution morphology is the arc between components
$b$ and $c$. 
What are possible origins for this curved structure, along with the rest 
of the aligned morphology? 

First, we consider the possibility that it is a gravitational arc; 
i.e., that it is either a distant
object lensed by some mass associated with 3C\,280, or a component of
3C\,280 lensed by an intervening mass.
It is, however, extremely unlikely that the arc ($d$) is 
a gravitational image of a distant object for several reasons: 
first, the presence of line emission
tracing the arc morphology at the redshift of the radio galaxy; second,
the edge brightening, plausibly associated with a cone of quasar
illumination approximately centered on
the radio axis; third,
the fact that the $K$ elliptical galaxy is centered at the peak of the 
optical emission (component $a$) rather than somewhere near the center of
curvature of the arc, and that there is no other likely candidate for the
lensing mass.  The second possibility, that of an intervening lensing
mass gravitationally lensing some 
feature associated with 3C\,280, would remove the
first objection, but not the second and third.  Indeed, the lack of
detection of a plausible lens mass becomes more acute as it is moved
to lower redshifts.

Another interpretation is that the arc could be a tidal tail from $c$.
A simulation by Mihos \markcite{mih95}(1995) of the detectability with 
WFPC2 of tidal tails
resulting from disk galaxy mergers indicate that from normal
spirals, tidal merger remnants such as tails would only 
be visible at $z \sim 1$ 
for about 200 Myr after the initial interaction. (Mihos assumes
a total exposure time with the WFC of 10,000 s, similar to our 3C\,280
exposure of 8800 s).  While the arc in
3C\,280 is not only detected, but could have been detected in a
considerably shorter exposure, its relatively high surface brightness
is almost certainly largely due to enhancement from external
ionization and scattering. 
In this view, the alignment of $c$ and $d$ with the radio axis would be
enhanced as a result of their lying within the ionization cone of
the active nucleus.

However, because of the close linear alignment of the peaks of $a$,
$b$, and $c$, this, like 3C\,324, is another case of an aligned structure that
is not consistent with scattering and ionization
filling an entire ionization cone. While the edge-brightened regions
of $d$ and $c$ do fill a fairly large angle ($\sim 30\deg$), the peaks
of $a$, $b$, and $c$ are aligned within degree or so, and raise
the possibility that this axis may correspond to the jet direction, and
that jet/cloud interactions are affecting the morphology.
Jet/cloud interactions and related shocks are considered a possible 
cause of the closely aligned components in 3C\,368 (Stockton et al.\
\markcite{sto96}(1996); Clark et al.\ \markcite{cla97}(1997), which 
also have large thermal continuum contributions.

In the near-infrared, the morphology and colors are consistent
with an elliptical galaxy. 
However, there may also be a contribution from an 
unresolved core, and the alignment of the
elliptical isophotes with the radio axis suggests 
that there is some contribution from an ``aligned'' component.  
The aligned components $c$ and $d$ would still be 
resolved from the central component $a$ at the ground-based 
resolution, but the component $b$ would not be.
The infrared morphology could not track the optical exactly,
as the central component is much rounder
than the combined $a$ and $b$ component in the HST image
when smoothed to ground-based
resolution.
However, the PA of the elliptical isophotes is close to the 
axis defined by $a$ and $b$, and the alignment of the 
observed infrared morphology 
could be most easily explained by the addition of the unresolved $b$
to the red resolved, round component that is 
much more dominant at $K$ than in the optical.
A second possibility is the contribution 
of some new red aligned component, perhaps a minor synchrotron 
component with a steep enough spectrum to be unobservable in the blue,
that is sufficient to align the infrared isophotes.
The third possibility is that even at high-resolution the infrared 
isophotes would be aligned, and the major axis of a 
nearly round ``elliptical galaxy''
is aligned with the radio axis. 
This question is interesting 
in light of our finding in several of our radio galaxies
that the infrared elliptical
isophotes of the central components are aligned with
the radio axis;
indeed, Dunlop \& Peacock \markcite{dun93}(1993) claim for their 
$z \sim 1$ sample that
the infrared morphologies are better aligned than the optical. 
In this particular example, the optical and infrared alignments
match well, and the simplest explanation is that the
infrared alignment comes from the unresolved contribution
of $b$. 

There appears to be a cluster around 3C\,280; we see some objects which
seem to have [\ion{O}{2}] emission at the redshift of the radio galaxy.
There is also
evidence for a cluster in the detection of extended X-ray emission
(Worral et al.\ \markcite{wor94}1994; Crawford \& Fabian \markcite{cra95}1995).

\subsection{3CR 289}
We show in Fig.\ \ref{3c289fig}$A$ and $B$ our HST and Keck $K'$ images
of the radio galaxy 3C\,289. This galaxy is unlike the rest of the galaxies
in our sample, in that it shows some discrete, 
elongated structure within the central few arcseconds
that is severely misaligned with the radio axis ($\Delta$PA=75$\deg$).
The components 
$a$, $b$ and the central component are collinear, at a PA 
of $0\deg$. 
This collinear feature looks superficially
similar to the ``chain'' of components seen in 3C\,212, but that feature
lies directly on the radio axis. 
At higher isophotal levels the central component has a 
PA within $20\deg$\ of the radio axis; PSF subtraction shows most of
the central component to be unresolved, but with a small bright asymmetry
as well
as the obvious low surface brightness material.
Using azimuthal averaging
to remove the asymmetric material in a manner
similar to that used for 3C\,2. we
derive a flux for this extended low surface brightness material 
and give this value 
in Table \ref{photomtab} under the label ``elliptical''. (This is the flux of
the symmetric optical material, minus that of the unresolved core).

In the infrared, the galaxy looks morphologically similar to its
appearance in the optical. There is a dominant central ``elliptical galaxy''
component; outside of the seeing radius, the isophotes are
well fit by a de Vaucouleurs profile with $r_e = 1\farcs2 \pm 0\farcs3
\sim7$ kpc.
The PA is within $10\deg$ of the PA
of the elliptical component in the optical ($30\deg$ from the radio
axis). These PAs are close, as well, to that of 
extension $c$. The [\ion{O}{2}] image shows a bright extension that
matches $c$ well; this is then a case in which the emission
line morphology is better aligned with the radio axis than is
the optical or infrared continuum. 
In the near infrared, the elliptical central component dominates 
the components $a$, $b$, and $d$, though $a$ and $d$ are also detected in the 
infrared. The optical and infrared fluxes for the symmetric material
are estimated by azimuthally averaging the images and therefore removing
any non-symmetric components, and 
are given in Table \ref{photomtab} with the label ``elliptical''. 
The optical flux is corrected for the unresolved nucleus, while
the infrared value is not corrected for any unresolved component
contribution.  We have only a limit to the infrared nuclear
component, but the contribution should be small (at most 20\%).
Component $d$ is unresolved on the HST image with FWHM
$\sim 0\farcs15$, while $a$ is marginal and $b$ is clearly resolved with
FWHM $\sim 0.3$. Component $d$ could be a faint, red foreground star, 
while $a$ and $b$ are potentially companions or components of 3C\,289.

\subsection{3C\,336}
\subsubsection{Morphology}
We show in Fig.\ \ref{3c336fig}$A$ and $B$ our HST and Keck $K'$ images
of the quasar 3C\,336; 
this is the source with the largest angular size in the radio of any
in our sample (28\arcsec). 
The radio structure of this quasar has been well-studied by Bridle et al.
\markcite{bri94}(1994).
We give the directions toward each radio lobe as arrows 
in panel $A$. 
In the HST image, we have resolved low-surface
brightness extension near to the nucleus; this can be seen 
well in panel $C$ where we display a magnified version of
the PSF-subtracted HST image, slightly Gaussian-smoothed. 
Also seen are PSF subtraction residuals: this quasar had the worst 
match between PSF and nucleus, particularly in the diffraction spikes.
(We chose the subtraction scaling we display here by requiring
approximate monotonicity in regions which exclude these diffraction
spike residuals.) 

Bridle et al.\ detect a radio jet to the SW; the knot 
nearest to the nucleus in this jet is at a radius of 0\farcs97
and PA $-162\deg$.
The brightest extension
we have resolved around 3C\,336 in the HST image is at a similar radius,
on the jet-side of the nucleus, but it is not coincident with the
radio knot. 
In fact, this radio knot position corresponds to
a break in the extension to the south of the quasar, to
the west of extension $a$, giving a ``channel''
effect similar to that seen in Cygnus A (Stockton, Ridgway \& Lilly
\markcite{sto94}1994). 

We display in panel $D$ the PSF-subtracted, continuum-subtracted [\ion{O}{2}] image.
The emission line region is very asymmetric with respect to
the nucleus, and it has low surface brightness structure to the northwest
that is approximately perpendicular
to the radio axis. There are relative peaks in the emission
at the positions of the continuum components $b$ and $a$.
The [\ion{O}{2}] peak near $a$ is at a radius of 1\farcs0 and 
PA $-172\deg$; though the [\ion{O}{2}] is at lower resolution than 
both the HST image and the radio map, this [\ion{O}{2}] knot seems 
to be better associated with the peak $a$ in the HST image (PA$\sim-176\deg$)
than with the position of the radio jet knot.

\subsubsection{Discussion}
We see no evidence for optical synchrotron emission in this quasar.
The large scale size of the radio source could mean that the
radio axis is closer to the plane of the sky than it is for
the other quasars in our sample, preventing our detection of
beamed optical synchrotron components  
like those we see in at least three of our other quasars.

We cannot be certain from our imaging data alone that the
extensions we see are necessarily
associated with the quasar; the spectrum of the quasar shows 
6 metal absorption line systems, indicating the presence of several
foreground objects along the line of sight (Steidel et al.\ 
\markcite{ste97}1997).
The field of this quasar has been quite well studied from
the ground.
Bremer et al.\ \markcite{bre92}(1992) made a spectroscopic study of 
the [\ion{O}{2}] extension and detected extended line emission
out to 5\arcsec\ north and 3\arcsec\ south. 
This is qualitatively consistent with our [\ion{O}{2}] image.
Hintzen, Romanishin, \& Valdes \markcite{hin91}(1991)
found an over-density of field objects, and suggested
that 3C\,336 was likely in a rich cluster. 
A spectroscopic survey of the field by Steidel \& Dickinson 
\markcite{sted92}(1992)
identified a galaxy 5\farcs7 to the southwest of the quasar 
as responsible for the $z=0.4722$ absorption
line system. 
Very recently, Steidel et al.\ \markcite{ste97}(1997) have completed
a comprehensive survey of galaxies in the 3C\,336 field, using HST WFPC2
and ground-based imaging, and Keck spectroscopy.  While their main
interest is in identifying galaxies responsible for the multiple
absorption-line systems seen in the spectrum of the quasar, they
find that the objects we have labelled $a$, $b$, and $c$ have,
respectively, redshifts of 0.927, 0.931, and 0.928, showing a clear
association with the quasar at a redshift of 0.927.  From both the
morphology and the close agreement in redshift, it appears likely
that $a$ and $c$ are either features in the quasar host galaxy or companions
in the process of merging with it.  Object $d$ has
a redshift of 0.892 and is responsible for one of the
absorption-line systems.  If we ignore $d$ (as a confirmed foreground
galaxy) and $b$ (as a discrete companion), the remaining objects $a$
and $c$, which appear to be embedded in diffuse surrounding emission 
and to be at essentially the same velocity as the quasar, are fairly well
aligned with the radio axis.
In addition to $a$, $b$, and $c$, Steidel et al.\ find 6 galaxies
with redshifts close to that of the quasar within a radius of 50\arcsec,
demonstrating that the quasar is in a significant group or cluster.

\section{Properties of the Sample}
\label{chap:propsamp}
Here we discuss the results from our isophotal and annular magnitude 
and moment analysis
of all the objects in our sample. 
To determine the properties of the extended material, we 
first exclude known or likely companions. 
We recognize from the start that it will be difficult 
with this small sample
to judge whether the  
radio galaxy and quasar subsamples could be drawn from the
same population, particularly
for properties such
as the total magnitude, color, and $\Delta$PA,
since the intrinsic dispersion in the properties in each subsample is
large.

\subsection{Magnitudes of the Extended Material}
\label{sec:magextend}

If radio galaxies
are unbeamed versions of quasars, the extensions of both classes
(if they are host
galaxies) should have about the same total luminosities. 
The photometry
for the extended material in the HST images is given in Tables \ref{qsohst} and
\ref{rghst}, and for the near-infrared images in Tables \ref{qsoir} and 
\ref{rgir} (all values are normalized to redshift 1).
We give in these tables both isophotal flux densities (down to a common
limiting isophotal level) and flux densities within an aperture, excluding 
a central
PSF-subtraction region for both the quasars and for the radio galaxies, 
as well as with no central exclusion for the radio galaxies. In addition,
for some of the 
quasars with sufficient extended material, we give flux densities with the
central subtraction region interpolated with a low-order polynomial.
The photometry generally excludes spectroscopically determined foreground 
objects; details of the
calculations are given in Appendix \ref{chap:moment}.

For inter-comparison, we use the flux densities that
exclude the region of PSF subtraction
(generally the inner $0\farcs45 \sim 2.7$ kpc radius for the HST images, 
and the inner $0\farcs85 \sim 5$ kpc for the near-infrared images),
and we also exclude values
below the lowest common normalized isophote above sky noise.

For the WFPC2 data, we
find that the average flux density (all values are in units of
10$^{-20}$ ergs s$^{-1}$ cm$^{-2}$ 
\AA$^{-1}$) in the annular aperture 
0\farcs45--3\farcs75 above our fiducial 
isophotal cutoff level is 206$\pm$137 for the quasars, and 124$\pm$95 for the 
radio galaxies,
where the errors given represent the dispersion in the flux over each set of 5
objects. For the total annular flux density, without application of the
isophotal cutoff level, the quasars have an average value of 398$\pm$116,
and the radio galaxies have an average flux density of 308$\pm$96. 

The optical flux densities of the two subsamples therefore match well
within the observed dispersion, despite 
the fact that we see a beamed
optical synchrotron component in the quasars but not in the radio
galaxies. 
We can estimate this contribution by explicitly subtracting
the flux from the probable
optical synchrotron components 3C\,2$a$, 3C\,245$a,b$, and 3C\,212$a$,$b$,
$c$, giving a mean total annular flux density for the quasars
of 360.6 $\pm$ 116. This change is negligible within the error.

In the $K'$ data, we find 
for the quasars, the average total isophotal
flux density is 168 $\pm$156, and the average total annular flux density is
253$\pm$168; for the radio galaxies, the average isophotal flux density 
198$\pm$133, and the average total annular flux density is 302$\pm$169.
(The synchrotron contribution is minor, and resolved in the infrared
only in 3C\,2.)
Our results are consistent with the unification hypothesis of AGNs,
in that we see no evidence for a significant luminosity difference between
the quasars and the radio galaxies for extended material that
plausibly could be identified with a host galaxy. The intrinsic 
dispersion in this small sample is large, however.

As the optical quasar extensions seem generally to have larger synchrotron
contributions,
the $K^\prime$ fluxes of the extensions are
more appropriate
to use in considering how these quasar extensions compare
to earlier quasar host and radio galaxy determinations.
First, we should try to correct, if possible, for flux obscured 
in the PSF-subtraction region.
The average flux added to the quasar total fluxes by interpolation over the 
central region is only 34.7, and the largest correction 
is that for 3C\,2, whose inner region 
is affected by the high surface brightness northern synchrotron component.
More relevant is the comparison of 
the total $K^\prime$ fluxes for the radio galaxies
with and without exclusion of the nuclear region. We find the average flux
difference is 194 $\pm$ 72, which corresponds to a 
difference of 0.55 mag. If we then apply this average
correction to the quasar hosts, we can estimate the average $K^\prime$
magnitude to be 17.4. Both the mean and the dispersion
of the quasar hosts are consistent with the $K$---$z$ relation
for radio galaxies ({\it e.g.,} Lilly \markcite{lil89}1989). 

We can further attempt to estimate how the radio
galaxies and quasar hosts compare with an $L_K^*$ galaxy at the present
epoch.
Mobasher, Sharples, \& Ellis \markcite{mob93}(1993) find $M_K^*=-25.1$
(for $H_0=50$), independent of galaxy type.  Converting to our assumed
$H_0=75$ km s$^{-1}$ Mpc$^{-1}$ and placing such a galaxy at our
fiducial redshift of $z=1$, we find that it would have a flux density
of 176 (in our standard units of $10^{-20}$ ergs cm$^{-2}$ s$^{-1}$ \AA$^{-1}$)
in the redshifted $K$ band.  To determine the flux density in our {\it
observed} $K'$ band, we must include the portion of the usual $k$-correction
due to the difference of the galaxy SED between the rest-frame and
observed bandpasses (we have already included the ($1+z$)$^{-1}$ bandwidth
factor in the above calculation).  The spectral index between 1 $\mu$m 
and 2.5 $\mu$m
differs very little among different stellar populations with ages greater
than $\sim1$ Gyr; we assume a 4-Gyr-old population, consistent with
the color we find for the elliptical component of 3C\,280, our best
studied case.  From a Bruzual-Charlot \markcite{bru93}(1993) model,
we find a flux-density ratio of 4.58, so our $L^*$ galaxy at $z=1$
would have an observed $K'$ flux density of 807, where we have ignored
any evolutionary corrections.  Taking our actual 
observed average values for our sample of quasar hosts and radio galaxies,
and correcting for both the average 0.55 mag loss in the masked central region
and an estimated 0.2 mag aperture correction, we find formal corrected flux
densities of $506\pm183$ for the quasar hosts and $604\pm184$ for the
radio galaxies.  Thus we find these galaxies to be roughly $L^*$ or a
little fainter.  Given potential systematic uncertainties in this result
and the apparent range in host-galaxy properties both in our sample and
in the low-redshift samples, it would be premature to claim that our result is
seriously at variance with the finding that low-redshift QSO hosts
average 0.4--0.7 mag {\it brighter} than $L^*$ (McLeod \& Rieke 
\markcite{mcL94b}\markcite{mcL95}1994$b$,1995; Bahcall et al.\ 
\markcite{bah97}1997). 

\subsection{Morphologies and Alignments of the Extended Material}
\label{sec:morphextend}
We might also expect the radio galaxy UV continuum 
``alignment effect'' to show itself in the quasars in some
form, but at a reduced amplitude relative to that seen in the radio
galaxies because of projection effects.
The amplitude of the aligned component in quasars, for an 
opening angle of $45\deg$,
should be about half of that seen in radio galaxies.
Using techniques similar to the magnitude derivation (excluding
the central region from both the radio galaxies and the quasars, and
using the flux only to normalized comparable surface brightness levels),
we derive $\Delta$PA (Radio PA $-~0.3\mu$m PA, and Radio PA
$-~1 \mu$m PA).
We use a simple, formal moment analysis as discussed 
in Appendix \ref{sec:moment}.
While it is straightforward to plot the dependence of the axial ratio
and PA on the isophotal cutoff and aperture,
the difficult (and subjective) part of this procedure is in choosing 
what aperture and isophotal cutoff (and resultant position angle)
one will select. 
We display in Fig.\ \ref{momfig} an example of the variation we see
in our moment analysis. We plot the isophotal cutoff on the $x$ axis
and the PA on the $y$ axis; the size of the circular
points are proportional to the size of the aperture used in the 
moment determination.

To choose one position angle to represent the overall alignment of
a complex source may not be possible in many cases.
We thus endeavor to understand any dependence of position angle on the aperture and isophotal cutoff
on a case-by-case basis. 
Adopting a single aperture and cutoff level
is perhaps the most objective criterion, but it will also introduce a greater 
level of dispersion into the measurement of alignment; moments will
still be skewed
by included foreground or unrelated companion 
objects, and unexcluded noisy sky pixels.
We have a small sample, and cannot afford this increase in dispersion.

In many cases, there is a
``dominant'' position angle associated with an 
isophotal-cutoff {\it vs.} PA plot. For the HST images,
once the isophotal cutoff is above the value chosen for
our magnitude analysis, the sky noise contribution is much
reduced. This can be seen in these plots as a stabilization of
the position angle with respect to changes in isophotal cutoff or
aperture size.  After such stabilization, the position angle 
in some cases will remain quite constant with aperture size 
and increasing isophotal cutoff. In these cases, 
the dominant position angle is fairly clear, and we adopt this value,
generally an average over a range in which the position angle is
relatively flat.
In other cases, the determination is not so simple: 
discrete companions may skew the position angle at larger apertures or lower
isophotal levels, a primary object may itself have different
dominant position angles at low and high isophotal levels, 
and, in at least one case, once above the
sky noise the isophotes twist smoothly, and
there does not appear to be any dominant position angle.
We adopt an aperture criterion similar to that of Dunlop \& Peacock 
\markcite{dun93}(1993)
to remove ambiguity in a few cases; in the others, we 
record both a ``high'' and ``low'' isophotal position angle;
our ``low'' isophotal cutoff level is generally the same as
that used for our magnitude 
calculation, and our ``high'' level is about 5 times this. 
The existence of this ``high'' and ``low'' position angle
is particularly important in light of our need to compare results 
between the radio galaxies and quasars; in some cases, the removal
of the center of the source removes any ``high'' isophotal level.

The $K'$ data are of lower resolution, and 
the radio galaxies have fairly round central components. 
We find that in most cases the variation 
in position angle is related to inclusion or exclusion 
of companions. We therefore once again use an aperture criterion 
similar to that of Dunlop \& Peacock \markcite{dun93}(1993),
and a moderate isophotal cutoff level of about 2 times
our magnitude isophotal cutoff level for $K'$.

In addition, we are often interested in the behavior of the 
most central component (or extension) alone. 
For example,
we have seen in \S \ref{sec:3c280} that the near-infrared central elliptical
component in 3C\,280 seems to have elliptical isophotes that 
are aligned preferentially with the radio axis. A similar 
moment analysis of the central component will be clearer if
we exclude the bright red companion. 
For this reason, for the infrared images, 
we have also calculated position angles excluding 
all companions that we 
interpret as discrete from our high-resolution HST images or that we know
are unrelated from spectroscopy.

In Figs.\ \ref{palofig} and \ref{pahifig}, we give histograms of the results of 
our alignment effect analysis for the HST images for the complete sample,
and for the two subsamples.
In Fig.\ \ref{kalignfig} we show histograms of the alignments derived
from the near-infrared imaging for the samples; the values shown are for
the central component, excluding companions.
The radio axis of 3C\,175.1 is bent; we adopted the average PA of
the two lobes as the radio PA for this histogram.

Application of a formal alignment analysis finds alignment in some of
the galaxies and in several of the quasars, and  
we find no significant difference between the two samples from this
analysis, for either the near-infrared or optical alignments. 

It is clear, however, from inspecting the details of the moment 
analyses, that much of the alignment detected in the quasar subsample
comes from discrete components that we have identified as 
high-surface-brightness optical synchrotron emission.
The quasars 3C\,2, 3C\,212, and 3C\,245 all show direct correspondences
between the optical and radio peaks. 
The alignment detected in the radio galaxy subsample, however,
is qualitatively different. In 3C\,280 and 3C\,237,
the most aligned galaxies, most of the extended material and components
are aligned in both the optical and near-infrared
without having a detailed optical-radio coincidence.

In the rest-frame $\sim1 \mu$m images, we find that four of the radio
galaxies are clearly resolved.  The three with sufficient resolution to
attempt a profile analysis (3C\,175.1, 3C\,280, and 3C\,289) are well fit
by de Vaucouleurs r$^{1\over4}$-law profiles with effective radii varying
from 5 to 7 kpc.  The two radio galaxies which show close alignment in the
optical (3C\,237 and 3C\,280) have IR central components whose major axes
are aligned with the optical/radio axis, presumably from unresolved
contributions from the aligned material (rather than from a global alignment
of an old stellar population).  The presence of highly-structured, often
aligned components in the rest-frame UV, coupled with a dominant elliptical
component at $1 \mu$m is entirely consistent with the view that most
powerful $z\sim1$ radio galaxies result from nuclear activity in already
mature stellar systems (the ``old-galaxy hypothesis;'' {\it e.g.,}
Rigler et al.\ \markcite{rig92}1992).  This view is also supported by
the colors of the elliptical components (see \S \ref{sec:colcomp}).

\subsection{Colors of Components}

\label{sec:colcomp}

In Table \ref{photomtab}, we give colors in terms of 
the optical--near-infrared spectral index $\alpha$
for a number of the identifiable components around both
the quasars and the radio galaxies. 
We divide these objects and components
into somewhat arbitrary categories: ``discrete,'' consisting primarily of
totally resolved companion objects and synchrotron components,
``nebulous,'' consisting of extended components, such as 
central radio galaxy components and the nebulosities
surrounding the quasars, and  ``nuclear,'' the quasar
nuclei with all extended material subtracted. 
Some components (such as $d$ of 3C\,212) we include in 
both the discrete and nebulous classes.
In Fig.\ \ref{colorfig},
we show histograms of these $\alpha$ values for each 
of the categories.
The top panel gives the values for the quasar nuclei (minus
any extension); the middle panel gives the values for discrete
companions, while the lower panel represents primarily nebulous
material. Once again, the objects that are intermediate between
these two classes are included in both middle and lower histograms.
We note certain characteristics that are common to a group 
of components.
The most obvious of these is
that the ``discrete'' components (consisting primarily of 
totally resolved companion objects and synchrotron components) are 
bluer in color than the ``nebulous'' components, which include
a number of the radio galaxies themselves as well as large scale
nebulosities surrounding the quasars. In addition, all of
the components that surround a particular quasar are of the same
color or redder than the nucleus of the quasar in the field.

This supports (in a general sense) the idea that the nebulosities
surrounding high-$z$ quasars are dominated by stars, and not by scattered 
quasar light.
For comparison, a Bruzual-Charlot model of
a stellar population with a Scalo IMF that has aged 4 Gyr since a
delta-function starburst has an optical spectral index (from 0.33$\mu$m
to 1 $\mu$m) $\alpha$= 2.56. 

What actual evidence is there for stars in these nebulous hosts?
In general, ``proof'' of the existence of stars is scarce in high-$z$
radio sources. In 3C\,65, a red, unaligned $z \sim 1.2$ radio galaxy, 
and in 53W091,
a low radio luminosity $z \sim 1.5$ radio galaxy, Keck LRIS spectroscopy
has identified stellar absorption lines directly and found ages $\sim$ 4 Gyr
(Stockton et al.\ \markcite{sto95}1995; Dunlop et al.\
\markcite{dun96}1996).
Additionally, 4000 \AA\ breaks have been
detected in a number of lower signal-to-noise spectra of radio galaxies
(Cimatti et al. \markcite{cim93}1993).
In our sample, we have only a few broad-band colors; we can only
say that the colors of the quasar nebulosities are inconsistent with
simple scattering models and broadly consistent with a stellar origin.
We find also that in those radio galaxies where we
can identify a common ``elliptical'' component in the observed infrared
and HST images, the colors of this component are consistent with those expected
for a 3--4 Gyr Bruzual-Charlot model. 

\subsection{Clusters and Companions}
The detailed clustering analysis of these fields will be presented
elsewhere (Ridgway \& Stockton \markcite{rid97b}1997$b$);
here we will simply mention the best candidates (normally within 2--3\arcsec)
for companion 
galaxies physically close to the 
active galaxies in this sample, 
and a few cases in which there is obvious evidence for a cluster. 

The only spectroscopically-confirmed close companions to objects in our 
sample are 3C\,245$e$ and 3C\,336$a$,$b$,$c$.  The 
bright companion to 3C\,245 is an elliptical 
with a 3000 km s$^{-1}$ velocity difference from the quasar redshift (LeF\`evre
\& Hammer \markcite{leF92}1992).
Its color index is $\alpha_{K,R}$ $\sim 2.6$ and its effective radius is
4.5 kpc.  Of the objects associated with 3C\,336, $a$ and $c$ may be part of the
host galaxy, but $b$ appears to be a companion with a velocity difference of
$\sim600$ km s$^{-1}$ (Steidel et al.\ \markcite{ste97}1997).

Any other companion identification we must base solely on 
morphology and colors. 3C\,2 has component $c$
which is resolved and
as previously discussed has colors of an old stellar population if it
is at the redshift of the quasar. 
3C\,217 includes component $a$, which is red and resolved, with $\alpha_{K,R}$
$\sim 3.0$. 3C\,289 has $d$, also resolved, and red with
$\alpha_{K,R}$ $\sim 2.5$. 
Not quite resolved from the 3C\,196 nuclear region is $a$, a blue component
($\alpha_{K,R}$ $\sim 1.3$), which could be part of the host or
other associated material, perhaps with contributions from scattered light 
or thermal continuum; 
it is less likely to be a discrete companion. 

Around 3C\,212, there is an over-density of objects within 10$\arcsec$, which 
could be due to a cluster at the redshift of the quasar.
Around 3C\,280,
we have more definite signs of a cluster, in that we detect a number
of objects in a continuum-subtracted
[\ion{O}{2}] image. We will present spectroscopic
follow-up to 
these objects in future papers. 
3C\,289 also has a likely over-density of objects in its immediate vicinity.

\section{Modelling the Quasars with the Radio Galaxies}
\label{chap:model}
In order to understand how the radio galaxy properties compare with those of the
quasars, we must address how the radio galaxies we have observed
would appear if they were superposed on quasar nuclei. 
Though we can estimate how much of the radio galaxy would
certainly have been lost in a nuclear subtraction by simply
excluding the central region as we have done in the magnitude
and moment analyses discussed above, this does not include the considerable
noise that may be introduced in the subtraction process. 
If the PSF is observed (as are the ones we use here) 
rather than modelled, then subtraction adds
sky noise (partially flat-fielding residuals, partially
Poisson) and Poisson noise from the overlying bright nucleus. 
In addition, the details of any observation (jitter, positioning
on the chip) will introduce PSF variations, so another potential
source of noise is the PSF subtraction residual. 

To model as closely as possible the quasars as we observed them,
we would like to have a PSF independent of the stellar PSFs to 
simulate the quasar nuclei. If we use the same stellar PSFs
to subtract that we use to model the nuclei (with independent
noise added, of course), we may succeed in
simulating the effect of the Poisson noise, but will not be able
to address the deviations caused by PSF mismatches.
The modelled Tiny Tim PSFs show large residuals when subtracted from
observed PSFs, and they are not an adequate 
representation of the actual PSFs in our fields.
We have therefore tried to obtain an
average PSF from the quasar nuclei themselves. 

We take the host galaxies and other
residual structure after the subtraction of the PSFs for each of our
quasars and interpolate over the inner region that is dominated by 
noise and PSF mismatch residuals.  We slightly smooth these
images and mask off regions that are essentially just sky.
We then subtract these images from the corresponding original
quasar images and combine these host-galaxy-subtracted PSFs
with {\it gcombine} and the ``rsigclip'' clipping algorithm,
giving us a fairly clean PSF that is largely independent of our stellar
PSFs.  

We scale this PSF to simulate each of our observed quasar
profiles, and we then superpose each
of these 5 PSFs on each of the 5 radio galaxies, adding in the appropriate
noise.
More details of
this modelling process are given by Ridgway \markcite{rid95}(1995).

We then run the same centering, scaling, and monotonic subtraction procedures
described in \S \ref{sec:psfsub} on these 25 images. These models
are on the conservative side; some of the radio galaxy
images had higher sky noise than the final quasar image we 
were attempting to simulate. But they give a good impression 
of some of the effects of the Poisson noise and PSF mismatches.
(For the opposite, optimistic limit, 
we have in essence simulated a near-perfect ``noiseless'' subtraction
by simply removing the central region, as discussed in
\ref{sec:moment}).

We display a montage of the subtracted images in Fig.\ \ref{modfig}.
The first row shows the observed radio galaxies;
they are oriented as observed on the WFC CCD. The second row shows
the unsubtracted models of 3C\,245,
consisting of an average PSF (scaled
to the magnitude of the actual 3C\,245 nucleus)
with each of the observed radio galaxies added. 3C\,245 is the brightest
of our quasars, and this row shows therefore the case for which it is
most difficult to recover the extensions.
The next 5 rows show subtracted versions of the models of each of
the 5 quasars, constructed in a similar fashion. In each case, we subtract
the average stellar PSF appropriate for the filter in which the quasar
was actually observed, to give an idea of the range of possible residuals.
The distance
between tickmarks represent 1\arcsec.

Visual inspection of the models reveals some interesting points. 
As expected, 
it is easier to recover host galaxy structure from
fainter QSOs like 3C\,2 and 3C\,212. 3C\,2 also happens to have been fairly
deeply observed, and therefore to have lower sky noise.  
We also see an extra feature to the upper right in 
the 3C\,245 and 3C\,336 simulations; this may correspond to a poor subtraction
of the diffraction spike in this corner, but it is also visible
in the difference between the average stellar F622W PSF and F675W PSF.
We get an idea from these models out to what radius 
PSF residuals are likely to dominate our recovered data: we find that
the morphologies and fluxes are generally little affected outside our 
adopted radius of 0\farcs45.

This suite of models also allows us to make a better comparison of 
the alignments in the two samples.
We would be able to detect closely aligned
structure under most of the model nuclei in only the two most
aligned radio galaxies in our sample, 3C\,280 and 3C\,237. 

In our sample of actual observed quasars, if we exclude the contributions 
of optical synchrotron emission, we see similar aligned continuum structures
in 3C\,196
and in 3C\,336 (if we disregard the nearby components spectroscopically
determined to be unassociated with the 3C\,336 host galaxy).
We thus find that the incidence of ``standard'' aligned material
is certainly consistent between the two samples. 

\section{Discussion}
\subsection{Alignment Effect}
A clear prediction of unified models for quasars and FR II radio galaxies
is that the quasar hosts should show an alignment effect, but with
reduced amplitude and with larger misalignment because of projection.

Though the alignment effect is certainly a statistically significant 
property of large samples of high-radio-luminosity, high-redshift 
radio galaxies, individual objects seem to exhibit different kinds
of aligned morphologies and these structures probably differ in
origin. It is therefore often necesssary to consider
the observational evidence relevant to each individual object.

The morphologies observed by HST in the UV for many of the 
$z \sim 1$ radio galaxies 
(Longair et al.\ \markcite{lon95}1995, Best et al.\ 
\markcite{bes96}1996, Dickinson et al.\ \markcite{dicm95}1995) argue strongly
for the interaction of the jet with the ambient medium being the primary 
cause of the structures observed. The curved, linear, discrete structure 
of 3C\,324 which tracks the jet direction leaves little room for
doubt. Still, however, it is debatable exactly what material is 
involved, what the 
jet interaction caused, and what the actual emission mechanism for the aligned
UV light is. 
Jet-induced star formation as the source of the aligned UV lumps in
most of these smaller scale radio sources is basically consistent with 
the observed HST morphologies; however, 
some of these
same sources (such as 3C\,324; Cimatti et al.\ \markcite{cim96}1996) 
show evidence for a significant level of
polarization. This polarization percentage is consistent with 
the bulk of the UV light coming from dust scattering, with dilution 
coming from an evolved stellar population showing a significant 4000 \AA\
break.

In addition, the thermal nebular continuum has been shown to be a 
major contributor to the aligned component of
3C\,368 (Dickson et al.\ \markcite{dicr95}1995; Stockton et al.\ 
\markcite{sto96}1996),
which we now know to be unpolarized (Dey et al.\ \markcite{dey97}1997). 
This means that even among the 
``small-scale, very aligned'' examples, we have one in which 
scattered light is probably the dominant contributor of 
UV light (3C\,324) and 
one in which nebular continuum is dominant (3C\,368). 

In our sample, we see evidence 
for a number of the mechanisms that are considered possible contributors
to the alignment effect. Most obviously, we see a number of quasars (but
no radio galaxies) in which we almost certainly have beamed optical synchrotron 
emission: 3C\,2, 3C\,212, and 3C\,245.  In the quasars 3C\,196 and 3C\,336,
and in the radio galaxies 3C\,237 and 3C\,280, we see examples of
alignment that apparently cannot be attributed to synchrotron emission.
In the three of these objects for which we have the necessary observations,
we detect aligned [\ion{O}{2}] emission.
We see evidence of illumination (scattering or photoionization) effects 
in the edge brightening 
observed in the arc in 3C\,280, and possibly in the very blue objects within
the ``scattering cone'' of 3C\,196. The arc and peak
structure in 3C\,280 
has a strong contribution from thermal continuum emission from
ionized gas.
The only major proposed alignment mechanism for which
we have no direct evidence is jet-induced star formation. 

In summary, in our sample, we find evidence for most of 
the major mechanisms for alignment, and we find neither support nor
counter-evidence for 
jet-induced star formation.  From these purely morphological 
studies we cannot, however, pin down certain origins for most
of the aligned components.
Follow-up studies of key objects and components, such as the NW component
in 3C\,212, may prove vital for understanding the kind of multi-component
morphology observed in many radio galaxies. 

\subsection{Unification Hypothesis}

Past comparisons of low-redshift radio galaxy and quasar host properties
have been inconclusive and contradictory, primarily because
of large uncertainties introduced by the PSF subtraction, and lack of
complete samples. 
An example of such a contradiction is in the studies of 
Dunlop et al.\ \markcite{dun93b}(1993), who find that $z \sim 0.3$ 
radio galaxy and
quasar host magnitudes are the same, and of Smith \& Heckman 
\markcite{smi89}(1989) who find
that host magnitudes in similar samples differ significantly.
Our magnitude and alignment effect 
results are basically consistent with a standard FR II unification
model but do not provide a strong test of it because of the limited size of
our radio galaxy and quasar samples. 
Indeed, we expect an intrinsic dispersion in properties such as
opening angle, and this variation could 
be enough to hide a statistical difference in observed properties
between radio galaxy and quasar samples.

We also detect faint unresolved cores in all of 
our HST radio galaxy images. The brightnesses of these cores
seem to roughly correlate with the radio
core brightnesses ({\it e.g.,} 3C\,217 has the faintest optical
core and is undetected in the radio). These optical cores
may be the obscured active nuclei.

\subsection{Optical Synchrotron Emission}
\label{sec:optsynch}
In at least three of our sources, we have found
detailed correspondences between optical and radio structure that
can plausibly be interpreted as optical synchrotron emission.
In one case, 3C\,2, the emission comes from the north-eastern
lobe at a projected separation of
$\sim$6 kpc; in the others, we have evidence for an optical counterparts to 
radio jets on the northeastern side of the core of 3C\,212 and the 
western side of the core of 3C\,245. 
In 3C\,212, the physical separation from the farthest ``jet knot'' to
the nucleus is $\sim$20 kpc; in 3C\,245, the jet extends to a length of
$\sim10$ kpc.

Likely examples of optical synchrotron emission 
have been detected previously in about a dozen other extragalactic objects. 
(Crane et al.\ \markcite{cra93}1993 and references
therein;
Meisenheimer et al.\ \markcite{mei89}1989 and references therein).
Their radio/optical spectral properties
have many similarities, and they generally fit a common pattern
often found for astrophysically observed synchrotron emission:
that of a power law with $\alpha \sim 
0.5$--0.8 (where $S_\nu \propto \nu^{-\alpha}$), with a steepening of the 
power law spectrum at
higher frequencies above a certain cutoff frequency $\nu_c$.

An alternative mechanism for generating optical emission from a
distribution of relativistic electrons is the inverse Compton effect,
which can upconvert low-frequency photons to the optical region via
scattering from the relativistic electrons (Daly \markcite{dal92b}1992$b$ 
and references
therein).  This process is normally important at low redshifts only in
the context of contributing to the cooling of the lower-energy electrons
in compact sources with very high radiation densities.  However, since
the energy density of the microwave background increases as
($1+z$)$^4$, at a sufficiently high redshift this energy density will
be comparable to that of the magnetic field in typical radio lobes or
jets, and the inverse Compton process may dominate the observed
emission (Daly \markcite{dal92b}1992$b$).

We consider the case of the north-eastern lobe of 3C\,2, for which
resolved radio data at multiple frequencies is available (Saikia et al.\ 
1987).
Estimating the magnetic field from the minimum-energy condition, using
the usual (but not necessarily correct) assumptions of unity filling factor,
an equal energy division between electrons and protons, uniform
magnetic field, and source depth along the line of sight equal to 
the observed source width ({\it e.g.,} Miley \markcite{mil80}1980), we
obtain $B_{\perp}\approx 10^{-4}$ Gauss.  The NE lobe of 3C\,2 dominates
the total flux from the source at all radio frequencies for which it was
resolved, and the total spectral index is nearly constant over a range
of 3 decades in frequency.  This constancy in spectral index allows us
to use Eqn 8$b$ of Daly \markcite{dal92b}(1992$b$) to estimate the flux density due to
inverse Compton scattering in the NE lobe.  We obtain 
$f_{IC}=2.2\times10^{-21}$ erg cm$^{-2}$ s$^{-1}$ \AA$^{-1}$.
Our observed value of $\sim10^{-18}$ erg cm$^{-2}$ s$^{-1}$ \AA$^{-1}$
indicates that the optical emission is strongly dominated by
synchrotron radiation rather than inverse Compton scattering of
microwave background photons.  

Further support for this result comes from the break we see between
the spectral index in the radio and optical regions (Fig.\ \ref{3c2radfig}).
The inverse Compton process should produce optical/IR radiation with
a spectral index like that of the synchrotron radiation from the
(low-energy) relativistic electrons responsible for the upscattering,
{\it i.e.,} typically that at $\sim1$ MHz ({\it e.g.,} Daly
\markcite{dal92b}1992$b$).
In general, radio spectral indices at such frequencies are unlikely to be
steeper than those at centimeter to meter wavelengths, so our steeper
optical/IR spectral index is evidence for optical synchrotron emission,
where the synchrotron spectrum has a high-frequency break normally
resulting from the aging of the electron energy distribution due to
synchrotron losses.  This effect is shown clearly in the Fermi
acceleration models of Meisenheimer et al.\ \markcite{mei89}(1989).

De Koff et al.\ \markcite{deK97}(1997) note that, for their $z < 0.2$ 3C galaxy sample,
the angular radio sizes of those galaxies showing optical synchrotron
emission
are less than those in which no such detection is made. They suggest
that this is due to beaming effects; {\it i.e.,} that the radio galaxies
with optical synchrotron detections have jet axes that are preferentially
aligned closer to our line of sight.  This interpretation is consistent
with our finding evidence for optical synchrotron emission in most of our
quasars, but in none of our radio galaxies.

\section{Summary}

With our HST imaging going nearly as deeply as our best previous
ground-based imaging (even on low-surface-brightness features),
HST's improved resolution over the ground-based observations makes a 
significant difference in our interpretation of the extensions observed
around the quasars. Often structure that seems nebulous and 
galaxy-like with ground-based seeing is revealed to consist of
discrete components that may have more to do with the AGN
than with stellar populations.
An example of this is the NE lobe of 3C\,2; without the high-resolution
image it would likely be considered a portion of the host galaxy,
and perhaps evidence for a second nucleus, rather than identified
as optical synchrotron emission.
For the radio galaxies, on the other hand, this improvement 
in resolution in some cases may give entirely new information,
as for 3C\,280, while in others it mainly
confirms what could be surmised about the radio galaxy structure 
from lower-resolution data ({\it e.g.,} 3C\,217 or 3C\,289).

Our samples are small, but some results are 
clearcut:

We find evidence for continuum structure around all of our
quasars in the high resolution WFPC2 data, although generally 
much of the extension does not
resemble a ``normal'' galaxy host. 3C\,2 seems, however, to have a
bright elliptical galaxy-like nebulosity. 
We resolve structure around 4 of the 5 quasars in the near-infrared
images. 

We observe a high incidence of morphological oddities,
all sometimes ascribed to 
interactions: a high incidence of nearby companion galaxies, lumpy 
morphology, and asymmetric emission-line gas. In addition, we see an
arc in 3C 280 which bears morphological resemblance to a tidal tail.

We see morphological and color evidence for illumination 
effects from the active nucleus, {\it i.e.},
scattered quasar light or
photoionization. In 3C\,196 and 3C\,280, we see very blue and/or
edge-brightened structures that lie within the probable
quasar opening angle.

We see evidence for aligned emission
in all of our quasars.
In 3C\,212, we see an object that lies $beyond$ the radio lobe but
looks morphologically quite similar to a radio hotspot and tail; 
this object is bright in the infrared and has a steep spectral gradient
across the tail. If this is truly a result of the radio jet,
it is a unique object that may prove vital to understanding the
relationship between the UV continuum alignment effect and the radio source. 
We have detected optical counterparts to radio jets in the quasars 3C\,212 and 
3C\,245, and an optical counterpart to a radio lobe in 3C\,2. All of
these structures have such a detailed, high-resolution, point-to-point
correspondence with the radio structures that they are very likely the result
of optical synchrotron radiation, and the spectral indices in 3C\,2 are 
consistent with this interpretation.  In 3C\,196 and 3C\,336, we see
aligned structure that does not appear to correlate directly with radio
emission and that seems to be similar to the aligned UV components seen
in radio galaxies.

The colors of the extended components around the quasars are in all cases
similar to or redder than that of the quasar nucleus.
In some cases at least, nearby companions or overall nebulosity have colors
consistent with a stellar population of 4 Gyr or more.
Thus, the emission we see around the quasars and in the radio galaxies
can probably be ascribed to the following mechanisms:

1) optical synchrotron in three quasars:  3C\,2, 3C\,212, and 3C\,245

2) stellar host galaxies, particularly seen in the near-infrared

3) scattered quasar light or thermal continuum in some components and companions

4) true nearby companion galaxies

5) ``standard'' UV aligned emission, for which the origin is uncertain. 

We find that the optical and infrared flux densities of the extended
material are consistent in the radio galaxy and
quasar subsamples within the dispersion of
our small sample, and that this result is not dependent on the contribution
of the beamed optical synchrotron components detected in the quasars
and not in the radio galaxies.
The mean and dispersion of the infrared magnitudes
of the quasar extensions (after correction for the missing flux from
the obscured central region) are consistent with
the $K$-$z$ relation found for radio galaxies. We find
the quasar and radio galaxy hosts have $K'$ luminosities of $L^*$
or slightly less.

Our elliptical profile fits to the $K'$ images of some of the
radio galaxies indicate that their surface-brightness distributions
are well fit by an r$^{1\over4}$ law.  These profiles, and the red
colors we have found for these components, suggest that at rest-frame
1 $\mu$m these radio galaxies are often dominated by an old (3--4 Gyr)
stellar population.

From the correspondence between the total magnitudes in the
radio galaxies and quasars,
and the detection of aligned components
in quasars, we conclude that this study provides general
support for the unification hypothesis of radio galaxies and quasars.
These issues are confused by the wide range of morphological
types found among the aligned components and 
by the small size of our sample.
Nevertheless, 4 of the 5 quasars show some extended material that can
plausibly be attributed to stars, and 3C\,196 and 3C\,336
have aligned components 
that are unlikely to be due to optical synchrotron radiation, and
appear qualitatively similar to those often seen in high-redshift radio
galaxies.

We are currently obtaining WFPC2 imaging of an additional
5 3CR quasars from our larger complete sample (\S \ref{sec:sample}).
These observations will double the number of quasars for which we
have WFPC2 imaging, 
and this increase in sample size
will allow us to reduce the uncertainties caused
by the small number of objects in the current work.
With additional 
ground-based imaging and deep spectroscopy of individual fields, we
hope to make further progress in understanding the properties
of quasars at $z \sim 1$ and their relationship with radio galaxies.

\acknowledgments {We thank Ken Chambers, Esther Hu, Mark Lacy, and Neil Trentham
for reviewing and commenting on various versions of the paper.
We are very grateful to Mark Lacy and Katherine
Blundell for guidance in the reduction our VLA data. 
We also thank Neil, Mark, Katherine, Steve Rawlings,
Joss Bland-Hawthorne and Ruth Daly for the
time they have spent and interest they have shown
in discussing the alignment effect in general and 3C\,212 in particular with us.
Comments from the referee, Patrick McCarthy, helped us improve both
the content and the presentation of this paper.
We thank Patrick Leahy for a critical evaluation of the best radio maps 
of our sources in the literature, and we thank Sylvia Baggett at STScI for her
help in recalibrating the HST data.
Jonathan Higa wrote the original versions of the IRAF scripts we used
for IR image reduction and PSF subtraction.
The authors were visiting astronomers at the NASA Infrared Telescope
Facility, operated by the University of Hawaii under contract with NASA.
Support for this work was provided by NASA through grant number 
GO-05401.01-93A from the Space Telescope Science Institute, which
is operated by AURA, Inc., under NASA contract NAS 5-26555.
Some of the ground-based observations were supported by NSF grant
AST 92-21909, and
SER was supported by a PPARC Postdoctoral Fellowship while finishing this
work.} 

\appendix
\section{Magnitude and Moment Analysis of the Images}
\label{chap:moment}
\subsection{Magnitudes}
We discussed in \ref{sec:mag1} how we created isophotal masks 
from the images (with flux densities normalized to $z = 1$) in 
order to allow us to compare the imaging results for
our objects less biassed by the details of our observations.
Here we will give more details and discuss the differences
in the way we treated the ground-based data from the HST images.

For the WFPC2 images, we first define a single set of isophotal
levels; these are chosen based on our knowledge of
the average normalized 1 $\sigma$ sky in our observed frames,
given in Table \ref{HSTlog}.
For each image, multiplied by the $z=1$ correction factor,
we create masks at these isophotal levels.  We then smooth
the mask contours slightly by convolving with a Gaussian with
a sigma of 1 rebinned pixel (0\farcs05) and excluding any pixels whose
``smoothed'' values are less than 0.5. This smoothing tends to
eliminate any isolated sky pixels which might have been picked up in
the mask and gives continuous contours; it therefore allows us to consider
magnitudes to slightly lower flux limits than would have been otherwise
possible. Otherwise, choosing a flux level at which all images will not have 
noise-dominated profiles would exclude much of the interesting
features in the quasars. (As it is, we still will not include in the
objective magnitude comparison some 
low-level nebulosity which falls below our best ``common'' isophotal
level).

We then calculate the flux densities in the unsmoothed image
after multiplication by these smoothed masks (in a series
of apertures).
We find this smoothing
process works well on radio galaxies, but tends to lose flux from linear 
features such as those seen in 3C\,212 and 3C\,245.
For this reason we reduced the smoothing as much as possible and checked 
the smoothed lowest isophotal masks visually to see how much
of the structure of interest we are preserving.
From this visual inspection, we also select the lowest 
isophotal level at which all of our normalized image masks
have little or no sky contribution. We take the lowest
common level to be $4.56 \times$ 
10$^{-19}$ ergs cm$^{-2}$ s$^{-1} $\AA$^{-1}$ arcsec$^{-2}$),
about 2.5 times the average normalized sky $\sigma$.
For the $K'$ data, we follow a very similar procedure, 
except that we do not need to smooth the masks,
prior to applying them to the images. 
Our adopted ``common'' isophotal level for
the $K'$ data is $1.4 \times$
10$^{-19}$ ergs cm$^{-2} $s$^{-1} $\AA$^{-1}$ arcsec$^{-2}$.

This magnitude calculation 
procedure is straightforward for the radio galaxies; for the quasars 
we must deal with the interior region of the 
subtracted image which is dominated by residuals from the profile
subtraction. As a first pass, we mask out the
profile subtraction region
in the quasars to exclude them from our moment
and magnitude calculations; a circular region with radius 0\farcs45
is generally necessary to remove the region dominated by PSF 
subtraction rediduals in the HST images. 
We also mask out an identical region
in the radio galaxies to make an accurate comparison, and to 
understand how this subtraction process
affects the final magnitudes we calculate. 
We have also PSF-subtracted the radio galaxies, and verified that
outside of this masked-off central region, the wings of the PSF do not
significantly contribute to the radio galaxy annular fluxes.
For the near-infrared radio galaxy images, the extended flux
is mostly quite dominant, and the best PSF-subtraction is 
just an estimate of where the galaxy profile obviously 
is no longer monotonic.
We follow the entire masking and magnitude- and 
moment-determination procedure for both subtraction
limits for those quasars in which the monotonic limit
differed significantly from the $\chi^2$ minimum fit.
These $\chi^2$ minimum subtractions give us a probable
lower-limit to the flux we observe around the quasars and 
an idea of whether changing our PSF-subtraction technique would
affect our results significantly.  We then carry out a similar set
of calculations, where, instead of masking out the central region,
we interpolate across it with a low-order 2-dimensional
fit.

We include companions within 3\arcsec\ for our objective 
comparison; however, in some cases
(like those of 3C\,280 and 3C\,289) we found it instructive to consider 
the central component alone as well.
We also exclude from our moment and magnitude analyses known foreground objects,
such as the probable $z=0.437$ absorbing spiral galaxy in front 
of 3C\,196 (BB), the star in the 3C\,175.1 field, the probable foreground 
galaxy in the 3C\,280 field, (for most purposes) the $z=1.013$ elliptical
galaxy companion to 3C\,245 (LeF\`evre \& Hammer \markcite{leF92}1992)
and the foreground galaxy $d$ in the 3C\,336 field.

\subsection{Position Angle Determination}
\label{sec:moment}

To quantify the alignments of the objects in our sample,
we adopt a technique similar to that 
of Dunlop \& Peacock \markcite{dun93}(1993), with some exceptions.
First, we explicitly mask out any 
known foreground or companion objects, particularly those for which likely
redshifts exist. 
We then make our moment calculations in the isophotal masks we have
generated, with a common range of apertures (from 2\arcsec\ up to 10\arcsec\
in diameter) that should span the probable range of physical scales
of interest for all
of the objects. 

To find the principal axes of the flux distribution, we
calculate the first and second flux-weighted moments.
Where $\mu_i$ is the 
sky-subtracted intensity of the $i$'th pixel at position ($x_i$,$y_i$), 
we calculate $\sum_i(\mu_i)$ and the first moments $\sum_i (x_i \mu_i) /
\sum_i(\mu _i)$ and
$\sum (y_i \mu_i) / \sum(\mu_i)$.
We then calculate the second moments around the center that 
most closely matches a probable AGN location. For our quasars,
this is simply the quasar center; for our radio galaxies, 
we detect unresolved optical contributions in all of the
HST images which match the best infrared centers. We therefore
use in all cases the WFPC2-determined ``nuclear'' center.
(In all of our objects the optical and near-IR 
cores match well, thus we use the higher resolution HST data 
to determine the best center. Such small shifts in centering affect the
moments little).
We derive the second moments 
$I_{xx} = \sum_i (x_i^2 \mu_i) / \sum_i(\mu _i)$, 
$I_{yy} = \sum_i (y_i^2 \mu_i) / \sum_i(\mu _i)$, 
$I_{xy} = \sum_i (x_iy_i \mu_i) / \sum_i(\mu _i)$, where ($x_i,y_i$)
is the position of the $i$'th pixel relative to the best determined
center. 
We use these to determine the position
angle $\theta$ and  the ratio of the minor to major axis
$a/b$
of the equivalent ellipse.

\newpage
\begin{table}
\dummytable\label{HSTtab}
\dummytable\label{HSTlog}
\dummytable\label{obslog}
\dummytable\label{psftab}
\dummytable\label{photomtab}
\dummytable\label{3c2sedtab}
\dummytable\label{qsohst}
\dummytable\label{rghst}
\dummytable\label{qsoir}
\dummytable\label{rgir}
\end{table}
\clearpage
\begin{figure}
\caption{The results of subtracting our 
observed stellar PSFs from each other with the centering and 
scaling technique described in text. 
The images are to the same scale as
the magnified panels in Figs\ 3--12;
the distance between tick marks is 1\arcsec\ and a scale bar is shown in panel (d).
Panel (a) gives the 3C\,212 PSF minus the 3C\,245 PSF,
both observed in the F675W filter. Panel (b) shows the 3C\,2 PSF minus the 
3C\,212 PSF, in the F675W filter. Panel (C) shows the difference of the
3C\,196 PSF minus the 3C\,336 PSF, in the F622W filter. Note that the diffraction
spikes subtract poorly close to the stellar profile in this case. In Panel (D),
we show the average F622W PSF, from the addition of the 3C\,196 PSF and the 3C
336 PSFs.}
\label{psffig}
\end{figure}
\begin{figure}
\caption{PSF Subtraction Method: We show one example
of how we determine the (QSO/PSF) scalings in our PSF subtraction
procedure. 
We show in the upper panel
the $\chi^2$ of the PSF subtraction residuals in an annulus
versus the scaling of the subtracted PSF for the 3C\,2 $K'$ image.
The minimum in this plot gives the scaling for the lower-limit quasar
subtraction. 
In the lower panel we plot the mean flux in an inner and outer annulus versus
the PSF scaling; the circles are the outer annulus, the squares are the 
inner circle. At the intercept of these two lines, the flux in the residual
extension will be constant (monotonic) across the entire inner region. 
This example (3C\,2 at $K'$) is the quasar image with the greatest
contribution of extended material.}
\label{psfsub}
\end{figure}
\begin{figure}
\caption{3C\,2 (quasar). {\it (A)}  PSF-subtracted HST WFC image (inset shows
the same image prior to PSF subtraction).  The white
cross shows the position of the southern radio hotspot.
{\it (B)}  PSF-subtracted CFHT Redeye $K'$ image.  Again, the inset
  shows the unsubtracted version.  {\it (C)}  HST image, enlarged $2\times$.
  The inset shows the unsubtracted version with radio
  contours (2 and 32 mJy beam$^{-1}$) from the 2-cm VLA map of Saikia 
  et al.\ 1987. This does not include the peak of the 
  core flux.
 {\it (D)}  Same as {\it (C)}, but higher contrast.
  For this and following illustrations, N is at the top and E to the left.
  Tick marks are at 1\arcsec\ intervals, and long tick marks indicate the
  source center (optical peak for quasars, $K'$ peak for radio galaxies).}
\label{3c2fig}
\end{figure}
\begin{figure}
  \caption {3C\,175.1 (galaxy).  {\it (A)}  HST WFC image.  The crosses
  indicate the positions of the radio hotspots.  {\it (B)}  Keck
  NIRC $K'$ image.  {\it (C)}  HST image, enlarged $2\times$, and
  slightly smoothed to show low-surface-brightness features better.
  Inset shows lower-contrast image without smoothing.  {\it (D)}
  Same as {\it (A)}, but smoothed and overlayed with contours from our
  3.6 cm VLA superposed.  The levels shown 
  are 0.5, 2, 8, 32, and 128 mJy beam$^{-1}$.}
\label{3c175fig}
\end{figure}

\begin{figure}
  \caption{3C\,196 (quasar).  {\it (A)}  PSF-subtracted HST WFC image.
  The spiral galaxy {\it c} is in the foreground, at $z=0.437$ (Boisse
  \& Boulade 1990).  The crosses indicate the positions 
  of the hotspots, and the inset shows the unsubtracted image.  {\it (B)} 
  PSF-subtracted Keck NIRC $K'$ image, with the inset showing the unsubtracted
  version.  {\it (C)} HST image, enlarged $2\times$.  The inset
  shows the same, with lower contrast.  {\it (D)}  Image obtained with the
  CFHT SIS fast guider through a $\sim30$ \AA\ interference filter centered
  on the [O\,II] $\lambda$3727 line.  The image has been continuum subtracted.
  The inset shows the same image at lower contrast.} 
\label{3c196fig}
\end{figure}

\begin{figure}
\caption{3C\,212 (quasar).  {\it (A)}  PSF-subtracted HST WFC image.  Note
  the alignment of objects $a, b, c, f,$ and $g$ with the radio axis, indicated
  by the crosses on the radio hotspot positions.
  The inset shows the
  unsubtracted version.  {\it (B)}  PSF-subtracted Keck NIRC $K'$ image.
  {\it (C)}  HST image, enlarged $2\times$. {\it (D)}  Same as
  {\it (C)}, but smoothed and at higher contrast.}
\label{3c212fig}
\end{figure}

\begin{figure}
 \caption{3C\,217 (galaxy).  {\it (A)}  HST WFC image.  The cross indicates
  the approximate position of the E radio hotspot, and the arrow
  shows the direction to the W hotspot, which lies outside the frame.
  3C\,217 is the only
  source in our sample that does not have a detected radio core component,
  so we have assumed the optical position given by Pedelty
  et al.\ (1989).  {\it (B)}  Keck NIRC $K'$ image.
  {\it (C)}  HST image, enlarged $2\times$, and at lower contrast than
  {\it (A)}.  The inset shows the image at even lower contrast, to show
  the faint unresolved core.  {\it (D)}  Same as {\it (C)}, but
  smoothed and at higher contrast.} 
\label{3c217fig}
\end{figure}

\begin{figure}
\caption{3C\,237 (galaxy).  {\it (A)}  HST WFC image.  {\it (B)}  Keck
  NIRC $K'$ image.  {\it (C)}  HST image, enlarged $2\times$, and at lower
  contrast than {\it (A)}.  The inset shows the positions of the radio
  hotspots.  {\it (D)}  Same as {\it (C)}, but higher contrast.}
\label{3c237fig}
\end{figure}

\begin{figure}
\caption {3C\,245 (quasar).  {\it (A)}  PSF-subtracted HST WFC image.  
  The crosses show the positions of the radio hotspots, 
  although the E lobe is very diffuse, covering much of the region over
  which low-surface-brightness optical emission is seen.  The inset shows
  the unsubtracted version of the image.  {\it (B)}  PSF-subtracted Keck
  NIRC $K'$ image, with the inset giving the unsubtracted version.
  {\it (C)}  HST image, enlarged $2\times$ and smoothed.  The inset shows
  the enlarged, unsmoothed version at lower contrast.  It also shows,
  displaced 1\arcsec\ N, radio contours from the 6 cm map of Laing 
  (1989).
  Contour levels are 5, 10, 15, 20, 25, 50, 100, 125, 250, and 375 mJy 
  beam$^{-1}$.  Note the close correspondence between the optical and
  radio jet structures.  {\it (D)}  Continuum-subtracted [O\,II] $\lambda$3727
image through a $\sim30$ \AA\ interference filter at the UH 2.2m Telescope.}
\label{3c245fig}
\end{figure}

\begin{figure}
\caption {3C\,280 (galaxy).  {\it (A)}  HST WFC image.  The cross shows the
  position of the E radio hotspot, and the arrow shows the
  direction to the W radio hotspot, which lies outside the
  frame.  The inset shows the same image, smoothed and at higher contrast,
  in order to show better the low-surface-brightness material in the 
  direction of the E radio lobe.  {\it (B)}  Keck NIRC $K'$ image.  The
  inset shows the same image at lower contrast.  {\it (C)}  HST image,
  enlarged $2\times$.  The inset shows the same at lower contrast.
  {\it (D)}  Image obtained with the
  CFHT SIS fast guider through a $\sim30$ \AA\ interference filter centered
  on the [O\,II] $\lambda$3727 line.  The image has been continuum subtracted
  and slightly deconvolved.  The inset shows the same image at 
  lower contrast.}
\label{3c280fig}
\end{figure}

\begin{figure}
\caption{3C\,289 (galaxy).  {\it (A)}  HST WFC image.  The crosses show
  the position of the radio hotspots.  {\it (B)}  Keck
  NIRC $K'$ image.  {\it (C)} HST image, enlarged $2\times$.  The inset
  shows the same at lower contrast.  {\it (D)} Image obtained with the
  UH 88-inch Telescope through a $\sim30$ \AA\ interference filter centered
  on the [O\,II] $\lambda$3727 line.  The image has been continuum subtracted.}
\label{3c289fig}
\end{figure}

\begin{figure}
\caption {3C\,336 (quasar).  {\it (A)}  PSF-subtracted HST WFC image.  The 
  arrows show the
  directions to the radio hotspots, both of which lie outside
  the frame.  The inset shows the unsubtracted image.  {\it (B)}  
  PSF-subtracted Keck NIRC $K'$ image, with the unsubtracted image shown
  in the inset.  {\it (C)}  HST image, enlarged $2\times$ and smoothed.
  {\it (D)}  Image obtained with the
  CFHT SIS fast guider through a $\sim30$ \AA\ interference filter centered
  on the [O\,II] $\lambda$3727 line.  The image has been continuum subtracted.
  The inset shows the same image at lower contrast.}
\label{3c336fig}
\end{figure}

\begin{figure}
\caption{
Photometry in $F_\lambda$ of the various components of 3C\,2
versus $\lambda_0$, normalized
by $F_\lambda(3300 $\AA).
Filled circles are the QSO nucleus,
asterisks are the nebulosity in an annulus from 1\farcs5 to 3\arcsec,
unfilled squares are the companion 
$c$, filled triangles are
the 
northen lobe $a$, and unfilled circles are the faint
extension $b$. In panel $A$, the components $a$, $b$, and $c$ do not have 
the local background subtracted;
in panel $B$,
$a$, $b$ and $c$ photometry is given minus an estimated local background
contribution.
The QSO nucleus and nebulosity photometry are identical
in the two panels. 
In panel $B$, we plot models: at the top middle of the figure, we
show the SED of M31 from Coleman, Wu \& Weedman (1984) with
the alternately short- and long-dashed line;
the bluer model (with the same line pattern) is their S0 SED.
Dotted lines are Bruzual \& Charlot models: the upper model is 
a 4 Gyr stellar population, while 
the next two (in order of brightness at 1$\mu$m) 
are 2 Gyr and 1 Gyr old. 
A short-dashed line is a power-law
with $\alpha = 1.36$.}
\label{3c2sedfig}
\end{figure}
\begin{figure}
\caption{Photometry
of the northern lobe (component $a$) with 
the radio photometry of Saikia et al.\ (1987) of the same component (crosses).
The error bars shown on our data are 2$\sigma$ errors.
We plot a linear fit to our data (weighted instrumentally); this gives
a power-law with $\alpha$=1.36, where $S_\nu \sim \nu^{-\alpha}$. 
A linear fit to the Saikia et al.\ points gives $\alpha$=0.83. }
\label{3c2radfig}
\end{figure}

\begin{figure}
\caption{Our 3C\,212 WFPC2 image, with contours
from our 3.6 cm VLA map superposed.
The contours are at $-$1, 1, 4, 16 and 256 mJy
beam$^{-1}$.} 
\label{3c212xop}
\end{figure}

\begin{figure}
\caption{The results of fitting elliptical isophotes to the $K'$ 3C\,280 
image, with companion objects removed from the fit.
In the upper
panel, we give the mean surface brightness of each isophote versus
the semi-major axis.
If a de Vaucouleurs
profile is fit to the linear portion of the data, the effective radius
is $r_e = 0\farcs84 \sim5$ kpc 
corresponding to  $r_e^{ \frac{1}{4}} = 0.96$.
The inner portion of the profile is smoothed by atmospheric seeing, 
with seeing FWHM $\approx$ 0\farcs6.
In the lower panel, we show the PAs of the fit
ellipses. } 
\label{3c280ellfig} 
\end{figure}
\clearpage

\begin{figure}
\caption{Method used to determine the position angle: the example shown 
is the position 
angle on the sky for the Keck $K'$ images of 3C\,175.1 ($y$ axis) given
as function of the isophotal cutoff level ($x$ axis; the units are
$10^{-19}$ ergs cm$^{-2}$ s$^{-1}$ \AA$^{-1}$ arcsec$^{-2}$) and as a function
of the aperture size used (proportional to the size of the circular points;
radii vary from 1\arcsec\ to 3\farcs75).  The bright foreground star has
been excluded; all other companion objects are included.  The two companions
centered $\sim3\farcs5$ to the NE contribute to the moment calculations
at low isophotal levels and in apertures of 3\arcsec\ or more.  We adopt
70\deg\ for this position angle by choosing to exclude the companions; the
position angle then flattens out at $\sim$70\deg, before rising again at
higher isophotal-level cutoffs.}
\label{momfig}
\end{figure}

\begin{figure}
\caption{Histograms of the
difference between the object PA in the HST image 
(at a low isophotal level)
and the radio axis. The top panel is the sum 
of the radio galaxy and quasars samples; the middle panel
is the quasars alone, and the bottom panel
is the radio galaxies alone.}
\label{palofig}
\end{figure}

\begin{figure}
\caption{Histograms of the
difference between the object PA in the HST image (at a high isophotal level,
where possible)
and the radio axis. The top panel is the sum 
of the radio galaxy and quasars samples; the middle panel
is the quasars alone, and the bottom panel
is the radio galaxies alone.}
\label{pahifig}
\end{figure}

\begin{figure}
\caption{Histograms of the
difference between the object position angle in the $K'$ image
and the radio axis. The top panel is the sum 
of the radio galaxy and quasars samples; the middle panel
is the quasars alone, and the bottom panel
is the radio galaxies alone.}
\label{kalignfig}
\end{figure}

\begin{figure}
\caption{Histograms of the 
optical--near-infrared spectral index $\alpha$
for the identifiable components whose 
photometry we present in Table 5. 
The top panel gives the values for the quasar nuclei (minus
any extension); the middle panel gives the values for discrete 
companions, while the lower panel represents primarily nebulous 
material. Many objects (such as $d$ of 3C\,212) are intermediate between
these two classes and are included in both middle and lower histograms.}
\label{colorfig}
\end{figure}

\begin{figure}
\caption{Simulations of the quasars, manufactured by adding a mean 
quasar ``nuclear'' PSF (scaled to each of the quasar magnitudes) to each of the
observed radio galaxies. \label{modfig} }
\end{figure}


\begin{references}
\reference{abr92} Abraham, R.G., Crawford, C.S., \& McHardy, I.M. 1992,
\apj, 401, 474
\reference{aku91}Akujor, C.E., Spencer, R.E., Zhang, F.J., Davis, R.J.,
Browne, I.W.A., \& Fanti, C. 1991, \mnras, 250, 214
\reference{ant93} Antonucci, R. 1993 \araa, 31, 473
\reference{ant90} Antonucci, R., \& Barvainis, R. 1990, \apjl, 363, L17
\reference{ant94} Antonucci, R., Hurt, T., \& Kinney, A. 1994, \nat, 371, 313
\reference{are95} Arextaga, I., Boyle, B.J., \& Terlevich, R.J. 1995, \mnras,
275, L27
\reference{bah94} Bahcall, J.N., Kirhakos, S., \& Schneider, D.P. 1994, \apjl,
435, L11
\reference{bah95a} Bahcall, J.N., Kirhakos, S., \& Schneider, D.P. 1995a, 
\apjl, 447, L1
\reference{bah95b} Bahcall, J.N., Kirhakos, S., \& Schneider, D.P. 1995b, 
\apj, 450, 486
\reference{bah96} Bahcall, J.N., Kirhakos, S., \& Schneider, D.P. 1996, \apj, 
457, 557
\reference{bah97} Bahcall, J.N., Kirhakos, S., Saxe, D.H., \& Schneider,
D.P. 1997, \apj, 479, 642
\reference{bar88} Barthel, P.D., Miley, G.K. 1988, \nat, 333, 319
\reference{bar89} Barthel, P.D. 1989, \apj, 336, 606
\reference{bau89} Baum, S.A., \& Heckman, T. 1989, \apj, 336, 681
\reference{beg89} Begelman, M.C., \& Cioffi, D.F. 1989, \apjl, 345, L21
\reference{ben62} Bennett, A.S. 1962, \memras, 68, 163
\reference{bes96} Best, P.N., Longair, M.S., \& R\"ottgering, H.J.A. 1996,
\mnras, 280, L9
\reference{bir91} Biretta, J.A., Stern, C.P., \& Harris, D.E. 1991, \aj, 
101, 1632
\reference{blu94} Blundell, K.M., \&  Alexander, P. 1994, \mnras, 267, 241
\reference{boi90} Boisse, P., \& Boulade, O. 1990, \aap, 236, 291 (BB)
\reference{bre97} Bremer, M.N. 1997, \mnras, 284, 126
\reference{bre92} Bremer, M.N., Crawford, C.S., Fabian, A.C., \& 
Johnson, R. M.  1992, \mnras, 254, 614 
\reference{bri94} Bridle, A.H., Hough, D.H., Lonsdale, C.J., Burns, J.O., \&
Laing, R.A. 1994, \aj, 108, 766
\reference{bro85} Brodie, J.P., Bowyer, S. \& McCarthy, P. 1985, \apj, 293, L59
\reference{bru93} Bruzual, G \& Charlot, S. 1993, \apj, 405, 538 
\reference{bur84} Burstein, D., \& Heiles, C. 1984, \apjs, 54, 33
\reference{car88} Carilli, C.L., Perley, R.A., \& Dreher, J.H. 1988, 
\apjl, 334, L73
\reference{cas92} Casali, M. \& Hawarden, T. 1992, JCMT---UKIRT Newsletter,
Aug. 1992, 33 
\reference{cha88} Chambers, K.C., Miley, G.K., \& Joyce, R.R. 1988, \apjl,
329, L75
\reference{cha87} Chambers, K.C., Miley, G., \& van Breugel, W. 1987, \nat,
329, 609
\reference{cha90} Chambers, K.C., \& McCarthy, P.J. 1990, \apjl, 354, L9
\reference{cla97} Clark, N.E., Tadhunter, C.N., Axon, D.J., \& Robinson, A.
1997, in the on-line proceedings of the Sheffield
workshop on ``Jet-cloud interactions in active galaxies'',
http://www.shef.ac.uk/~phys/research/astro/conf/ 
\reference{cim93} Cimatti, A., di Serego Alighieri, S., Fosbury, R.A.E.,
Salvati, M., \& Taylor, D. 1993, \mnras, 264, 421
\reference{cim96} Cimatti, A., Dey, A., van Breugel, W., Antonucci, R., \& 
Spinrad, H. 1996, \apj, 465, 145
\reference{coh96} Cohen, R.D., Beaver, E.A., Diplas, A., Junkkarinen, V.T., \&
Lyons, R.W. 1996, \apj, 456, 132.
\reference{col80} Coleman, G.D., Wu, C.C., \& Weedman, D.W. 1980, \apjs, 
43, 393
\reference{cra93} Crane, P. et al.\ 1993, \apjl, 402, L37  
\reference{cra95} Crawford, C.S., \& Fabian, A.C. 1995, \mnras, 273, 827
\reference{dal90} Daly, R.A. 1990, \apj, 355, 416
\reference{dal92a} Daly, R.A. 1992a, \apjl, 386, L9
\reference{dal92b} Daly, R.A. 1992b, \apj, 399, 426
\reference{deK97} de Koff, S., Baum, S.A., Sparks, W.B., Biretta, J.,
Golombek, D., Macchetto, F., McCarthy, P., \& Miley, G.K. 1997, \apjs, in press
\reference{deY89} de Young, D.S. 1989, \apjl, 342, L59
\reference{dey94} Dey, A. \& van Breugel, W.J.M. 1994, \aj, 107, 1977.
\reference{dey96} Dey, A., Cimatti, A., van Breugel, W., Antonucci, R., 
\& Spinrad, H. 1996, \apj, 465, 157
\reference{dey97} Dey, A, van Breugel, W., Graham, J., Spinrad, H., 
Cimatti, A., Hurt, T., \& Antonucci, R. 1997, in preparation
\reference{dicr95} Dickson, R., Tadhunter, C., Shaw, M., Clark, N., 
\& Morganti, R.  1995, \mnras, 273, L29
\reference{dicm95} Dickinson, M., Dey, A., \& Spinrad, H. 1995, in Galaxies 
in the Young Universe, edited by H. Hippelein (Springer-Verlag, Heidelberg),
p.\ 164
\reference{diS94} di Serego Alighieri, S., Cimatti, A., \& Fosbury, R.A.E. 
1994, \apj, 431, 123
\reference{dis95} Disney, M.J., Boyce, P.J., Blades, J.C., Boksenberg, A., 
Crane, P., Deharveng, J.M., Macchetto, F., Mackay, C.D., Sparks, W.B., 
\& Phillipps, S. 1995, \nat, 376, 150 
\reference{djo87} Djorgovski, S., Spinrad, H., Pedelty, J., Rudnick, L., \&
Stockton, A. 1987, \aj, 93, 1307
\reference{dun93} Dunlop, J.S., \& Peacock, J.A. 1993, \mnras, 263, 936
\reference{dun96} Dunlop, J.S., Peacock, J.A., Spinrad, H., Dey, A., 
Jiminez, R., Stern, D. \& Windhorst, R.  1996, \nat, 381, 581
\reference{dun93b} Dunlop, J.S., Taylor, G.L., Hughes, D.H., \& Robson,
E.I. 1993, \mnras, 264, 455
\reference{eal97} Eales, S.A, Rawlings, S., Law-Green, D. Cotter, G., 
\& Lacy. M. 1997, \mnras, in press 
\reference{eli82} Elias, J.H., Frogel, J.A., Matthews, K. \& Neugebauer, G.
1982, \aj, 87, 1029
\reference{fab89} Fabian, A.C., 1989, \mnras, 238, 41P
\reference{fan90} Fanti, R., Fanti, C, Schilizzi, R.T., Spencer, R.E., 
Nan, R.D., Parma, P., van Breugel, W.J.M., \& Venturi, T. 1990, \aap, 231, 333
\reference{gar88} Garrington, S.T.,  Leahy, J.P., Conway, R.G., 
\& Laing, R.A. 1988, \nat, 331, 147 
\reference{gar91} Garrington, S.T. \& Conway, R.G. 1991, \mnras,
250, 198
\reference{ghi93} Ghisellini, G., Padovani, P., Celotti, A., \& 
Maraschi, L. 1993, \apj, 407, 65
\reference{han87} Hansen, L, N{\o}rgaard-Nielsen, H.U, \& J{\o}rgensen, 
H.E. 1987, \aaps, 71, 465
\reference{hec90} Heckman, T.M. 1990, in IAU Colloquium 124, Paired
and Interacting Galaxies, edited by J.W. Sulentic, W.C. Keel, and C. Telesco
(NASA CP-3098), p. 359
\reference{hec91} Heckman, T.M., Lehnert, M.D., van Breugel, W., \& 
Miley, G.K. 1991, \apj, 370, 78
\reference{hec92} Heckman, T.M., Chambers, K.C., \& Postman, M. 1992, 
\apj, 391, 39
\reference{her92} Herbig, T., \& Readhead, A.C.S. 1992, \apjs, 81, 83
\reference{hil96} Hill, G.J., Goodrich, R.W., \& DePoy, D.L. 1996, \apj,
462, 163
\reference{hin91}Hintzen, P., Romanishin, W., \& Valdes, F. 1991, \apj, 366, 7 
\reference{hol95} Holtzman, J.A., Burrows, C.J., Casertano, S., Hester. J.J.,
Trauger, J.T., Watson, A.M., \& Worthey, G. 1995, \pasp, 107, 1065
\reference{hut94a} Hutchings, J.B., Holtzman, J., Sparks, W.B., Morris, S.C., 
Hanisch, R.J., \& Mo, J. 1994a, \apjl, 429, L1
\reference{hut94b} Hutchings, J.B., Neff, S.G., Weadock, J., Roberts. L,
Ryneveld, S., \& Gower, A.C.  1994b, \aj, 107, 994
\reference{jed87} Jedrzejewski, R.I. 1987, \mnras, 226, 747 
\reference{kuh81} K\"uhr, H., Witzel, A., Pauliny-Toth, I.I.K., \& 
Nauber, U. 1981, \aaps, 45, 367
\reference{lac97} Lacy, M. 1997, in the on-line proceedings of the Sheffield
workshop on ``Jet-cloud interactions in active galaxies'',
http://www.shef.ac.uk/~phys/research/astro/conf/ 
\reference{lac95} Lacy, M., Rawlings, S.,  Eales, S., \& Dunlop, J. S. 
1995, \mnras, 273, 821.
\reference{lai83} Laing, R.A., Riley, J.M., \& Longair, M.S. 1983, 
\mnras, 204, 151
\reference{lai88} Laing, R. 1988, \nat, 331, 149
\reference{lai89} Laing, R. 1989, in Hotspots in Extragalactic Radio Sources,
(Springer-Verlag, Heidelberg), p. 27
\reference{leb96} Le Brun, V., Bergeron, J., Boiss\'e, P., \& Deharveng, J.M.
1997, \aap, in press
\reference{leF92} LeF\`evre, O., \& Hammer, F. 1992, \aap, 254, L29 
\reference{leh92} Lehnert, M.D., Heckman, T.M., Chambers, K.C., \& 
Miley, G.K. 1992, \apj, 393, 68
\reference{lil89} Lilly, S.J. 1989, \apj, 340, 77
\reference{liu91} Liu, R., \& Pooley, G.G. 1991, \mnras, 253, 669
\reference{liu92} Liu, R., Pooley, G.G., \& Riley, J.M. 1992, \mnras, 257, 545
\reference{lon95} Longair, M.S., Best, P.N., \& Rottgering, H.J.A. 1995,
\mnras, 275, L47
\reference{low95} Lowenthal, J.D., Heckman, T.M., Lehnert, M.D., \&
Elias, J.H., 1995, 439, 588
\reference{mas88} Massey, P., Strobel, K., Barnes, J.V., \& 
Anderson, E. 1988, \apj, 328, 315
\reference{mas90} Massey, P. \& Gronwall, C. 1990, \apj, 358, 344
\reference{mcC93} McCarthy, P.J. 1993, \araa, 31, 639
\reference{mcC95} McCarthy, P.J., Spinrad, H., van Breugel, W. 1995,
\apjs, 99, 27
\reference{mcC89} McCarthy, P.J., \& van Breugel, W.J.M. 1989, in Epoch of 
Galaxy Formation, edited by C.S. Frenk, R.S. Ellis, T. Shanks, A.F. Heavens,
and J.A. Peacock (Kluwer, Dordrech), p. 57
\reference{mcC87} McCarthy, P.J., van Breugel, W., Spinrad, H., \& 
Djorgovski, S. 1987, \apjl, 321, L29
\reference{mcL94a} McLeod, K.K., \& Rieke, G.H. 1994a, \apj, 420, 58
\reference{mcL94b} McLeod, K.K., \& Rieke, G.H. 1994b, \apj, 431, 137
\reference{mcL95} McLeod, K.K., \& Rieke, G.H. 1995, \apjl, 454, L77
\reference{mei89} Meisenheimer, K., Roser, H.-J., Hiltner, P.R., Yates, M.G., 
Longair, M.S., Chini, R., \& Perley, R.A. 1989, \aap, 219, 63
\reference{mih95} Mihos, J.C. 1995, \apjl, 438, L75
\reference{mil80} Miley, G.K. 1980, \araa, 18, 165
\reference{mil92} Miley, G.K., Chambers, K.C., van Breugel, W.J.M., \& 
Macchetto, F. 1992, \apjl, 401, L69
\reference{mob93} Mobasher, B., Sharples, R.M., \& Ellis, R.S. 1993,
\mnras, 263, 560
\reference{nef95} Neff, S.G., Roberts, L., \& Hutchings, J.B. 1995, 
\apjs, 99, 349 
\reference{ped89} Pedelty, J.A., Rudnick, L., McCarthy, P.J., \& Spinrad, H.
1989, \aj, 97, 647
\reference{ree89} Rees, M.J. 1989, \mnras, 239, 1P
\reference{rei95} Reid, A., Shone, D.L., Akujor, C.E., Browne, I.W.A.,
Murphy, D.W., Pedelty, J, Rudnick, L. \& Walsh, D. 1995, \aaps, 110, 213
\reference{rid95} Ridgway, S.E. 1995, Ph.D. thesis, University of Hawaii. 
\reference{rid97a} Ridgway, S.E., \& Stockton, A. 1997a, in preparation.
\reference{rid97b} Ridgway, S.E., \& Stockton, A. 1997b, in preparation.
\reference{rig92} Rigler, M.A., Lilly, S.J., Stockton, A., Hammer, F., \& 
LeF\`evre, O. 1992, \apj, 385, 61
\reference{sai95} Saikia, D.J., \& Kulkarni, V.K. 1995, \mnras, 270, 897
\reference{sai87} Saikia, D.J., Salter, C.J., \& Muxlow, T.W.B. 1987, 
\mnras, 224, 911
\reference{sch87} Scheuer, P. 1987, in Superluminal Radio Sources,
edited by J. Zensus and T. Pearson (Cambridge Univ. Press, Cambridge), p. 104
\reference{smi89} Smith, E.P., \& Heckman, T.M. 1989, \apj, 341, 658
\apjl, 361, L41
\reference{spi85} Spinrad, H., Djorgovski, S., Marr, J. \& Aguilar, L. 
1985, \pasp, 97, 932
\reference{sted92} Steidel, C.C. \& Dickinson, M. 1992, \apj, 394, 81
\reference{ste97} Steidel, C.C., Dickinson, M., Meyer, D.M.,
Adelberger, K.L., \& Sembach, K.R. 1997, \apj, 480, in press
\reference{sto90} Stockton, A.  1990. in Dynamics and Interactions of Galaxies,
edited by R.~Wielen (Springer-Verlag, Heidelberg), p. 440
\reference{sto94} Stockton, A., Ridgway, S.E., \& Lilly, S.J. 1994, 
\aj, 108, 414
\reference{sto95} Stockton, A., Kellogg, M., \& Ridgway, S.E. 1995, 
\apjl, 443, L69
\reference{sto96} Stockton, A., Ridgway, S.E., \& Kellogg, M. 1996, 
\aj, 112, 902
\reference{sto97} Stockton, A., \& Ridgway, S.E. 1997, in preparation
\reference{tad94} Tadhunter, C., Shaw, M., Clark, N., \& Morganti, R. 1994,
\aap, 288, L21
\reference{urr95} Urry, C.M. \& Padovani, P. 1995, \pasp, 107, 803
\reference{vanB85} van Breugel, W., Filippenko, A.V., Heckman, T., \& 
Miley, G. 1985, \apj, 293, 83
\reference{vanB92} van Breugel, W.J.M., Fanti, C., Fanti, R., Stanghellini, C.,
Schilizzi, R.T., \& Spenser, R.E. 1992, \aap, 256, 56
\reference{wai92} Wainscoat, R.J., \& Cowie, L.L. 1992, \aj, 103, 332
\reference{wes94} West, M.J. 1994, \mnras, 268, 79
\reference{wil91} Williams, A.G. 1991, in Beams and Jets in Astrophysics,
edited by P.A. Hughes (Cambridge Univ. Press, Cambridge), p. 342
\reference{wil85} Williams, A.G., \& Gull, S.F. 1985, \nat, 313, 34
\reference{wor94} Worrall, D.M., Lawrence, C.R., Pearson, T.J., \& Readhead,
A.C.S. 1994, \apjl, 420, L17
\end{references}
\end{document}


\pagestyle{empty}
\begin{deluxetable}{l c c c c c c c c c}
\tablenum{1}
\footnotesize
\tablewidth{0pt}
\tablecaption{Properties of Sample}
\tablehead{
\colhead{Source} & \colhead{Type} & 
\colhead {RA (J2000)} & \colhead{\phs DEC (J2000)} &
\colhead{ Redshift} & 
\colhead{ E(\bv)\tablenotemark{a}} &
\colhead{$\rm S_{178}(Jy)$} &
\colhead{$\alpha_{2700}$\tablenotemark{b}}&
\colhead{ LAS } &
\colhead{Reference}}

\startdata

3C\,2& Q & 00 06 22.59 & $-$00 04 24.69  & 1.037 & 0.013 & 14.9 
& 0.73 & \phn5\arcsec\phn&  \phn9 \nl
3C\,175.1 & G &07 14 04.62 &\phs 14 36 20.55 & 0.920 & 0.128 & 11.4 &0.93 &
\phn7\arcsec\phn&  \phn6\nl
3C\,196 & Q & 08 13 36.03 & \phs 48 13 02.58& 0.871 & 0.040 & 68.2 &
0.92 & \phn6\arcsec\phn & \phn8 \nl
3C\,212 & Q & 08 58 41.43&\phs 14 09 43.97 & 1.049 & 0.020 & 15.1
& 0.83& 10\arcsec\phn& \phn1 \nl
3C\,217 & G &09 08 50.63 & \phs 37 48 18.19& 0.898 & 0.005 & 11.3 &1.10 
&12\arcsec\phn & \phn7\nl
3C\,237 & G &10 08 00.01&\phs 07 30 16.38& 0.877 & 0.000 & 20.9 &0.93 
& \phn1\farcs3 & 10 \nl
3C\,245 & Q & 10 42 44.58&\phs 12 03 31.18&1.029& 0.005 & 14.4&
0.72&\phn8\arcsec\phn& \phn3 \nl
3C\,280 & G &12 56 57.06&\phs 47 20 19.32 & 0.996 & 0.000 &23.7 &0.85 
&13\arcsec\phn & \phn4 \nl
3C\,289 & G & 13 45 26.21&\phs 49 46 32.13 & 0.967 & 0.000 &12.0 &0.98 &
10\arcsec\phn& \phn5 \nl
3C\,336 &Q &16 24 39.02&\phs 23 45 12.72&0.927& 0.035 &11.5&
0.91&28\arcsec\phn& \phn2 \nl
\tablenotetext{a}{$E(\bv)$ values are derived from Burstein \& Heiles (1984); 
errors in these values are $\sim0.015$. }
\tablenotetext{b}{Spectral index at 2700 MHz, where 
$S_{\nu}\propto\nu^{-\alpha}$.  Values have been taken from Herbig \&
Readhead (1992) for all sources except 3C\,2 and 3C\,237, which were
calculated from data compiled by K\"uhr et al. (1981).}
\tablerefs{Sources for radio maps and LAS values: (1) Akujor et al.\ 1991; (2) Bridle et al.\ 1994;
(3) Laing 1989; (4) Liu \& Pooley 1991; (5) Liu, Pooley, \& Riley 1992;
(6) Neff, Roberts, \& Hutchings 1995; (7) Pedelty et al.\ 1989;
(8) Reid et al.\ 1995; (9) Saikia, Salter, \& Muxlow 1987; 
(10) van Breugel et al.\ 1992.} 
\enddata
\end{deluxetable}
\pagestyle{empty}

\begin{deluxetable}{l l c c l c }
\tablenum{2}
\tablewidth{0pt}
\tablecaption{HST WFPC2 Observations}
\tablehead{
\colhead{Source} & 
\colhead{Filter} & \colhead{Mean $\lambda_{0}$($\rm \AA$)} &
\colhead{$\Delta \lambda_{0}$($\rm \AA$) } & 
\colhead{Exposure} & \colhead{Normalized 1$\sigma$}  \nl
\colhead {} & \colhead {} & \colhead {} &
\colhead {} & \colhead {} &
\colhead {Detection Limit\tablenotemark{a}} }

\startdata

3C\,2  & F675W  & 3298.6 & 437 & 16 $\times$ 1100 s & 1.82 \nl
3C\,175.1 & F622W & 3221.4 & 487 & 16 $\times$ 1000 s & 2.23 \nl 
3C\,196 & F622W & 3305.8 & 500 & \phn8 $\times$ 900 s & 2.18 \nl
3C\,212 & F675W & 3279.2 & 434 & 16 $\times$ 1100 s & 2.08 \nl 
3C\,217 & F622W & 3258.8 & 493 & \phn8 $\times$ 900 s & 1.56 \nl
3C\,237 & F622W & 3295.2 & 498 & 16 $\times$ 1100 s & 2.60 \nl
3C\,245 & F675W & 3311.6 & 438 & 16 $\times$ 1100 s & 2.18 \nl
3C\,280 & F622W & 3098.8 & 446 & \phn8 $\times$ 1100 s & 2.12 \nl
3C\,289 & F622W & 3144.5 & 476 & \phn8 $\times$ 1000 s & 1.98 \nl
3C\,336 & F622W & 3209.7 & 485 & \phn8 $\times$ 900 s & 2.35 \nl
\tablenotetext{a} {The 1$\sigma$ detection limit in our images, normalized 
to $z = 1$ and $E(\bv)=0$,
measured in 10$^{-20}$ ergs cm$^{-2}$s$^{-1}$\AA$^{-1}$
arcsec$^{-2}$. }
\enddata
\end{deluxetable}
\pagestyle{empty}

\begin{deluxetable}{l l l l l c l c}

\small
\tablenum{3}
\tablewidth{0pt}
\tableheadfrac{0.05}
\tablecaption{Log of Ground-Based Imaging Observations}
\tablehead {
\colhead{Source} & 
\colhead{Date} & \colhead{Telescope} &
\colhead{Detector} & \colhead {Pixel Scale } & \colhead{Filter} &
\colhead{Exposure} & \colhead{Data Quality}}
\startdata

3C\,2 & 1992 Nov 26 & \phn\phn UH88 & Tek2048 & \phn\phn\phn0\farcs22 & 
7000/900\phn & \phn10 $\times$
600 s & 0\farcs86 \nl
 & 1993 Sep 21 & \phn\phn UH88 & Tek2048 & \phn\phn\phn0\farcs22 & 8964/1063 &\phn21 $\times$ 300 s& 0\farcs76 \nl
 & 1993 Nov 27 & \phn\phn CFHT & Nic256& \phn\phn\phn0\farcs2& $K^\prime$ & \phn27 $\times$ 180 s &
 0\farcs70 \nl
 & 1993 Nov 28 & \phn\phn CFHT & Nic256& \phn\phn\phn0\farcs2& $K^\prime$ & \phn38 $\times$ 180 s & 0\farcs63 \nl 
 & 1993 Dec 23 & \phn\phn IRTF & InSb256& \phn\phn\phn0\farcs15& $K^\prime$ & \phn53 $\times$ 50 s & 0\farcs67 \nl
 & 1994 Aug 10 & \phn\phn IRTF & InSb256 & \phn\phn\phn0\farcs15 & $H\phm{'}$ & \phn44 $\times$ 90 s & 0\farcs78 \nl
 &1994 Nov 21 & \phn\phn UH88 & Qrc1024 &\phn\phn\phn0\farcs18 & $H\phm{'}$ & \phn23 $\times$ 180 s &0\farcs97 \nl

3C\,175.1 &1993 Nov 28 & \phn\phn CFHT & Nic256 &\phn\phn\phn0\farcs2& $K^\prime$ 
& \phn36 $\times$ 180 s&0\farcs83 \nl
 &1994 Nov 19 & \phn\phn UH88 & Qrc1024 &\phn\phn\phn0\farcs18& $K^\prime$ & \phn13 $\times$ 180 s& 
 1\farcs07 \nl
 &1995 Mar 20 & \phn\phn Keck & InSb256& \phn\phn\phn0\farcs15 &$K^\prime$ & \phn23 $\times$ 120 s&
 0\farcs64 \nl

3C\,196 &1992 Apr 14 & \phn\phn UH88 & Nic256 &\phn\phn\phn0\farcs37& $K^\prime$ &\phn29 $\times$ 90 s& 0\farcs68 \nl
 &1992 May 07 & \phn\phn UH88 & Tek2048& \phn\phn\phn0\farcs22 & 6968/28\phn\phn & \phn\phn2 $\times$ 2400 s& 1\farcs08\nl
 &1992 May 08 & \phn\phn UH88 & Tek2048 & \phn\phn\phn0\farcs22& 6475/900\phn & \phn\phn8 $\times$ 600 & 0\farcs91\nl 
 &1992 May 11 & \phn\phn UH88 & Tek2048 & \phn\phn\phn0\farcs22&6968/28\phn\phn & \phn\phn2 $\times$ 2400 s& 
1\farcs30 \nl 
 &1992 May 10 & \phn\phn CFHT & Nic256& \phn\phn\phn0\farcs3 &$K^\prime$ & \phn30 $\times$ 90 s& 0\farcs71\nl
 &1993 Apr 19 & \phn\phn UH88 & Tek2048&\phn\phn\phn0\farcs22&6475/900\phn& \phn\phn9 $\times$ 600 s& 0\farcs83
\nl
 &1994 Mar 27 & \phn\phn UH88 & Nic256&\phn\phn\phn0\farcs37& $K^\prime$ & \phn51 $\times$ 90 s & 0\farcs97 \nl
 &1994 Apr 05 & \phn\phn UH88 & Tek2048 &\phn\phn\phn0\farcs22& 6120/960\phn & \phn\phn6 $\times$ 600 s&1\farcs15 \nl
 &1995 Feb 24 & \phn\phn CFHT & Orb2048 &\phn\phn\phn0\farcs087& 6983/32\phn\phn &\phn\phn3 $\times$ 2400 s& 0\farcs60 \nl 
 &1995 Mar 19 & \phn\phn Keck & InSb256 &\phn\phn\phn0\farcs15& $K^\prime$ & \phn27 $\times$ 96 s& 0\farcs85\nl

3C\,212&1992 May 11& \phn\phn CFHT& Nic256 &\phn\phn\phn0\farcs3& $K^\prime$ & \phn37 $\times$ 90 s& 0\farcs61\nl
 &1993 Apr 17 & \phn\phn UH88 & Tek2048 &\phn\phn\phn0\farcs22& 7000/900\phn & \phn\phn7 $\times$ 900 s& 0\farcs81  \nl 
 &1993 Dec 23 & \phn\phn IRTF & InSb256 &\phn\phn\phn0\farcs15 & $K^\prime$ & 126 $\times$ 50 s &  0\farcs83 \nl 
 &1994 May 04 & \phn\phn UH88 & Nic256& \phn\phn\phn0\farcs12 & $K^\prime$ & \phn10 $\times$ 360 s &0\farcs88 \nl
 &1995 Jan 17 & \phn\phn UH88 & Qrc1024 &\phn\phn\phn0\farcs06& $K^\prime$& \phn25 $\times$ 300 s &0\farcs53 \nl
 &1995 Feb 24 & \phn\phn CFHT & Orb2048 &\phn\phn\phn0\farcs087& 7646/32\phn\phn & \phn\phn3 $\times$ 2400 s& 0\farcs80\nl
 &1995 Mar 20 & \phn\phn Keck & InSb256 &\phn\phn\phn0\farcs15& $K^\prime$ & \phn27 $\times$ 120 s& 0\farcs67\nl

3C\,217 &1995 Mar 19 & \phn\phn Keck & InSb256 &\phn\phn\phn0\farcs15& $K^\prime$ & \phn18 $\times$ 120 s
 & 0\farcs83\nl

3C\,237 &1994 Nov 22& \phn\phn UH88 & Qrc1024&\phn\phn\phn0\farcs18& $K^\prime$ & \phn14 $\times$ 180 s&0\farcs78 \nl
 &1995 Mar 20 & \phn\phn Keck & InSb256& \phn\phn\phn0\farcs15 & $K^\prime$ & \phn15 $\times$ 120 s&0\farcs85 \nl

3C\,245 &1992 May 11 & \phn\phn CFHT & Nic256 & \phn\phn\phn0\farcs3& $K^\prime$ &\phn18 $\times$ 90 s&0\farcs74 \nl
 &1993 Apr 12& \phn\phn UH88 & Nic256 &\phn\phn\phn0\farcs37& $K^\prime$ & \phn50 $\times$ 90 s& 1\farcs11\nl
 &1993 Apr 16& \phn\phn UH88 & Tek2048&\phn\phn\phn0\farcs22& 6475/900\phn & \phn10 $\times$ 600 s& 0\farcs80 \nl 
 &1994 Apr 06 & \phn\phn UH88 & Tek2048 &\phn\phn\phn0\farcs22& 7571/31\phn\phn & \phn\phn3 $\times$ 1800 s& 0\farcs75 \nl
 &1995 Mar 20 & \phn\phn Keck & InSb256&\phn\phn\phn0\farcs15 & $K^\prime$ & \phn18 $\times$ 120 s& 1\farcs33\nl

3C\,280 &1994 Apr 06 & \phn\phn UH88 &Tek2048 & \phn\phn\phn0\farcs22& 7449/31\phn\phn & \phn\phn3 $\times$ 1800 s&0\farcs73 \nl
&1994 May 04 & \phn\phn UH88 & Nic256 & \phn\phn\phn0\farcs12& $K^\prime$ & \phn23 $\times$ 360 s &0\farcs90 \nl
&1995 Feb 23 & \phn\phn CFHT & Orb2048 & \phn\phn\phn0\farcs087 & 7449/31\phn\phn & \phn\phn4 $\times$ 2700 s&0\farcs66 \nl
&1995 Mar 19 & \phn\phn Keck & InSb256 & \phn\phn\phn0\farcs15 &$K^\prime$ & \phn27 $\times$ 96 s& 0\farcs58\nl

3C\,289 &1994 Apr 06 & \phn\phn UH88 & Tek2048 &\phn\phn\phn0\farcs22 & 7388/29\phn\phn & \phn\phn3 $\times$ 1800 s &0\farcs81 \nl
&1995 Mar 19 & \phn\phn Keck & InSb256 & \phn\phn\phn0\farcs15 &$K^\prime$ & \phn20 $\times$ 120 s&0\farcs78 \nl
&1995 Mar 20 & \phn\phn Keck & InSb256 & \phn\phn\phn0\farcs15 &$K^\prime$ & \phn24 $\times$ 120 s&0\farcs80 \nl

3C\,336 &1992 Apr 13 & \phn\phn UH88 & Nic256 &\phn\phn\phn0\farcs37 & $K^\prime$ & \phn12 $\times$ 90 s&
0\farcs70 \nl    
& 1992 Apr 14& \phn\phn UH88 & Nic256 &\phn\phn\phn0\farcs37& $K^\prime$ & \phn30 $\times$ 90 s&0\farcs75 \nl        
&1992 May 07& \phn\phn UH88 & Tek2048 &\phn\phn\phn0\farcs22& 6475/900\phn & \phn\phn5 $\times$ 600 s&0\farcs94 \nl       
&1992 May 10 & \phn\phn CFHT & Nic256&\phn\phn\phn0\farcs3& $K^\prime$ & \phn33 $\times$ 90 s &0\farcs76 \nl
&1993 Apr 17& \phn\phn UH88 & Tek2048 &\phn\phn\phn0\farcs22& 6475/900\phn & \phn\phn8 $\times$ 600 s& 0\farcs91 \nl
&1994 Apr 05&\phn\phn UH88 & Tek2048 &\phn\phn\phn0\farcs22& 7192/28\phn\phn  & \phn\phn2 $\times$ 2100 s& 1\farcs03\nl
&1994 May 02 & \phn\phn IRTF & InSb256 &\phn\phn\phn0\farcs31& $K^\prime$ &144 $\times$ 36 s&
1\farcs30 \nl
&1994 May 03 & \phn\phn UH88 & Nic256& \phn\phn\phn0\farcs12 & $K^\prime$& \phn\phn9 $\times$ 360 s &0\farcs72 \nl
&1995 Feb 24& \phn\phn CFHT & Orb2048 & \phn\phn\phn0\farcs087 &7192/28\phn\phn & \phn\phn3 $\times$ 2100 s&0\farcs64 \nl
&1995 Mar 19 & \phn\phn Keck & InSb256 & \phn\phn\phn0\farcs15& $K^\prime$ & \phn27 $\times$ 96 s& 0\farcs49\nl
\enddata
\end{deluxetable}

\pagestyle{empty}

\begin{deluxetable}{l c c l c c }
\tablenum{4}
\tablewidth{0pt}
\tablecaption{PSF Star Observations}
\tablehead{
\colhead{Quasar Name} &
\colhead{Star Name} & \colhead{Filter} &
\colhead{Integration} &
\colhead{QSO Saturation} & \colhead{PSF Saturation} \nl
\colhead{} & \colhead{} & \colhead{} & \colhead {Time} &
\colhead{Diameter} & \colhead{Diameter}}

\startdata
3C\,2 & GSC4663-00565 & F675W & 12$\times$4 s & 0\farcs00 & 0\farcs25\nl
3C\,196 & GSC3411-02379 & F622W & 12$\times$4 s & 0\farcs20 & 0\farcs05 \nl 
3C\,212 & GSC0817-02062 & F675W & 12$\times$16 s & 0\farcs00 & 0\farcs05 \nl
3C\,245 & GSC0842-00459 & F675W & 12$\times$3 s & 0\farcs25 & 0\farcs00 \nl
3C\,336 & \nodata & F622W & 12$\times$50 s & 0\farcs20 & 0\farcs10 \nl
\enddata
\end{deluxetable}
\pagestyle{empty}
\begin{deluxetable}{l c c c l c c}
\footnotesize
\tablenum{5}
\tablewidth{0pt}
\tablecaption{Optical and Near-Infrared Component Photometry}
\tablehead{
\colhead{Source} &
\colhead{Component} &
\colhead{Aperture Type\tablenotemark{a}} &
\colhead{Aperture Area\tablenotemark{b}} &
\colhead{$F_\lambda$(0.33$\mu$m)\tablenotemark{c}} &
\colhead{$F_\lambda$(1$\mu$m)\tablenotemark{d}} &
\colhead{$\alpha$(opt)\tablenotemark{e} }}

\startdata
3C\,175.1 & $a$ & I(2.7) & \phn0.77 & \phn\phn\phn47.5$\;\pm\;$2.3\phn\phn & \nodata & \nodata\nl
& Core & C(4.5) & 63.6\phn & \phm{000}80.0$\;\pm\;$8 & $<$150 & \nodata \nl
3C\,196 & QSO & C(5) & 80.1\phn &17649.0$\;\pm\;$350 & 4285.0$\;\pm\;$70\phm{0.} & 0.84 \nl
 & $a$ & R(3.2,1.65) & \phn5.3\phn & \phn\phn276.2$\;\pm\;$5 &\phn115.4$\;\pm\;$6\phm{00.}& 1.29 \nl 
 & $b$ & C(0.25) &\phn0.20 & \phn\phn\phn40.4$\;\pm\;$2 & \nodata & \nodata \nl
 & $d$ & C(1.0) & \phn3.1\phn & \phn\phn107.0$\;\pm\;$10 &\phn\phn\phn9.8$\;\pm\;$2\phm{0.0}& \nodata \nl
 & $e$ & C(1.0) & \phn3.1\phn & \phn\phn\phn73.6$\;\pm\;$8 & \phn\phn\phn8.7$\;\pm\;$1.4\phm{0}& \nodata \nl
3C\,212 & QSO & C(4) &50.3\phn & \phn4726.0$\;\pm\;$60 & 2365.0$\;\pm\;$30\phm{0.} & 1.39 \nl 
& $a$ & C(0.25)& \phn0.20 & \phm{000}11.4$\;\pm\;$1.9&\nodata & \nodata \nl
& $b$ & C(0.35)& \phn0.38 & \phm{000}20.1$\;\pm\;$3.5&\nodata & \nodata \nl
& $c$ & C(0.30)& \phn0.28 & \phm{0000}8.2$\;\pm\;$2.1&\nodata & \nodata \nl
& $b + c$ & R(1.55,1.7) & 26.4\phn & \phm{000}48.5$\;\pm\;$1.8 & \phm{00}25.0$\;\pm\;$6\phm{0.0} & 1.42 \nl
& $d$ & R(2.05,1.2) & \phn2.46 & \phm{000}39.6$\;\pm\;$3.5 & \phm{00}18.8$\;\pm\;$2.7\phm{0}&1.35 \nl
& $e$ & C(0.45)& \phn0.64 & \phm{0000}8.9$\;\pm\;$1.8&\phm{00}12.1$\;\pm\;$1.8\phm{0}&2.27 \nl
& $f$ & R(2.65,3.6) & \phn9.54 & \phm{00}240.7$\;\pm\;$10.5 & \phn146.6$\;\pm\;$20.9& 1.57 \nl
& $g$ & C(1.5) &\phn7.1\phn&\phm{000}37.6$\;\pm\;$3 &\phm{00}71.2$\;\pm\;$4.2\phm{0}& 2.56\nl
& $h$ & I(0.4)\tablenotemark{f} & \phn2.5\phn & \phm{000}40.7$\;\pm\;$2& \phm{000}8.6$\;\pm\;$0.9\phm{0}&0.64 \nl 
& Extension&A(1,2.65)&18.9\phn&\phm{00}182.2$\;\pm\;$10 &\phn128.6$\;\pm\;$20\phm{0.}&1.69 \nl

3C\,217 & $a$ & C(1.0) & \phn3.1\phn & \phm{000}16.1$\;\pm\;$1.3& \phm{00}52.7$\;\pm\;$4.3\phm{0}&2.97 \nl
     & $c$ & C(1.0) & \phn3.1\phn & \phm{000}22.2$\;\pm\;$6.5 & \phm{000}2.6$\;\pm\;$1.9\phm{0}&0.25 \nl
& Core & C(4.5) & 63.6\phn & \phm{000}70.2$\;\pm\;$5 & $<$168 & \nodata \nl

3C\,237 & East & R(0.95,0.85) & \phn0.81 & \phm{00}187.8$\;\pm\;$1.2 & \nodata & \nodata \nl
& West & R(0.95,0.9) & \phn0.86 & \phm{00}213.0$\;\pm\;$1.3 & \nodata & \nodata \nl
& Core & C(4.5) & 63.6\phn & \phm{00}162.1$\;\pm\;$4.3 & $<$199 & \nodata \nl

3C\,245 & QSO & C(4) & 50.3\phn & 21228.0$\;\pm\;$140 & 3468.0$\;\pm\;$200\phm{.}& 0.41 \nl
& $a$ & C(0.25) &\phn0.20 &\phm{000}22.3$\;\pm\;$1.1 & \nodata & \nodata \nl
& $b$ & C(0.25) &\phn0.20 &\phm{0000}8.7$\;\pm\;$1.1 & \nodata & \nodata \nl
& $a+b$ (jet) & I(3.6) &\phn0.42& \phm{000}46.1$\;\pm\;$1.9 & \nodata & \nodata \nl
& $c$ & I(3.6) & \phn0.22 & \phm{000}10.9$\;\pm\;$1 & \nodata  & \nodata \nl
& $d$ & I(3.6) & \phn0.79 & \phm{000}38.2$\;\pm\;$2.8 & \nodata & \nodata \nl
& $e$ & C(3.5) & 15.9\phn & \phm{00}423.2$\;\pm\;$6 & \phn879.0$\;\pm\;$47\phm{0.} & 2.64 \nl

3C\,280 & $a + b$ &I(5.2)& \phn0.56 & \phm{00}197.2$\;\pm\;$2 & \nodata & \nodata \nl
& $c$ &I(5.2) & \phn0.56 &  \phm{000}96.0$\;\pm\;$2 & \nodata  & \nodata \nl
& $d$ &I(5.2) & \phn0.78 &  \phm{000}96.7$\;\pm\;$2 & \nodata & \nodata \nl
& Elliptical & C(4) & 50.3\phn & \phm{00}191.0$\;\pm\;$40 &\phn608.0$\;\pm\;$30\phm{0.} & 2.95 \nl 
& Core & C(4.5) & 63.6\phn & \phm{00}164.3$\;\pm\;$5 & $<$69 & \nodata \nl
3C\,289 & $a$ & C(0.35) & \phn0.38 & \phm{0000}9.6$\;\pm\;$2 & \nodata & \nodata \nl
&$b$ & C(0.35) & \phn0.38 & \phm{0000}7.4$\;\pm\;$2 & \nodata & \nodata \nl
&$c$ & I(5.2) & \phn0.42 & \phm{000}17.2$\;\pm\;$1.6 &\nodata & \nodata \nl
&$d$ & C(1) & \phn3.1\phn&\phm{000}11.8$\;\pm\;$3& \phm{00}22.6$\;\pm\;$5.4\phm{0} & 2.53 \nl
& Elliptical & C(3.2) &32.2\phn & \phm{00}259.0$\;\pm\;$31 & \phn694.0$\;\pm\;$60\phm{0.} & 2.89 \nl
& Core & C(4.5) & 63.6\phn & \phm{00}117.1$\;\pm\;$8 & $<$140 & \nodata \nl
3C 336 & QSO & C(4) & 50.3\phn & 20705.0$\;\pm\;$414 & 2216.0$\;\pm\;$60\phm{0.} & 0.17 \nl
& $d$ & R(1.4,2.95)&\phn4.1\phn &\phm{000}79.6$\;\pm\;$9.6 &\phm{00}63.7$\;\pm\;$9.6\phm{0} & 1.82 \nl
& $a$ & R(1.15,0.75) & \phn0.86 & \phm{000}44.1$\;\pm\;$4.8 & \phm{00}10.4$\;\pm\;$1.4\phm{0} & 0.82 \nl
& $b$ & C(0.6) & \phn1.1\phn & \phm{000}26.3$\;\pm\;$3.6 &\phm{00}11.4$\;\pm\;$1.2\phm{0} & 1.32 \nl
& $c$ & C(0.6) & \phn1.1\phn & \phm{000}36.1$\;\pm\;$4.8 & \phm{00}13.2$\;\pm\;$1.8\phm{0} & 1.18 \nl

\tablenotetext{a} {Aperture type; $i.e.$, C($r$) is circular with radius $r$;
A($r1$,$r2$) is annular with inner radius $r1$ and outer radius $r2$; R($x$,
$y$) is rectangular with dimensions $x$ and $y$;
and I($\mu$) is an isophotal aperture with surface brightness cutoff level
$\mu$; 
$r$,$r1$,$r2$,$x$, and $y$
are all in arcseconds and $\mu$ is in
10$^{-19}$ ergs cm$^{-2}$s$^{-1}$\AA$^{-1}$ arcsec$^{-2}$,
at the wavelength of the HST image. This value is normalized by the 
correction to $z = 1$ factor.}
\tablenotetext{b} {Aperture areas in square arcseconds.}
\tablenotetext{c} {Background subtracted flux densities from the HST image,
in units of $10^{-20}$ ergs cm$^{-2}$ s$^{-1}$ \AA$^{-1}$.
The HST rest wavelength 
for each object is given in Table 2.  
Flux densities have been corrected for galactic extinction,
but not normalized to $z = 1$.} 
\tablenotetext{d} {Background subtracted flux densities from the $K'$ image,
in units of $10^{-20}$ ergs cm$^{-2}$ s$^{-1}$ \AA$^{-1}$.
Fluxes have not been normalized to $z = 1$.}
\tablenotetext{e}{Spectral index from 0.33 $\mu$m
to 1 $\mu$m; $F_\nu \propto \nu^{-\alpha}$.}
\tablenotetext{f}{Cutoff level is in Gaussian-smoothed HST image.}

\enddata
\end{deluxetable}
\pagestyle{empty}

\begin{deluxetable}{l c c c r r r r r}

\tablenum{6}
\small
\tablewidth{0pt}
\tablecaption{3C\,2 Component Photometry}
\tablehead{
\colhead {} & \colhead {} & \colhead {} &
\colhead {} & \colhead {} &
\multicolumn{2}{c}{Subtracted\tablenotemark{c}} &\multicolumn{2}{c}
{Unsubtracted\tablenotemark{d}}\nl 
\colhead{Component} &
\colhead{$\Delta$RA} & \colhead{$\Delta$DEC} &
\colhead{Aperture\tablenotemark{a}} &
\colhead{$\lambda_{0}$($\rm \AA$)\tablenotemark{b}} &
\colhead{$F_\lambda$} & \colhead{$\sigma$($F_\lambda$)} &
\colhead{$F_\lambda$} & \colhead{$\sigma$($F_\lambda$)} }

\startdata

$a$ (NE ``lobe'') & \phs0\farcs17&\phs 0\farcs80 & C(0.75)  &
3299 & 101.3 &   5.\phn & 125.0 &   5.\phn \nl
& & & & 4401 &  60.2 &  12.\phn &  75.7 &  12.\phn \nl
& & & & 8100 &  57.5 &   9.\phn &  90.3 &   9.\phn \nl
& & & &10309 &  48.4 &   5.\phn &  77.6 &   5.\phn \nl
$b$ (NW Extension) &$-$0\farcs53 &\phs 0\farcs13 &R(0.75,0.75) &
3299 & 15.6&  2.7& 29.2&  2.7\nl
& & & & 4401 &  6.7&  3.6& 12.8&  3.6\nl
& & & & 8100&  3.1&  6.4& 17.3&  6.4\nl
& & & & 10309 & 10.7&  2.7& 24.6&  2.7\nl
$c$ (SE Companion) &\phs 0\farcs80 &$-$1\farcs06 & C(0.75)&
3299 & 21.0&  2.7& 30.1&  1.8\nl
& & & &4401 & 45.6&  9.1& 85.8&  9.1\nl
& & & &8100 & 57.5&  6.4& 80.3&  6.4\nl
& & & &10309 & 55.7&  4.6& 94.9&  4.6\nl
Symmetric Nebulosity & \nodata&\nodata & A(1.75,3.0) &
3299 & 76.6&  6.4& \nodata & \nodata \nl
& & & &4401 & 95.8& 31.9 & \nodata & \nodata \nl
& & & &8100 &147.8& 41.1& \nodata & \nodata \nl
& & & & 10309 &104.9&  7.3 & \nodata & \nodata \nl
QSO (without extensions)  &0\farcs00 &0\farcs00 & C(4) & 3299 &1579.4 &18.\phn & \nodata &\nodata \nl
&&&&4401 &1367.0 &  59.\phn &\nodata &\nodata \nl
&& & &8100 & 609.5&  46.\phn&\nodata &\nodata \nl
&& & &10309 & 514.0&  27.\phn&\nodata & \nodata \nl

\tablenotetext{a} {Aperture type; $i.e.$, C($r$) is circular with radius $r$;
A($r1$,$r2$) is annular with inner radius $r1$ and outer radius $r2$; R($x$,
$y$) is rectangular with dimensions $x$ and $y$;
and $r$,$r1$,$r2$,$x$, and $y$
are all in arcseconds.}
\tablenotetext{b}{This is the rest wavelength in \AA, assuming each 
component has the same redshift as the narrow-line 3C\,2 redshift of 1.037.}
\tablenotetext{c} {For the components $a$, $b$, and $c$, these fluxes
have had the
best estimate of the 
local background due to the nebulosity subtracted; see the text for details.
For the QSO itself, all extensions have been subtracted. 
Flux densities are in units of $10^{-20}$ ergs cm$^{-2}$ s$^{-1}$ \AA$^{-1}$
and have been corrected for galactic extinction.}
\tablenotetext{d} {
These fluxes have had no local background subtracted
(other than standard sky subtraction).
Flux densities are in units of $10^{-20}$ ergs cm$^{-2}$ s$^{-1}$ \AA$^{-1}$
and have been corrected for galactic extinction.}
\enddata
\end{deluxetable}
\pagestyle{empty}
\begin{deluxetable}{l c c c c l }

\small
\tablenum{7}
\tablewidth{0pt}
\tablecaption{Quasar Optical Photometry}
\tablehead{
\colhead{Source\tablenotemark{a}} &
\colhead{Aperture\tablenotemark{b}} &
\colhead{Normalized Cutoff\tablenotemark{c}} &
\colhead{$F_\lambda$(mon)\tablenotemark{d}} &
\colhead{$F_\lambda$($\chi^2$ min)\tablenotemark{e}} &
\colhead{Notes\tablenotemark{f}}}

\startdata

3C\,2 & A(0.3,3.75) & I(4.56) & 243.1 & 230.2 & $c$ included; $\Delta F=23$ \nl
3C\,2 & A(0.3,3.75) &  None  & 484.8 & 469.3 &  $c$ included; $\Delta F=21$ \nl
3C\,2 (int) & A(0,3.75) & I(4.56) & 275.5  & 253.8 & $c$ included; $\Delta F=23$ \nl
3C\,2 (int) & A(0,3.75) & None & 528.7 & 505.2 & $c$ included; $\Delta F=21$ \nl
3C\,196 & A(0.45,3.75) & I(4.56) & 457.2 & 352.1 & foreground $c$ excluded \nl
3C\,196 & A(0.45,3.75) & None & 578.2 &  504.7& foreground $c$ excluded \nl
3C\,196 (int) & A(0,3.75) & I(4.56) & 535.9 & 427.3 & foreground $c$ excluded \nl
3C\,196 (int) & A(0,3.75) & None & 699.5 & 579.9 & foreground $c$ excluded \nl
3C\,212 & A(0.45,3.75) & I(4.56) & 108.7 & \nodata & $h$ included; $\Delta F=5$ \nl
3C\,212 & A(0.45,3.75) & None & 338.2 & \nodata & $h$ included; $\Delta F=21$ \nl
3C\,212 (int) & A(0,3.75) & I(4.56) & 136.4 & (\phn86.9)\tablenotemark{g} & $h$
included; $\Delta F=5$ \nl
3C\,212 (int) & A(0,3.75) & None & 371.1 &(362.2)\tablenotemark{g}& $h$ included; $\Delta F=21$ \nl
3C\,245 & A(0.45,3.75) & I(4.56) & 138.6 & (\phn88.9)\tablenotemark{h} &
 $e$ excluded \nl
3C\,245 & A(0.45,3.75) & None  & 325.6 & (272.1)\tablenotemark{h} &
 $e$ excluded \nl
3C\,245 (int) & A(0,3.75) & I(4.56) & 202.4 & \nodata & $e$ excluded\nl
3C 245 (int) & A(0,3.75) & None & 389.4 & (313.5)\tablenotemark{f} & $e$ excluded \nl
3C\,336 & A(0.55,3.75) & I(4.56) & \phn84.6 & \phn43.9 & foreground $d$
excluded\nl
3C\,336 & A(0.55,3.75) & None & 263.5 & 140.6 & foreground $d$
excluded\nl
3C\,336 (int) & A(0.55,3.75) & None & 291.7 & \nodata &foreground $d$ 
excluded\nl

\tablenotetext{a} {Quasar name; if followed by (int), this indicates that
the central region of the quasar has been interpolated with a low-order 
polynomial fit.}
\tablenotetext{b} {Aperture size; 
A($r1$,$r2$) signifies annular aperture with 
inner radius $r1$ and outer radius $r2$, where $r1$ and $r2$ are in arcseconds.}
\tablenotetext{c}{Isophotal cutoff level; 
I($\mu$) signifies a surface brightness cutoff level of $\mu$ in 
10$^{-19}$ ergs cm$^{-2}$s$^{-1}$\AA$^{-1}$ arcsec$^{-2}$,
at the wavelength of the HST image. This surface brightness 
value has been normalized to
$z = 1$.}
\tablenotetext{d} {Flux densities from the monotonic PSF subtraction limit,
in units of $10^{-20}$ ergs cm$^{-2}$ s$^{-1}$ \AA$^{-1}$.
Flux densities have been corrected for galactic extinction, and normalized to $z = 1$.} 
\tablenotetext{e} {Flux densities from the $\chi^2$ minimum
PSF subtraction limit,
in units of $10^{-20}$ ergs cm$^{-2}$ s$^{-1}$ \AA$^{-1}$.
Flux densities have been corrected for galactic extinction, and
normalized to $z = 1$.
Where the $\chi^2$ minimum and the monotonic subtraction limit agree, this 
value is placed in the monotonic-limit column only. Values in parentheses
are ``lower limit'' flux densities calculated by another method, and are
placed here for convenience. See notes associated with individual values.}
\tablenotetext{f}{We note here whether nearby companions were included
or not. We included all companions except the foreground spiral $c$ in 
the field 3C\,196, the elliptical $e$ in 3C\,245, and the foreground $d$
in the field of 3C\,336. For those objects in which 
we included a nearby (perhaps discrete) companion, we give the change
in flux density that would result from subtracting that companion's 
contribution.}
\tablenotetext{g}{This lower-limit value was derived by replacing the central
1\farcs25 $\times$ 0\farcs95 region with a constant; this excluded high
surface brightness regions immediately exterior to the PSF subtraction annulus.}
\tablenotetext{h}{This lower-limit value was derived by excluding the central
0\farcs65 instead of the standard 0\farcs45;
this certainly excludes real flux,
but also excludes PSF residuals near the edge of the PSF subtraction region.}

\enddata
\end{deluxetable}
\pagestyle{empty}
\begin{deluxetable}{l c c  c c }

\tablenum{8}
\tablewidth{0pt}
\tablecaption{Radio Galaxy Optical Photometry}
\tablehead{
\colhead{Source} &
\colhead{Aperture\tablenotemark{a}} &
\colhead{$F_\lambda$(isophotal)\tablenotemark{b}} &
\colhead{$F_\lambda$(total)\tablenotemark{c}} &
\colhead{Notes}}

\startdata
3C\,175.1 & A(0,3.0) & 154.5 & 355.3 &\tablenotemark{d}\nl
& A(0.45,3.0) & \phn38.1 & 236.4 &\tablenotemark{d}\nl
3C\,217 & A(0,3.75)& 211.9 & 373.7 & \nl
& A(0.45,3.75) & \phn83.9 & 244.8 & \nl
& A(0,3.75) & 191.9 & 343.1 & \tablenotemark{e} \nl  
& A(0.45,3.75) & \phn63.8 & 215.0  & \tablenotemark{e} \nl  
3C\,237 & A(0,3.0) & 472.5 & 607.2 & \nl
& A(0.45,3.0) & 183.6 & 316.5 & \nl
3C\,280 & A(0,6.0) & 581.9 & 915.9 & \tablenotemark{f} \nl
& A(0,3.75) & 513.5 & 682.0 & \tablenotemark{f,}\ \tablenotemark{g}\nl
& A(0.45,3.75) & 280.3 & 494.7 & \tablenotemark{f,}\ \tablenotemark{g}\nl
3C\,289 & A(0,3.75) & 169.8 & 391.6 & \nl
& A(0,3.75) & 160.8 & 374.2 & \tablenotemark{h} \nl 
& A(0.45,3.75) & \phn36.1 & 247.4 & \tablenotemark{h}\nl

\tablenotetext{a} {Aperture size; 
A($r1$,$r2$) signifies annular aperture with 
inner radius $r1$ and outer radius $r2$, where $r1$ and $r2$ are in arcseconds.}
\tablenotetext{b}{Flux densities in units of $10^{-20}$ ergs cm$^{-2}$ s$^{-1}$
\AA$^{-1}$, above an isophotal cutoff level 
of 4.56 $\times$ 10$^{-19}$ ergs cm$^{-2}$ s$^{-1}$ \AA$^{-1}$ arcsec$^{-2}$,
at the wavelength of the HST image, normalized to E(\bv)=0,
$z = 1$. }
\tablenotetext{c} {Total flux densities in aperture, with no isophotal cutoff,
in units of $10^{-20}$ ergs cm$^{-2}$ s$^{-1}$ \AA$^{-1}$. 
Flux densities have been corrected for galactic extinction, and normalized to $z = 1$.} 
\tablenotetext{d} {Foreground star removed.}
\tablenotetext{e} {Components $a$ and $c$ excluded.}
\tablenotetext{f} {Foreground galaxy removed.}
\tablenotetext{g} {Companions, eastern lobe region excluded.}
\tablenotetext{h} {Component $d$ excluded.}

\enddata
\end{deluxetable}
\pagestyle{empty}
\begin{deluxetable}{l c c c c l }

\small
\tablenum{9}
\tablewidth{0pt}
\tablecaption{Quasar Near-Infrared Photometry}
\tablehead{
\colhead{Source\tablenotemark{a}} &
\colhead{Aperture\tablenotemark{b}} &
\colhead{Normalized Cutoff\tablenotemark{c}} &
\colhead{$F_\lambda$(mon)\tablenotemark{d}} &
\colhead{$F_\lambda$($\chi^2$ min)\tablenotemark{e}} &
\colhead{Notes\tablenotemark{f}}}

\startdata

3C\,2 & A(0.4,3.75) & I(1.40) & 427.3 & 391.3 & $c$ included; $\Delta F=39$ \nl
3C\,2 & A(0.4,3.75) &  None  & 515.6 & 470.1 &  $c$ included; $\Delta F=39$ \nl
3C\,2 (int) & A(0,3.75) & None & 554.1 & \nodata & $c$ included; $\Delta F=39$ \nl
3C\,2 (int) & A(0,3.75) & I(1.40) & 465.8 & \nodata & $c$ included; $\Delta F=39$ \nl
3C\,196 & A(0.9,3.75) & I(1.40) & \phn73.5 & \phn67.8 & foreground $c$ excluded \nl
3C\,196 & A(0.9,3.75) & None & 131.8 &  123.9 & foreground $c$ excluded \nl
3C\,196 (int) & A(0,3.75) & I(1.40) & \phn75.1 & \phn67.8 & foreground $c$ excluded \nl
3C\,196 (int) & A(0,3.75) & None & 148.8 & 142.3 & foreground $c$ excluded \nl
3C\,212 & A(0.8,3.75) & I(1.40) & 143.6 & 130.0 & $h$ included; $\Delta F=9$ \nl
3C\,212 & A(0.8,3.75) & None & 278.7 & 265.7 & $h$ included; $\Delta F=9$ \nl
3C\,336 & A(0.85,3.75) & I(1.40) & \phn26.4 & \phn25.4 & foreground $d$ excluded\nl
3C\,336 & A(0.85,3.75) & None & \phn84.5 & \phn82.1 & foreground $d$ excluded\nl
3C\,336 (int) & A(0,3.75) & None & \phn93.7 & \nodata&foreground $d$ excluded\nl
\tablenotetext{a} {Quasar name; if followed by (int), this indicates that
the central region of the quasar has been interpolated with a low-order 
polynomial fit.}
\tablenotetext{b} {Aperture size; 
A($r1$,$r2$) signifies annular aperture with 
inner radius $r1$ and outer radius $r2$, where $r1$ and $r2$ are in arcseconds.}
\tablenotetext{c}{Isophotal cutoff level; 
I($\mu$) signifies a surface brightness cutoff level of $\mu$ in 
10$^{-19}$ ergs cm$^{-2}$s$^{-1}$\AA$^{-1}$ arcsec$^{-2}$.
This surface brightness value has been normalized to $z = 1$.}
\tablenotetext{d} {Flux densities from the monotonic PSF subtraction limit,
in units of $10^{-20}$ ergs cm$^{-2}$ s$^{-1}$ \AA$^{-1}$.
Flux densities have been corrected for galactic extinction, and normalized to $z = 1$.} 
\tablenotetext{e} {Flux densities from the $\chi^2$ minimum
PSF subtraction limit,
in units of $10^{-20}$ ergs cm$^{-2}$ s$^{-1}$ \AA$^{-1}$.
Flux densities have been corrected for galactic extinction, and
normalized to $z = 1$.}
\tablenotetext{f}{We note here whether nearby companions were included
or not. We included all companions except the foreground spiral $c$ in 
the field 3C\,196 and the foreground $d$ in the field of 3C\,336.
For those objects in which 
we included a nearby (perhaps discrete) companion, we give the change
in flux density that would result from subtracting that companion's 
contribution.}

\enddata
\end{deluxetable}
\pagestyle{empty}
\begin{deluxetable}{l c c  c c }

\tablenum{10}
\tablewidth{0pt}
\tablecaption{Radio Galaxy Near-Infrared Photometry}
\tablehead{
\colhead{Source} &
\colhead{Aperture\tablenotemark{a}} &
\colhead{$F_\lambda$(isophotal)\tablenotemark{b}} &
\colhead{$F_\lambda$(total)\tablenotemark{c}} &
\colhead{Notes}}

\startdata
3C\,175.1 & A(0,3.5) & 325.7  & 447.0 &\tablenotemark{d}\nl
& A(0.85,3.5) & 128.3 & 249.6 &\tablenotemark{d}\nl
3C\,217 & A(0,3.75)& 177.5 & 253.1 & \nl
& A(0.85,3.75) & \phn66.2 & 141.5 & \nl
& A(0,3.75) & 136.6 & 212.0 & \tablenotemark{e} \nl  
& A(0.85,3.75) & \phn25.2 & 100.9  & \tablenotemark{e} \nl  
3C\,237 & A(0,3.75) & 294.9 & 342.0 & \nl
& A(0.85,3.75) & \phn99.2 & 146.0 & \nl
3C\,280 & A(0,3.75) & 598.5 & 711.2 & \tablenotemark{f} \nl
& A(0,3.75) & 493.6 & 605.9 & \tablenotemark{f,}\ \tablenotemark{g}\nl
& A(0.85,3.75) & 272.0 & 384.7 & \tablenotemark{f} \nl
& A(0.85,3.75) & 167.1 & 279.4 & \tablenotemark{f,}\ \tablenotemark{g}\nl
3C\,289 & A(0,3.75) & 569.0 & 735.0 & \nl
& A(0.85,3.75) & 422.4 & 589.0 & \nl

\tablenotetext{a} {Aperture size; 
A($r1$,$r2$) signifies annular aperture with 
inner radius $r1$ and outer radius $r2$, where $r1$ and $r2$ are in arcseconds.}
\tablenotetext{b}{Flux densities in units of $10^{-20}$ ergs cm$^{-2}$ s$^{-1}$
\AA$^{-1}$, above an isophotal cutoff level 
of 4.56 $\times$ 10$^{-19}$ ergs cm$^{-2}$s$^{-1}$\AA$^{-1}$ arcsec$^{-2}$,
normalized to E(\bv)=0, $z = 1$. }
\tablenotetext{c} {Total flux densities in aperture, with no isophotal cutoff,
in units of $10^{-20}$ ergs cm$^{-2}$ s$^{-1}$ \AA$^{-1}$.
Flux densities have been corrected for galactic extinction and normalized to $z = 1$.} 
\tablenotetext{d} {Foreground star removed. Companions to NE included; 
their contribution to $F_\lambda$(total)=30, and to $F_\lambda$(isophotal)=6. }
\tablenotetext{e} {Components $a$ and $c$ excluded.}
\tablenotetext{f} {Foreground galaxy removed.}
\tablenotetext{g} {Companions, eastern lobe region excluded.}

\enddata
\end{deluxetable}